\documentclass[12pt,twoside,final]{thesis}

\usepackage{epsfig, mydefs, deluxetable,graphicx,lscape} 
\usepackage{natbib}
\usepackage{amsmath}
\usepackage{aastex_hack}

\bibstyle{aa}

\begin{document}

\title{The Mass-Loss Return From Asymptotic Giant Branch Stars to The Large Magellanic Cloud Using Data From The SAGE Survey}
\author{Sundar Srinivasan}
\doctorphilosophy
\dissertation 
\copyrightnotice
\degreemonth{October}
\degreeyear{2009}
\maketitle

\begin{frontmatter}

\begin{abstract}
\dssp
The asymptotic giant branch (AGB) phase is the penultimate stage of evolution for low- and intermediate-mass stars. Nucleosynthesis products transported from the helium-fusing shell to the outer, cooler regions form gas molecules and dust grains whose chemistry depends on whether oxygen or carbon is more abundant on the surface. Radiation pressure causes the oxygen- or carbon-rich dust to flow outward, dragging the gas along.  Such outflows inject a significant amount of material into the interstellar medium (ISM), seeding new star formation. AGB mass loss is thus a crucial component of galactic chemical evolution. The Large Magellanic Cloud (LMC) is an excellent site for AGB studies. Over 40,000 AGB candidates have been identified using photometric data from the Spitzer Space Telescope {\it Surveying The Agents of a Galaxy's Evolution} (SAGE) mid-infrared (MIR) survey, including about 35,000 oxygen-rich, 7000 carbon-rich and 1400 ``extreme" sources. For the first time, SAGE photometry reveals two distinct populations of O--rich sources in the LMC: a faint population that gradually evolves into C--rich stars and a bright, massive population that circumvents this evolution, remaining O--rich.

This work aims to quantify the mass-loss return from AGB stars to the LMC, a rough estimate for which is derived from the amount of MIR dust emission in excess of that from starlight. I show that this excess flux is a good proxy for the mass-loss rate, and I calculate the total AGB injection rate to be (5.9-13) x 10$^{-3}$ \msunperyr. A more accurate determination requires detailed dust radiative transfer (RT) modeling. For this purpose, I present a grid of C--rich AGB models generated by the RT code 2DUST, spanning a range of effective temperatures, gravities, dust shell radii and optical depths as well as a baseline set of dust properties obtained by modeling a carbon star, data for which was acquired as part of the spectroscopic follow-up to SAGE. AGB stars are the best laboratories for dust studies, and the development of a model grid will reinforce future research in this field.

Advisor: Margaret Meixner.
\end{abstract}
\tableofcontents
\listoftables
\listoffigures
\begin{dedication}
{\em \Large \begin{center}
For Dayanand Pujari and Padmini Iyer\\ 
\end{center}}
\end{dedication}
\end{frontmatter}

\chapter{Introduction} 
\label{ch:intro}
\section{On the Asymptotic Giant Branch}
Low- and intermediate-mass stars (0.8 -- 8 \msun) spend most of their life in the core hydrogen-burning phase (the so-called Main Sequence). This is followed by exhaustion of the core hydrogen accompanied by shell H-burning (the ``Red Giant" phase) and a brief core He-burning stage (the ``Horizontal Branch"). The star is now left with an inert C/O core surrounded by a He-burning shell and a dormant hydrogen shell, and it has entered the asymptotic giant branch (AGB). The He-burning stage (``early-AGB" or E-AGB) lasts for $\sim10^7$ yr \citep{VW93} and eventually reignites the H shell. From this point onward, the star's energy output is due to long periods of quiescent shell H-burning punctuated by brief, violent flashes of shell He-burning (``thermal pulses") that extinguish the H layer by pushing it to outer, cooler regions. The star is now on the thermally-pulsing AGB or TP-AGB. Although brief \citep[$\sim5\times10^{4}$ yr,][]{VW93}, this is a very important chapter in the star's life, and is responsible for a variety of amazing phenomena such as thermal pulsations on timescales of hundreds of thousands of years, stellar pulsations on scales of up to hundreds of days, and the mixing of nuclear-processed material to the outer regions of the star.

AGB star radii are typically a few hundred \rsun. The low surface gravity makes it easy for stellar pulsations to levitate material to cooler regions leading to the formation of gas molecules and dust grains. Once an appreciable density is reached, the dust grains efficiently interact with the incident stellar radiation and are driven away from the star, dragging gas molecules along in slow ($\sim 10$ \kms) winds. The wind-forming region typically reaches 10--20 stellar radii, and the circumstellar envelope (CSE) formed by the wind can extend to $\gtapp 10^5$ \rsun. These outflows produce large mass-loss rates ranging from $10^{-7}$--$10^{-4}$ \msunperyr\ (compare this to the current mass-loss rate from the sun, $10^{-14}$ \msunperyr). The dust grains in the CSE absorb stellar radiation at UV and optical wavelengths and reradiate it in the infrared (IR), making AGB stars some of the brightest sources in the IR sky. Increased formation of dust results in progressively more obscuration of starlight until eventually the star cannot be detected at visible wavelengths. As the star evolves along the TP-AGB, products of shell-burning are deposited onto the inert C/O core, increasing its mass. If the core mass exceeds the Chandrasekhar limit, the star must explode in the form of a supernova type 1.5. Since these have never been observed, it follows that stars with initial masses greater than the Chandrasekhar limit must lose mass at rates faster than the nuclear reaction rates. Calculations of the mass-loss rates from the brightest AGB stars have shown this to be true \citep{vL1999}.

The carbon-rich nuclear processed material is mixed into the outer layers of the star during the {\em third dredge-up} process, which increases the relative abundance of carbon compared to the oxygen that is already present in the AGB atmosphere. The less abundant element of the two is locked up in the extremely stable CO molecule, and the more abundant species decides the chemistry of the CSE. Based on the C/O ratio, we can have O--rich AGB stars (M stars, silicate dust), C--rich AGB stars (C stars, carbonaceous dust) and S stars, where the C/O ratio is close to unity. As the star evolves along the AGB, its C/O ratio gradually increases, goes from an M--star to an S--star and finally ends up as a C--star. In stars more massive than about 4 \msun (the value is metallicity dependent), the temperature at the bottom of the convective envelope is high enough to convert the carbon formed in the CNO cycle into N and O, destroying it before it reaches the atmosphere, thus causing it to never become a C--rich AGB star. This process is called Hot-Bottom Burning \citep{BoothroydSackmann1992}.

AGB mass loss injects nuclear processed material into the surrounding interstellar medium (ISM), thus enriching the environment. AGB stars thus directly effect the chemical evolution of galaxies as well as contributing to their integrated spectra. Since the mass-loss rate effectively determines the lifetime on the AGB \citep{Willson2000}, it is important for understanding stellar evolution along the AGB. Mass loss also has important consequences for the luminosity function (LF) \examp\ of carbon stars. LF information derived from optical data would miss the most luminous (and most obscured in the visual) stars. The contribution of these missing stars could be determined by measuring the mass-loss rates for a large number of stars with a wide range of luminosities. The rate of dust production depends on the metallicity of the progenitor AGB stars. It is important to measure the mass-loss rates of the gas and dust and therefore the gas-to-dust ratio in environments of varied metallicites in order to understand AGB mass loss and quantify the rate of mass-loss return to the ISM.

\section{The SAGE survey}
The Large Magellanic Cloud (LMC) is one of our nearest neighbors, at a distance of about 50 kpc \citep{Feast99} and a favorable viewing angle of 35$^\circ$ \citep{Nikolaevetal2004}. These factors permit a detailed view of the resolved stellar populations and ISM clouds, making it a great laboratory for studying the life cycle of baryonic matter. Studies of these populations in the LMC are not hampered by crowding along the line of sight or high extinction due to ISM, unlike the Milky Way or the Small Magellanic Cloud. The sub-solar metallicity of the LMC is similar to that of the ISM during the epoch of peak star formation in the Universe \citep[at z$\sim$1.5. See, \examp, ][]{Madauetal1996,Peietal1999} which makes it a proving ground for star formation and galaxy evolution models in the early Universe \citep[see ][]{BekkiChiba2005}. In the recent past, the LMC has been the subject of three major IR surveys. The Infrared Astronomical Satellite \citep[{\it IRAS},][]{IRAS} surveyed the LMC in four bands (12, 25, 60 and 100 \mic), identifying 1823 point sources \citep{IRASPSC}. The Midcourse Space Experiment \citep[{\it MSX},][]{MSX} imaged the LMC in one band only (8 \mic) and identified 1806 point sources \citep{MSXLMC}. More recently, the 2 micron All Sky Survey \citep[{\it 2MASS},][]{2MASS} surveyed the LMC \citep{2MASSLMC} in the J, H and \ks\ bands (1.24, 1.66 and 2.16 \mic), and when combined with the I (0.8 \mic), J, H and \ks\ Deep Near-Infrared Survey of the Southern Sky \citep[{\it DENIS},][]{DENIS}, it produced a catalog of about a million point sources. The SAGE \citep[{\it Surveying the Agents of a Galaxy's Evolution},][]{sage1} survey improved on these observations with a $7^\circ\times7^\circ$ image of the LMC using the IRAC (filter bands centered at 3.6, 4.5, 5.8 and 8 \mic) and MIPS (24, 70 and 160 \mic) cameras on board the Spitzer Space Telescope \citep[{\it Spitzer},][]{Spitzer}. The SAGE Epoch 1 Catalog contained about 4 million point sources, including $\sim$ 45\,000 dust evolved stars. The 8 \mic\ data was sensitive enough to be able to detect the stars with mass-loss rates down to $\sim10^{-8}$ \msunperyr. \citet{sage2} presented color-magnitude diagrams for the evolved stars in the survey, identifying about 17\,500 O--rich and 7000 C--rich sources, and 1200 sources classified as ``extreme" AGB stars based on their brightness and extremely red colors (these correspond to the heavily obscured AGB stars). The color-magnitude diagrams for these sources can be used to assess their relative importance to the mass loss budget in the LMC, followed with detailed model calculations for their mass-loss rates. A detailed spectroscopic followup to the SAGE survey has been performed (SAGE-Spec, Kemper et al. 2009, in preparation) that will help further constrain the mass-loss rates of these sources. Some of the bright LMC AGB stars have already been analyzed \citep[see, \examp, ][]{vL1999,vL2005,Zijlstra06}.

\section{This thesis}
The main aim of this thesis is to quantify the mass-loss return from AGB stars to the LMC. An accurate measurement of the mass-loss return requires detailed radiative transfer modeling of the circumstellar dust shell around each AGB star. This can be achieved, for instance, by developing a grid of AGB star models that samples the relevant range of stellar and dust shell parameters. Towards this end, I present a model grid for carbon stars in this thesis. A companion study by Sargent et al. (in preparation) addresses oxygen-rich AGB stars. The thesis is organized as follows: in Chapter \ref{ch:excesses}, I calculate infrared excesses for the AGB star candidates identified from the SAGE survey, and present empirical relations showing that these mid-IR excesses can be used as a proxy for the mass-loss rate. Furthermore, I derive a rough estimate for the dust injection rate from AGB stars into the LMC ISM. In Chapter \ref{ch:cagbmodel}, I describe a radiative transfer model for the circumstellar dust around \ogles, a variable carbon star observed in the SAGE and SAGE-Spec studies, using the \twodust\ code. This is done in order to derive a baseline set of dust properties for use in the carbon star model grid. In this chapter a simple treatment for the circumstellar \acet\ feature is also provided. Chapter \ref{ch:modelgrid} describes the details of a grid of carbon star models spanning a wide range in the relevant parameter space. I provide synthetic photometry from these models in the 2MASS and {\it Spitzer} bands as well as the AKARI and WISE passbands, and present color-color and color-magnitude diagrams in order to compare the models with SAGE observations. I also revisit the model for \ogles\ in order to test the applicability of the model grid. In the future, this model grid will be fit to the entire SAGE AGB candidate list in order to derive mass-loss rates for the entire population, therefore enabling a better determination of the AGB dust injection rate. Data from the AKARI and WISE missions will help tailor our models further towards simulating more realistic AGB dust shells, and we will be able to provide a general purpose fitter for galaxy-wide data sets of AGB stars.

\chapter{The Infrared Excesses of LMC AGB Stars}
\label{ch:excesses}

\centerline{Note: The results from this chapter are summarized in \citet{excesses}}

\section{Introduction}
\noindent
AGB mass loss is believed to be a two-step process: pulsations first levitate material above the photosphere, where the cool temperatures result in the formation of dust grains. Radiation pressure then drives the dust grains (which are collisionally coupled with the gas) outward in an efficient stellar wind \citep[][]{Wick66,GS76,BowenWillson91,Wachteretal02}. An increase in luminosity (and hence radiation pressure) must therefore, in general, be accompanied by an increased mass-loss rate (MLR). This inference is supported by various observations and model predictions \citep[see, {\it e.g.},][]{Reimers75,VW93,Blocker95,BowenWillson91,vL1999,Wachteretal02,vL2005}. 

The formation of a circumstellar envelope (CSE) is a direct result of the AGB mass-loss process. The flux from the CS dust shell appears as mid-infrared (MIR) emission in excess of that due to the central star alone. This MIR excess is directly related to the rate of mass loss, and is therefore expected to increase with increasing luminosity of the central star. Various studies in the past have demonstrated the relationship between the MIR excess and the rate of mass loss. \citet{SW88} showed that the MLR derived from CO rotational transitions was proportional to the strength of the 9.5 $\mu$m silicate dust feature in optically thin O--rich stars. \citet{Jura87} derived a relation between the MLR and 60 $\mu$m excess for O-- and C--rich AGB stars in the solar neighborhood. \citet{KGW92} found that the 12 $\mu$m emission from elliptical galaxies was proportional to the measured MLR. \citet{Athey02} compared the MIR excess emission from 9 galaxies to that of Galactic and LMC AGB stars and derived a proportionality relation between the MIR excess and MLR.

While the study of Galactic AGB stars is hindered by obscuring Galactic dust along the line of sight and high uncertainty in distance estimates, the Large Magellanic Cloud (LMC) is ideal for such observations owing to its proximity \citep[$\sim$50 kpc,][]{Feast99} and favorable viewing angle \citep[35$^\circ$,][]{vdM01}. 
Early LMC surveys \citep[][]{West78,West81,Rebeirot83,BBM80,BM83,FB90} looked for AGB stars  at optical and near-infrared (NIR) wavelengths, and hence preferentially detected sources with optically thin CS dust shells. The Infrared Astronomy Satellite \citep[IRAS;][]{IRAS} survey of the LMC \citep{IRASPSC} in conjunction with ground-based NIR confirmation \citep{Reid90} helped identify several of the brightest mass-losing evolved stars \citep{Loup97,Zijlstra96}. \citet{Trams99} performed follow-up mid-infrared (MIR) photometry and spectroscopy of 57 sources using the  Infrared Space Observatory \citep[ISO;][]{ISO} in order to determine the chemical composition of the CSEs, and 
\citet{vL1999} used radiative transfer modeling to derive luminosities and MLRs for these sources. More recently, the LMC has been surveyed in the optical by the Magellanic Clouds Photometric Survey \citep[MCPS;][]{MCPS}, and in the NIR by DENIS \citep{DENIS} and 2MASS \citep{2MASS}. The LMC data from these two surveys can be found in \citet{Cioni} and \citet{2MASSLMC}, respectively. A mid-infrared survey using the Midcourse Space Experiment \citep[MSX;][]{MSX}, which was four times more sensitive than IRAS, was performed by \citet{MSXLMC}.

As part of the SAGE survey \citep[{\it Surveying the Agents of a Galaxy's Evolution};][]{sage1}, a $\sim$7$^\circ\times 7^\circ$ area of the LMC was imaged in the Spitzer Space Telescope \citep[{\it Spitzer};][]{Spitzer} IRAC (3.6, 4.5, 5.8 and 8.0 $\mu$m) and MIPS (24, 70 and 160 $\mu$m) bands. One of the main goals of SAGE was to detect all the evolved stars with MLRs $> 10^{-8}$ \msunperyr, to characterize the total rate at which material is returned to the ISM by dusty evolved stars, and to better understand the physics governing mass loss among evolved stars in the LMC. When complete, the SAGE data will be $\sim$1000 times more sensitive than MSX, and they will allow a detailed quantitative derivation of the global mass-loss budget from all stellar populations when combined with existing and future MIR spectroscopic observations of evolved stars \citep[see, {\it e.g.},][]{vL1999,vL2005,Zijlstra06}. Continuing the analysis begun by \cite{sage1}, \citet{sage2} identified about 32,000 color-selected evolved stars brighter than the tip of the red giant branch (11.85 mag in the IRAC 3.6 $\mu$m band), including 17,500 oxygen--rich (O--rich), 7000 carbon--rich (C--rich), and 1200 ``extreme" AGB stars, and presented color-magnitude diagrams (CMDs) of the SAGE epoch 1 data.

Our aim is to investigate the MLRs of the AGB candidates selected from the SAGE survey, which will require detailed radiative transfer modeling of the CSE around each such star. In order to simplify this effort to model $\sim 10^4$ stars, we would like to constrain the range of some of the input model parameters, such as the total source luminosity, the typical temperature of the CS dust and the optical depth at a given wavelength. As a first step towards this modeling goal, therefore, in this chapter we take an empirical approach and calculate MIR excess fluxes due to dust from these stars. The distance independence of the MIR colors of AGB sources makes them a good choice for tracing the mass loss \citep[see, {\it e.g.},][]{Whitelocketal2006}, but our study based on the MIR excesses for sources in the LMC is aided by the fact that the distance to the stars in the LMC is essentially the same. Data from the SAGE survey also provide an unprecedented opportunity to include a large number of AGB star candidates. We will show that the MIR excess can be used as a proxy for the MLR. In a very statistical way, we study the overall trends of excess and derived quantities such as color temperature of the dust and MIR optical depth with source luminosity. In subsequent papers, we will also be presenting the luminosity function and detailed radiative transfer calculations for the AGB star candidates selected in this paper. The chapter is organized as follows: in \S \ref{sec:excesses:data} we describe the SAGE database and our observational sample of AGB stars. Our procedure for calculating IR excesses is explained in \S \ref{sec:excesses:excess}. We present our results in \S \ref{sec:excesses:results}, and in \S \ref{sec:excesses:discuss} we compare our results to previous work and discuss their implications for future AGB studies in the LMC.

\section{Data}
\label{sec:excesses:data}
\noindent The SAGE catalog and archive point source lists from both epochs of observations have been delivered to the Spitzer Science Center (SSC), and are available for download at the SSC\footnote{http://ssc.spitzer.caltech.edu/legacy/sagehistory.html}. The SAGE epoch 1 point source catalog is discussed by \citet{sage1} and \citet{sage2}. In this study, we select sources from the IRAC epoch 1 archive instead of the catalog. The archive accepts fainter sources than the catalog. Faint limits for both epochs are 18.5, 17.5, 15, and 14.5 mag for IRAC 3.6, 4.5, 5.8, and 8.0 \mic\, respectively, compared to 18, 17, 15, and 14 mag for the catalog. A source is excluded ({\it i.e.}, culled) from the archive if it has neighbors within a 0.5\arcsec~radius, whereas this radius is 2\arcsec~for the catalog. This culling procedure ensures the creation of reliable lists of point sources at the expense of completeness \citep[see][]{sage1}. In order for two sources to appear blended, they both have to be detected, and they have to be bright enough. The probability of such blending in the 24 \mic\ band is very low, since the 24 \mic\ catalog has very few sources  ($\sim 40,000$) compared to the IRAC catalog. However, we do find a few ($\sim$40 out of $\sim 10^4$) IRAC sources that are matched to more than one MIPS 24 \mic\ source which we corrected in our results and in the main catalog. In addition to this culling, a flux value for each archive source in any of the IRAC bands is non-null only if its signal-to-noise ratio (S/N) is greater than 5. For the catalog, the S/N is 6 for the IRAC [3.6], [4.5], and [5.8] \mic\ bands, and 10 for the [8.0] \mic\ band. As a result of these criteria, the archive has slightly more sources and more flux values than the catalog, allowing the inclusion of more faint AGB candidates. As of the second delivery, version S13, the archive has $\sim$4.5 million sources compared to $\sim$4.3 million in the catalog. See the SAGE data delivery document\footnote{The file SAGEDataDescription\_Delivery2.pdf is available at\\
http://data.spitzer.caltech.edu/popular/sage/20080204\_enhanced/documents/} for details of the source selection in the catalog and archive.  The fact that the fluxes of these fainter AGB candidates are more uncertain when using the archive list is mitigated by our requirement of 2MASS detections thus affirming that the point source is real. The AGB source lists are extracted from a universal table which lists the nearest neighbor in the MCPS and IRAC epoch 1 archive to each source in the MIPS24 epoch 1 full catalog ($\sim$40,000 sources). The matching radius is 3\arcsec, but the nearest neighbor is used in the match. The IRAC archive sources are also bandmerged with the 2MASS catalog.\\

We classify sources as low- or moderately-obscured O--rich/C--rich AGB candidates based on their location in the \ks\ versus J-\ks\ CMD\footnote{The UBVI and JH\ks\ magnitudes are dereddened using the same procedure as in \citet{Cioni}} \citep{Cioni,sage2}, but we exclude stars without H-band fluxes from our list, as our procedure for calculating the excesses (see \S \ref{sec:excesses:excess}) relies on an H-band detection. This criterion probably excludes the faintest O--rich AGB stars, and thus does not significantly affect our results. As a separation into O--rich and C--rich chemistries based on near-IR colors is not possible for the heavily obscured (J--[3.6]$>$3.1, \citet{2MASSLMC}) ``extreme" AGB candidates, we follow a procedure similar to \citet{sage2} and select these sources based on their location in the [3.6] versus J--[3.6] CMD or (when no 2MASS counterpart exists) the [8.0] versus [3.6]--[8.0] CMD. A simple trapezoidal integration of the optical U through MIPS24 fluxes is performed to estimate the luminosities for all our sources.

The \citet{Kastneretal2008} list of objects with spectroscopic classifications contains 14 sources classified in our study as O--rich AGB stars. \citet{Kastneretal2008} classify 10 of these as red supergiants (RSGs). We also find that we have classified 21 of their sources as extreme AGB stars -- most of these have been identified as possible HII regions, none of them are RSGs. Since these are point sources, we interpret HII regions as compact HII regions or massive young stellar objects (YSOs). We thus realize that our sample of AGB stars may be contaminated by non-AGB objects such as RSGs and YSOs. We estimate this contamination using simple color-magnitude cuts. 

The figures in \citet{sage2} also show that the AGB stars and RSGs are not well separated in the MIR CMDs. It is possible that some of the most luminous sources in our list are RSGs. We find 556 objects (124 classified as O--rich, 17 as C--rich, and 415 as extreme AGB candidates) more luminous than M$_{\rm bol}=-7.1$, the classical AGB luminosity limit. It is not unlikely that some of these sources are AGB stars. Luminosities above the classical limit can be achieved by AGB stars at the peak of their pulsation cycle (as in the case of OH/IR stars) or by stars undergoing hot bottom burning \citep[HBB,][]{BoothroydSackmann1992}. Nevertheless, these numbers provide a very conservative upper limit for the RSG contamination in our sample.

The reddest sources in our list fall in the region of the MIR CMD also populated by YSOs. \citet{sage3} isolate regions in the [8.0] vs [8.0]--[24] CMD occupied more densely by YSO models (Figure 3 in their paper). This separation includes a stringent cut at [8.0]--[24]$\sim$2.2 (corresponding approximately to a 24 \mic\ flux $\ge$ the 8 \mic\ flux) to exclude AGB stars. We find that 5 O--rich and 24 extreme AGB star candidates in our list are part of the \citet{sage3} list of high-probability

\clearpage 
\thispagestyle{empty}
\begin{landscape}
\begin{deluxetable}{rrrrrrrrrrrrrrrr}
\setlength{\tabcolsep}{0.05in}
\tablewidth{0pt}
\tablecolumns{16}
\tablecaption{Source List of Color-Selected AGB Star Candidates \label{SourceList}\tablenotemark{1}}
\tabletypesize{\scriptsize}
\tablehead{
\colhead{Identifier\tablenotemark{a}} & \colhead{Type\tablenotemark{b}} &
\colhead{magU\tablenotemark{c}} & \colhead{$\delta$magU} & \colhead{magB} & \colhead{$\delta$magB} & \colhead{magV} & \colhead{$\delta$magV} & \colhead{magI} & \colhead{$\delta$magI} & \colhead{magJ} & \colhead{$\delta$magJ} & \colhead{magH} & \colhead{$\delta$magH} & \colhead{magK} & \colhead{$\delta$magK}
}
\startdata
SSTISAGE1A J054938.72--683458.2&O--rich&99.99&99.99&19.46&0.04&17.55&0.05&13.76&0.07&12.01&0.03&11.11&0.03&10.78&0.02\\
SSTISAGE1A J055530.35--684647.5&O--rich&20.46&0.19&18.11&0.03&16.18&0.03&13.95&0.04&12.59&0.02&11.72&0.02&11.48&0.02\\
SSTISAGE1A J055420.11--680449.5&O--rich&99.99&99.99&18.87&0.04&16.59&0.05&13.53&0.03&11.91&0.02&11.02&0.02&10.71&0.03\\
SSTISAGE1A J055729.20--684444.2&O--rich&17.97&0.06&17.47&0.04&16.32&0.08&16.38&0.30&11.93&0.02&11.03&0.02&10.75&0.02\\
SSTISAGE1A J055321.17--683114.7&O--rich&21.04&0.24&18.49&0.09&16.51&0.05&13.87&0.03&12.52&0.02&11.60&0.02&11.37&0.02\\
SSTISAGE1A J054522.57--684244.5&C--rich&99.99&99.99&20.68&0.06&16.45&0.03&13.82&0.04&12.49&0.02&11.17&0.02&10.38&0.02\\
SSTISAGE1A J055650.80--675030.5&C--rich&99.99&99.99&20.42&0.07&16.66&0.11&13.56&0.04&12.12&0.02&11.01&0.03&10.35&0.02\\
SSTISAGE1A J055835.31--682009.7&C--rich&99.99&99.99&20.13&0.05&16.28&0.03&13.55&0.04&12.30&0.02&11.02&0.03&10.20&0.02\\
SSTISAGE1A J055311.71--684720.9&C--rich&99.99&99.99&21.91&0.17&17.93&0.04&14.56&0.04&12.03&0.02&10.91&0.03&10.14&0.02\\
SSTISAGE1A J055036.67--682852.3&C--rich&21.61&0.44&19.31&0.05&16.68&0.03&14.42&0.04&12.87&0.03&11.92&0.03&11.52&0.02\\
SSTISAGE1A J052742.48--695251.5&Extreme&19.08&0.09&18.58&0.09&18.72&0.11&14.86&0.07&14.13&0.07&12.41&0.06&11.08&0.04\\
SSTISAGE1A J052714.19--695524.3&Extreme&20.04&0.12&19.65&0.07&18.36&0.05&14.45&0.05&13.74&0.03&12.13&0.03&10.85&0.02\\
SSTISAGE1A J053239.06--700157.5&Extreme&99.99&99.99&20.29&0.20&17.30&0.10&14.81&0.04&13.16&0.03&11.73&0.02&10.73&0.02\\
SSTISAGE1A J052950.52--700000.1&Extreme&18.80&0.09&18.28&0.09&17.40&0.14&16.08&0.05&13.96&0.04&12.30&0.04&10.80&0.03\\
SSTISAGE1A J053441.38--692630.7&Extreme&99.99&99.99&22.78&0.45&20.26&0.11&14.59&0.04&12.36&0.02&10.92&0.02&9.87&0.03\\
\enddata
\tablenotetext{1}{The electronic version of this table contains all the AGB candidates considered in this study.}
\tablenotetext{a}{SAGE Epoch 1 Archive (SAGE1A) designation, including position coordinates of IRAC source.}
\tablenotetext{b}{Sources are classified as O--rich or C--rich based on their J-\ks colors, or as ``Extreme" based on their 2MASS and IRAC colors. For more details, see \S \ref{sec:excesses:data}.}
\tablenotetext{c}{Magnitudes and errors in the UBVI, JH\ks, IRAC and MIPS 24 \mic\ bands. A value of 99.99 in any band represents either a saturation or a non-detection.}
\end{deluxetable}

\clearpage
\thispagestyle{empty}
\begin{deluxetable}{rrrrrrrrrr}
\addtocounter{table}{-1}
\tablewidth{0pt}
\tablecolumns{10}
\tablecaption{\em (Continued)}
\tabletypesize{\scriptsize}
\tablehead{
\colhead{mag36} & \colhead{$\delta$mag36} & \colhead{mag45} & \colhead{$\delta$mag45} & \colhead{mag58} & \colhead{$\delta$mag58} & \colhead{mag80} & \colhead{$\delta$mag80} & \colhead{mag24} & \colhead{$\delta$mag24}\\
}
\startdata
10.53&0.03&10.72&0.03&10.50&0.04&10.44&0.05&9.86&0.07\\
11.34&0.03&11.40&0.03&11.28&0.04&11.19&0.04&10.59&0.14\\
10.47&0.04&10.64&0.02&10.46&0.05&10.40&0.04&9.97&0.07\\
10.47&0.03&10.60&0.02&10.39&0.03&10.30&0.04&10.01&0.08\\
11.07&0.07&11.02&0.04&10.90&0.03&10.76&0.03&10.10&0.08\\
9.67&0.03&9.70&0.03&9.68&0.04&9.30&0.04&8.89&0.04\\
9.76&0.04&9.86&0.03&9.81&0.04&9.30&0.04&9.26&0.03\\
9.34&0.05&9.03&0.02&8.80&0.03&8.68&0.03&8.61&0.03\\
9.51&0.05&9.43&0.04&9.27&0.04&8.99&0.03&8.66&0.03\\
11.06&0.04&11.08&0.04&10.89&0.04&10.68&0.05&10.21&0.09\\
9.02&0.05&8.65&0.04&8.39&0.03&8.00&0.03&7.49&0.02\\
9.62&0.04&9.12&0.03&8.71&0.03&8.24&0.03&7.61&0.03\\
9.38&0.04&9.33&0.03&9.19&0.04&8.83&0.03&8.46&0.03\\
8.94&0.04&8.23&0.04&7.55&0.03&6.85&0.02&6.03&0.02\\
9.06&0.05&8.33&0.05&7.81&0.02&7.29&0.03&6.74&0.02\\
\enddata
\end{deluxetable}
\end{landscape}
\clearpage

YSOs. A more conservative estimate is obtained by looking for sources fainter than [8.0]=7 with $F_{24 \mu m}/F_{8 \mu m}\ge 1$, which puts the YSO contamination in our lists at about a hundred sources (61 O--rich, 14 C--rich, and 34 extreme AGB candidates). 

\begin{figure}[!htb]
\epsscale{0.83}\plotone{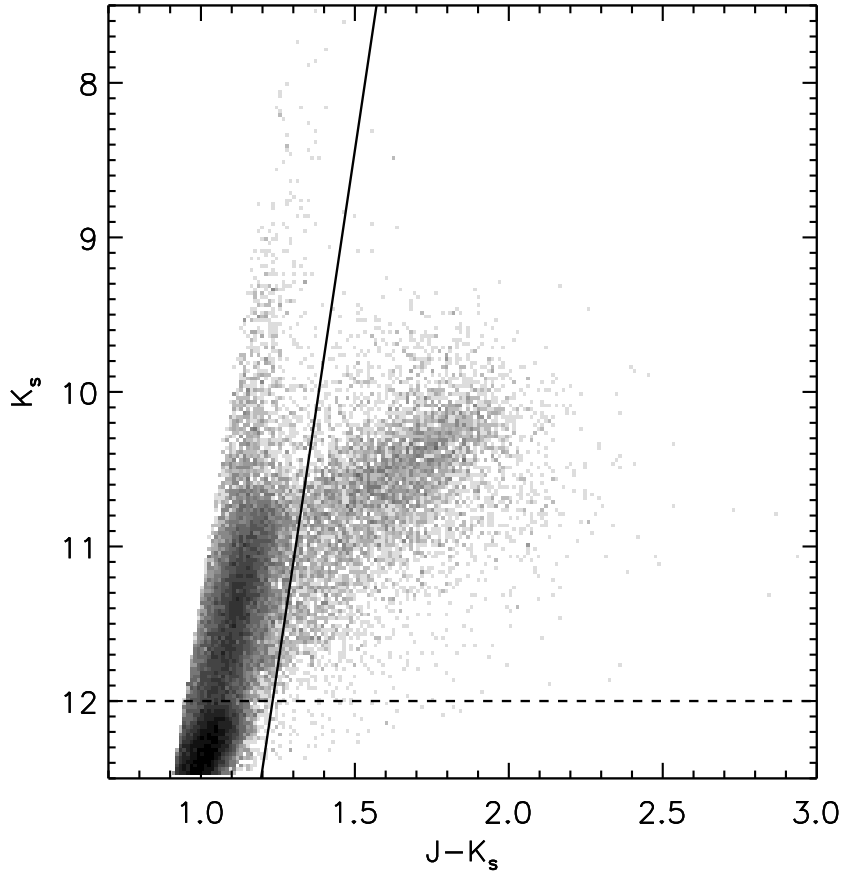}
\ssp{\caption[Near-infrared color-magnitude diagram for SAGE AGB candidates]{A Hess diagram showing the locus of AGB stars in the \ks\ versus J--\ks\ CMD. Stars brighter than the tip of the RGB (K$_{\rm s} =$12, dashed line) are selected as AGB candidates, and these are classified as O--rich and C--rich based on the CMD cuts in \citet{Cioni}. The O--rich sources are bounded by the lines K$_1$ and K$_2$ (shown as solid line in figure) in their paper, while sources redward of K$_2$ are classified as C--rich. The extreme AGB candidates (not shown in this figure) are classified based on their 2MASS and IRAC colors.\label{fig:jkcmd}}}
\end{figure}

\begin{figure}[!htb]
\plotone{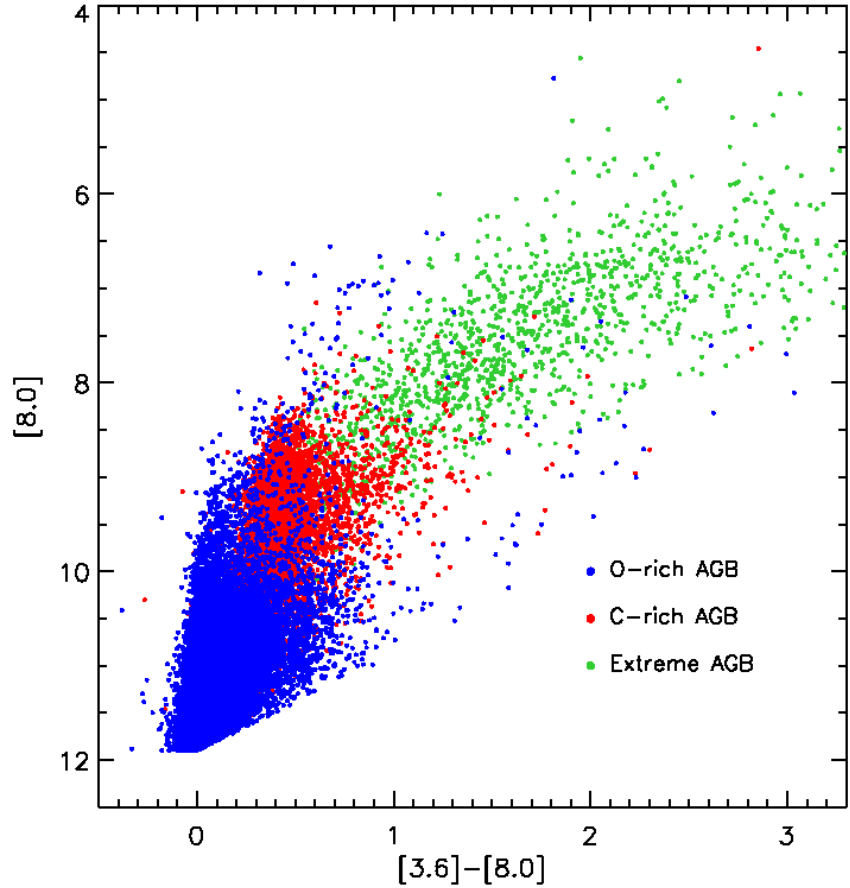}
\ssp{\caption[IRAC color-magnitude diagram for SAGE AGB candidates]{The three types of AGB candidates (O--rich: blue, C--rich: red, Extreme: green) on an [8.0] vs [3.6]--[8.0] CMD. The tip of the RGB is at [3.6]$\approx$[8.0]=11.9.\label{fig:iraccmd}}}
\end{figure}

\begin{figure}[!htb]
\epsscale{0.7}\plotone{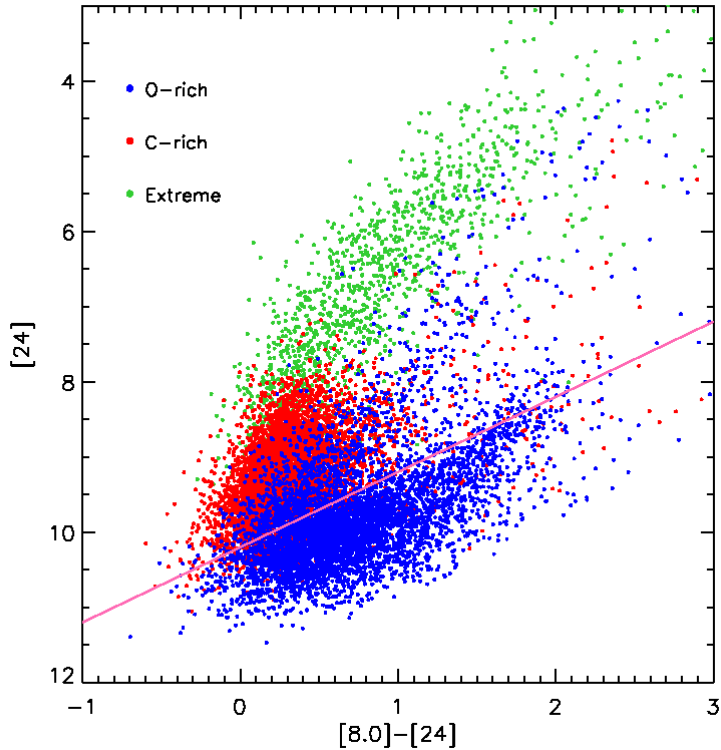}
\ssp{\caption[IRAC--MIPS color-magnitude diagram showing two O--rich AGB populations]{The O--rich (blue), C--rich (red) and extreme AGB (green) stars on the [8.0]--[24] versus [24] CMD. The sources below the solid line are the fainter, redder population mentioned in \citet{sage2} (The finger ``F" in their Figure 6). About 80\% of the O--rich stars in our sample belong to this population.\label{fig:twopops}}}
\end{figure}

Figure \ref{fig:jkcmd} shows the \ks\ versus J-\ks\ Hess diagram for the region populated by AGB stars. The IRAC and MIPS24 CMDs are shown in Figures \ref{fig:iraccmd} and \ref{fig:twopops}. \citet{sage2} noted the presence of a fainter, redder population of O--rich sources (finger ``F'' in their Figure 6). The same population can be seen in our IRAC-MIPS24 CMD (Figure \ref{fig:twopops}) at [8.0] magnitudes fainter than $\sim$10. We use a magnitude cut at 10.2, shown in the figure as a solid line. Throughout this chapter, we will differentiate between the bright and faint O--rich populations based on this magnitude cut. Almost 80$\%$ of the O--rich stars in our sample are fainter than [8.0]=10.2. Photometric information for a few sources of each type is shown in Table \ref{SourceList}. The entire list of AGB star candidates is available with the electronic version of \citet{excesses}.

\begin{figure}[!htb]
\plotone{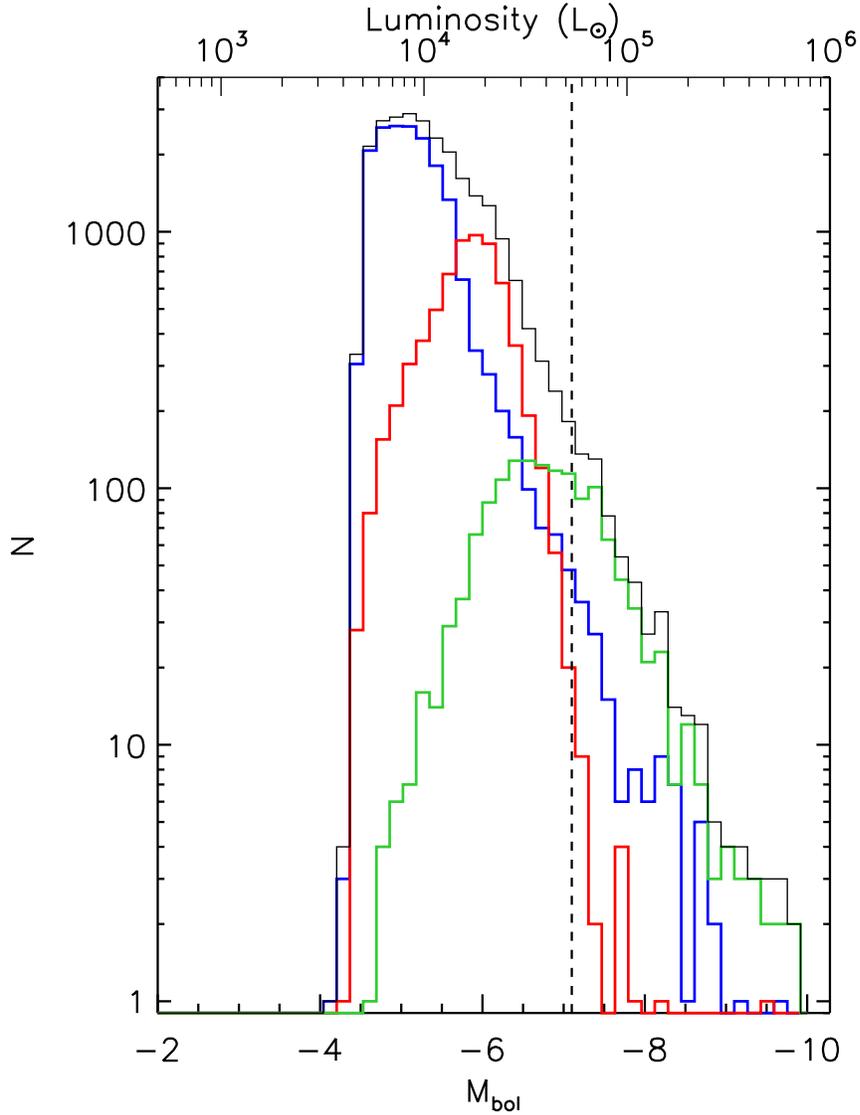}
\ssp{\caption[Luminosity function of SAGE AGB sample]{The luminosity function for our sample of AGB candidates (blue: O--rich, red: C--rich, green: Extreme, black: total). The vertical dashed line is the classical AGB luminosity limit.\label{fig:AGBLF}}}
\end{figure}

Figure \ref{fig:AGBLF} shows the luminosity function for all three classes of AGB candidates. The lower luminosity limit ($M_{bol}\approx -4$ ) for the O--rich and C--rich AGB candidates is set by our color-magnitude cuts which exclude sources fainter than the tip of the RGB. The C--rich sources have a tighter luminosity distribution than their O--rich counterparts (there are only a handful of C--rich sources brighter than $M_{bol}=-7.5$, whereas the O--rich distribution falls off around $M_{bol}=-8.5$). The range in luminosities for the extreme AGB candidates is $M_{bol}=-4.5$ to $M_{bol}\approx -10$. The extreme AGB candidates thus have the highest luminosities in the sample. However, we have insufficient information at this point to say anything concrete about the breakdown of these sources into O--rich and C--rich chemistries. The spectroscopic follow-up to SAGE will provide some information about the dust chemistry of these extreme AGB candidates. These distributions peak at $M_{bol}\approx -5$ (O--rich), $M_{bol}\approx -6$ (C--rich) and $M_{bol}\approx -6.5$ (Extreme). The vertical dashed line shows the classical AGB luminosity limit. While this limit can be exceeded by deeply embedded AGB stars \citep{Woodetal1992}, O--rich sources with luminosities higher than $M_{bol}=-7.8$ are more likely to be RSGs, while the brightest extreme AGB stars may be massive YSOs. We will discuss the astrophysical implications of these luminosity functions in a future paper.

\section{Procedure}
\label{sec:excesses:excess}
\noindent
We estimate the MIR excess emission of the low- and moderately-obscured O--rich and C--rich AGB star candidate populations by comparing their observed SEDs to an expected SED for the stellar photosphere. The IR excess in each wavelength band is calculated by comparing the total flux from the source observed in that band to the flux expected from the central star as prescribed by a ``best-fit" model photosphere. As the emission from CS dust dominates the MIR flux of extreme AGB stars, we set the MIR flux equal to the excess in each band for our extreme AGB star candidates. In this work, we are interested in describing overall trends in the AGB parameters such as IR excess as opposed to detailed models of each source. To this end we choose a single model photosphere for each type of AGB star -- one that best fits the SED shape of AGB stars with little or no dust. We use the plane-parallel C--rich MARCS models of \cite{Gautschy2004} and the spherical O--rich PHOENIX models of \cite{Omodels} to calculate the photospheric AGB star emission. The differences between plane-parallel and spherical models are insignificant as far as the resulting broad band AGB star photospheric information is concerned. The range of model parameters we search for a ``best-fit" are as follows: the thirty-two solar-metallicity C--rich AGB models have C/O ratios between 1.1 and 1.8, effective temperatures ranging from 2600 K to 3200 K, and surface gravity $\log{g}$ between about $-$0.76 and 0. The $\sim$200 O--rich AGB models of solar mass and solar metallicity have effective temperatures ranging from 2000 K to 4700 K, and $\log{g}$ between $-$0.5 and 2.5 in steps of 0.5. Synthetic photometry in the optical, near- and mid-infrared was obtained from each of these models by convolving their SEDs with the filter response curves of the Johnson-Kron-Cousins UBVI 
\footnote{The MCPS magnitudes were placed on the Johnson-Kron-Cousins UBVI system. The detector quantum efficiency curve was obtained from the Las Campanas Observatory website, 
http://www.lco.cl/lco/index.html. Filter profiles for the Johnson U, Harris B, V, and Cousins I filters were obtained from the references in Table 9 of \citet{Fukugita}.}
, 2MASS JH\ks
\footnote{The 2MASS filter relative
  spectral responses (RSRs) derived by \citet{2MASSRSRs} were obtained
  from the {\it 2MASS All-Sky Data Release Explanatory Supplement} on
  the worldwide web at http://www.ipac.caltech.edu/2mass/releases/allsky/doc/sec6$_{}$4a.html.}
, and Spitzer IRAC
\footnote{The IRAC RSRs are plotted in \citet{IRAC}, and were obtained from the {\it Spitzer Science Center} IRAC pages at http://ssc.spitzer.caltech.edu/irac/spectral$_{}$response.html}
 and MIPS 24 $\mu$m
\footnote{The MIPS \citep{MIPS} RSRs were obtained from the {\it Spitzer Science Center} MIPS pages at http://ssc.spitzer.caltech.edu/mips/spectral$_{}$response.html}
  filters. The oxygen-rich models had spectral information in the range 0.1--1000 \mic\ and the convolutions were done directly with the models. In order to compensate for the insufficient wavelength coverage ($\sim$0.5--25 \mic) offered by the carbon-rich models, the U and B band fluxes were dropped, and the flux in the MIPS24 band was extrapolated to 30 \mic\ assuming a Rayleigh-Jeans falloff.

\begin{figure}[!htb]
\epsscale{0.6}\plotone{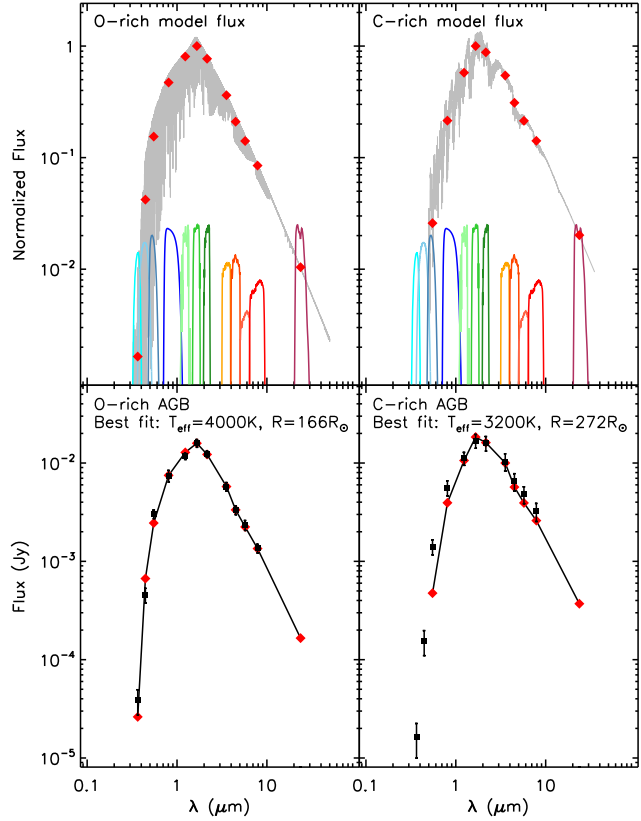}
\ssp{\caption[Demonstration of the excess best-fit procedure]{The panels on the top show the normalized SEDs for the best-fit O--rich (left) and C--rich (right) models in light gray. The response curves for the optical UVBI, 2MASS JH\ks, IRAC and MIPS24 filters are also shown. Folding the transmission profiles of each filter into the model SED results in a flux in every filter (red diamonds). The bottom panels show the median flux in each band for the hundred bluest (in V--\ks\ color) O--rich (left) and C--rich (right) sources (black squares) plotted over the flux of the corresponding best-fit model (red diamonds) scaled to the median flux in the H band. The H band magnitudes of these bluest sources are in the range 11.99 to 11.36 (O--rich) and 11.95 to 10.32 (C--rich). The O--rich model is a 1 M$_\odot$ model with $T$=4000 K and $\log{g}=0$, while the C--rich model has $T$=3200 K, a C/O ratio of 1.3, and $\log{g}=-0.43$.\label{fig:bestfitmodels}}}
\end{figure}

As we are most interested in obtaining the correct shape of the SED for a best-fit, the model SEDs are scaled to the H-band flux of the median SED of the hundred bluest sources in V$-$\ks\ color. We use the same model to calculate the excesses for both the bright and faint O--rich populations because the bluest O--rich candidates lack detections in the 24 \mic\ band and thus can not be separated into bright and faint populations. The model that comes closest to describing the oxygen--rich median SED has surface gravity $\log{g}$=0 and an effective temperature $T$=4000 K. The corresponding carbon--rich best-fit model parameters (for a C/O ratio of 1.3) are $\log{g}$=--0.43 and $T$=3200 K. Figure \ref{fig:bestfitmodels} demonstrates the filter-folding for these best-fit models (top), and shows the comparison to the median SED of the bluest sources (bottom). There are many more O--rich candidates that are fainter and bluer than most of the C--rich candidates, which might explain the considerable difference in the best-fit model temperature and radius of the two types of sources.

To determine the excess flux $X_\nu$ in a band centered around frequency $\nu$ due to CS emission, the best-fit model flux must first be scaled to the flux of each source at some ``pivot" wavelength and then the difference between the observed flux $F_\nu$ and the corresponding scaled model flux in that band must be calculated. Fitting the SEDs of all our sources to one best-fit model photosphere is a simple first approach towards modeling the CSE around each AGB star candidate. The effects of interstellar and CS extinction on our sources is minimal in the 2MASS JH\ks\ bands, which also contain the wavelength range corresponding to maximum emission from AGB star photospheres.\footnote{The most obscured sources will suffer from CS extinction even in the NIR bands, but choice of pivot wavelength is not an issue for these sources, as they are probably members of our extreme AGB list.} Almost all of our O--rich sources peak in the H band\footnote{This is partly due to the fact that the opacity of the H$^-$ ion reaches a minimum in the H-band.}, while over two-thirds of the C--rich sources peak in the \ks\ band. All of the photospheric models available to us exhibit H-band SED maxima. For these reasons, we place the pivot wavelength in the H band (centered at 1.65 \mic). The excess then depends on the source flux $F_\nu$ and model flux $F^{mod}_\nu$ according to
\begin{eqnarray}
X_\nu=F_\nu-\left(\frac{F^{mod}_\nu}{F^{mod}_{\rm H}}\right)F_{\rm H}
\end{eqnarray}
where the subscript H denotes H-band fluxes. This equation may overestimate the MIR excesses from the redder C--rich sources with $F_{K}> F_{H}$. On the other hand, using the same best-fit model with a \ks~-band pivot would produce underestimates for the excesses. For the reddest C--rich source in our sample, the difference in 8 \mic\ excess resulting from choice of pivot is $\sim 30\%$. This is to be regarded as an upper bound to the error introduced in the excess determination due to the choice of the H-band as pivot. While this will not alter the general trends we discuss in this chapter, it will affect our numerical results for higher excesses (mass-loss rates).

\begin{figure}
\epsscale{0.8}\plotone{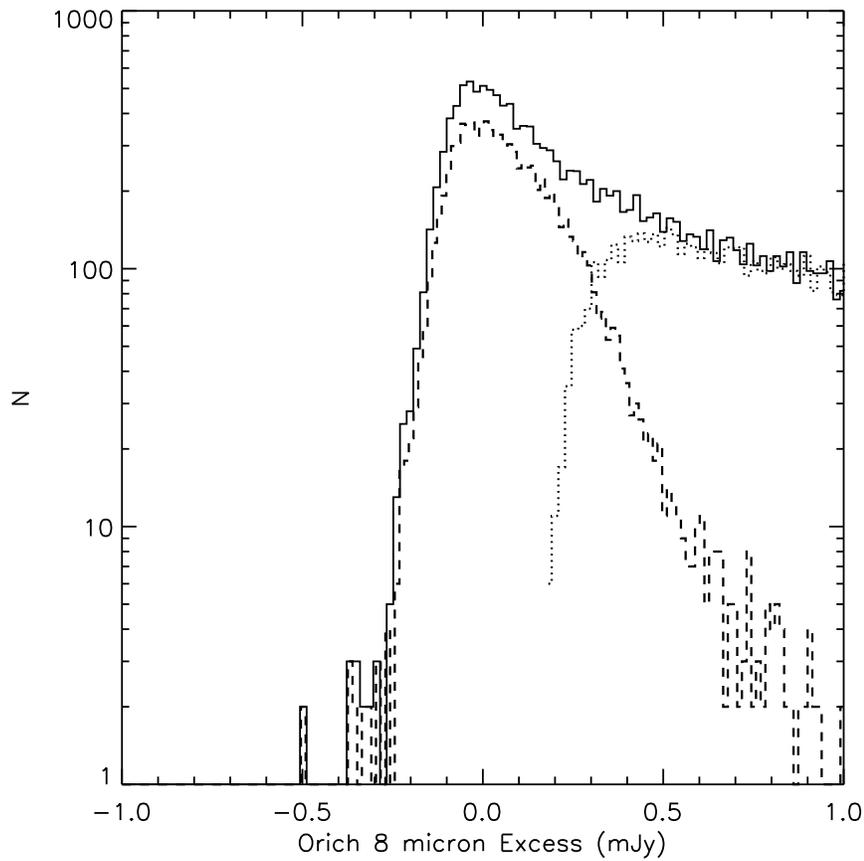}
\ssp{\caption[Histogram of 8 \mic\ excesses for faintest O--rich AGB candidates]{Histogram of the 8 \mic\ excesses for the faintest O--rich AGB candidates (and thus with the lowest S/N ratio excess emission). The solid line is for all sources. The dotted line shows sources with S/N$>$3. The dashed line shows sources with excesses considered unreliable and hence were excluded from our study.\label{fig:xshisto}}}
\end{figure}

The photometric errors associated with the source fluxes were used to estimate the error $\delta X$ in the calculated excess. An excess measurement in a wavelength band centered around frequency $\nu$ was deemed ``reliable" only if its signal-to-noise ratio was greater than 3. In other words,
\begin{eqnarray}
\label{qualcut}
{\delta X_\nu \over X_\nu} \leq {1 \over 3}
\end{eqnarray}
Figure \ref{fig:xshisto} illustrates the effect of selecting sources using this ``3-$\sigma$" criterion. The distribution of sources that are rejected based on this criterion is fairly symmetric about zero, with a slight asymmetry on the positive excess side, indicating that our cut is conservative. A similar distribution of rejected sources is seen for the C--rich candidates, but they are substantially fewer in number. In both cases, sources with excesses below $\sim$ 0.1 mJy are rejected.

Very little can be inferred about the chemical composition of the dust shells around the extreme sources without spectroscopic confirmation, although most of these objects are probably C--rich. We find that 60 of our extreme AGB candidates are identified as C--rich in the \citet{Kastneretal2008} study, while only 9 are classified as O--rich. Our procedure for calculating excesses relies on a classification into O--rich or C--rich sources, which is not possible for most of the stars in this list. However, the excess emission from their extremely dusty CSEs dominates over the photospheric emission in the MIR, so that we can set the excess equal to the MIR flux to a good level of approximation. We find $\sim$ 8200 O--rich, 5800 C--rich and 1400 extreme sources with reliable 8 \mic\ excesses, and about 4700 O--rich, 4900 C--rich and 1300 extreme sources with reliable 24 \mic\ excesses. Table \ref{ExcessList} shows the excesses calculated for the fifteen sample sources shown in Table \ref{SourceList}. The electronic version of \citet{excesses} contains all the sources with valid 8 \mic\ excesses.

We estimate the temperature of the CS dust and the optical depth from the 8 \mic\ and 24 \mic\ excesses in a manner similar to \citet{Thompsonetal2006} and \citet{Dayaletal1998}. The continuum dust emission is modeled as a blackbody at temperature $T_d$, with optical depth $\tau_\nu$,
\ben
\label{coltau}
X_\nu \propto \Omega_\nu (1-e^{-\tau_\nu}) B_\nu(\lambda, T_d) 
\een
where 
\ben
\Omega_\nu=\pi\left(\frac{R_{in}}{D}\right)^2
\een
is the geometrical dilution factor with $R_{in}$ the distance from the central star corresponding to maximum emission due to dust at both 8 and 24 \mic. For small optical depths, the excess is proportional to the opacity $\kappa_\nu$. If the emissivity of the dust can be modeled by a $\lambda^{-1}$ power law in the relevant range of wavelengths, we have
\ben
\label{coltemp}
X_\nu \propto \Omega_\nu\lambda^{-1}B_\nu(\lambda, T_d)
\een

Assuming a power-law emissivity in the MIR ignores any effects due to strong absorption or emission features from silicate dust for example, which can be significant for the more obscured O--rich sources. The power-law index will also depend on the dust species in general \citep[see, {\it e.g.},][]{Hofner2007}. The  $\lambda^{-1}$ power-law dependence adequately describes the emissivity at MIR wavelengths for the purpose of this chapter (the {\it Spitzer} 8 and 24 \mic\ bands will only detect the wings of silicate emission unless the sources are OH/IR stars.) The color temperature $T_d$ is calculated by constructing 

\clearpage
\begin{landscape}
\thispagestyle{empty}
\begin{deluxetable}{rrrrrrrrrrrr}
\tablewidth{0pt}
\tablecolumns{12}
\tablecaption{Mid-infrared Excess Fluxes \label{ExcessList}}
\tabletypesize{\scriptsize}
\tablehead{
\colhead{Identifier} & \colhead{Type} & \colhead{X36\tablenotemark{a}} & \colhead{$\delta$X36\tablenotemark{b}} & \colhead{X45} & \colhead{$\delta$X45} & \colhead{X58} & \colhead{$\delta$X58} & \colhead{X80} & \colhead{$\delta$X80}& \colhead{X24}& \colhead{$\delta$X24}\\
&&\colhead{(mJy)}&\colhead{(mJy)}&\colhead{(mJy)}&\colhead{(mJy)}&\colhead{(mJy)}&\colhead{(mJy)}&\colhead{(mJy)}&\colhead{(mJy)}&\colhead{(mJy)}&\colhead{(mJy)}
}
\startdata
SSTISAGE1A J054938.72--683458.2&O--rich&4.546&0.622&1.918&0.315&2.338&0.287&1.335&0.210&0.457&0.051\\
SSTISAGE1A J055530.35--684647.5&O--rich&0.843&0.287&0.727&0.171&0.726&0.135&0.450&0.091&0.213&0.053\\
SSTISAGE1A J055420.11--680449.5&O--rich&4.254&0.784&1.932&0.279&2.170&0.363&1.222&0.187&0.349&0.050\\
SSTISAGE1A J055729.20--684444.2&O--rich&4.351&0.531&2.413&0.292&2.740&0.230&1.693&0.190&0.326&0.050\\
SSTISAGE1A J055321.17--683114.7&O--rich&2.387&0.654&2.392&0.268&1.909&0.163&1.313&0.100&0.428&0.049\\
SSTISAGE1A J054522.57--684244.5&C--rich&18.980&1.265&12.830&0.720&8.010&0.533&7.247&0.488&1.283&0.078\\
SSTISAGE1A J055650.80--675030.5&C--rich&13.170&1.375&7.974&0.645&5.065&0.509&6.534&0.504&0.606&0.048\\
SSTISAGE1A J055835.31--682009.7&C--rich&29.860&2.606&31.540&0.989&26.200&0.957&16.000&0.617&1.761&0.076\\
SSTISAGE1A J055311.71--684720.9&C--rich&19.900&1.948&16.800&1.066&13.170&0.819&9.984&0.475&1.570&0.077\\
SSTISAGE1A J055036.67--682852.3&C--rich&1.080&0.494&1.209&0.257&1.308&0.207&0.953&0.160&0.238&0.051\\
SSTISAGE1A J052742.48--695251.5&Extreme&69.590&2.958&62.220&2.082&50.750&1.422&40.600&1.094&7.258&0.160\\
SSTISAGE1A J052714.19--695524.3&Extreme&39.680&1.490&40.240&1.117&37.570&1.141&32.440&0.895&6.457&0.154\\
SSTISAGE1A J053239.06--700157.5&Extreme&49.920&1.958&33.430&0.812&24.320&0.834&18.790&0.459&2.950&0.068\\
SSTISAGE1A J052950.52--700000.1&Extreme&74.320&3.098&91.780&3.533&110.200&2.910&117.100&2.549&27.820&0.396\\
SSTISAGE1A J053441.38--692630.7&Extreme&66.460&3.197&83.610&3.607&86.330&1.828&77.850&2.089&14.490&0.305\\
\enddata
\tablenotetext{1}{The electronic version of this table contains all the AGB candidates with valid 8 \mic\ excesses.}
\tablenotetext{a}{The MIPS 24 \mic\ and IRAC 8.0, 5.8, 4.5, and 3.6 \mic\ band excess fluxes and errors in mJy. }
\tablenotetext{b}{The errors in the excess fluxes have been calculated by propagating the photometric errors:
\ben
\nonumber
\frac{\delta X_\nu}{X_\nu}=\sqrt{\left(\frac{\delta F_\nu}{F_\nu}\right)^2+\left(\frac{F^{mod}_\nu}{F^{mod}_{\rm H}}\right)^2\left(\frac{\delta F_{\rm H}}{F_{\rm H}}\right)^2}
\een
}
\end{deluxetable}
\end{landscape}

\noindent
a look-up table with the excess ratio calculated as per Equation \ref{coltemp} for a wide range of temperatures, and selecting the value of temperature from this table that reproduces the observed ratio of excesses. Once the color temperature is calculated, Equation \ref{coltau} can be solved for the optical depth. The color temperature and optical depth values calculated using these simplified equations are primarily useful in investigating the trends of color temperature and optical depth with excess. The equations break down for sources with high optical depth ({\it i.e.}, for extreme AGB stars). The temperature derived from a ratio of two broad band fluxes is a simplification that will in turn affect the optical depth calculation.  We will model the CS shells of our AGB candidates considering details such as the wavelength dependence of the emissivity and the choice of dust species in a future paper to obtain more precise estimates for the temperatures and optical depths.

Since we are only interested in the overall trend of the optical depth with excess flux, we fix $R_{in}$ at ten times the stellar radius $R_*$, which is calculated by scaling the radii of the model photospheres to the luminosity of each star. According to models of dust condensation, most of the dust has condensed by $\sim$ 10$R_*$ \cite[see, {\it e.g.},][]{Hofner2007}. In practice, the dust condensation radii and $R_{in}/R_*$ values will vary, but this complication is ignored in this work as we are only interested in a comparative study of the MIR color temperatures and optical depths. While the temperature of CS dust varies with radius from $\sim$1000 to $\sim$100 K, the (single) color temperature calculated here will be dominated by the distance corresponding to the hottest and most optically thick dust emission in the CS shell, which is typically close to the inner radius of the dust shell. Our values for $\tau_\nu$ describe the optical depth of the high density inner region of the CS shell. This estimate is insensitive to large but cool optical depths. This optical depth will be the dominant component of the line of sight optical depth. Table \ref{TempTauList} shows the color temperatures and optical depths for the sources in Tables \ref{SourceList} and \ref{ExcessList}. Data for the entire set of sources with valid 8 \mic\ excesses is available with the electronic version of \citet{excesses}.

\section{Results}
\label{sec:excesses:results}
\noindent In this section, we investigate the relation of the MIR excess to other observed parameters for the sample including luminosity, color temperature of the CS dust, and opacity of the CS dust shell. We focus on the statistical nature of the sample and global trends of the results. This approach provides the range of physical parameters for the sample and is a first and necessary step to obtain appropriate ranges for model parameters for any detailed modeling that may follow for the sample.

\subsection{Excess-Luminosity Relations}
\begin{figure}[!htb]
\epsscale{0.8}\plotone{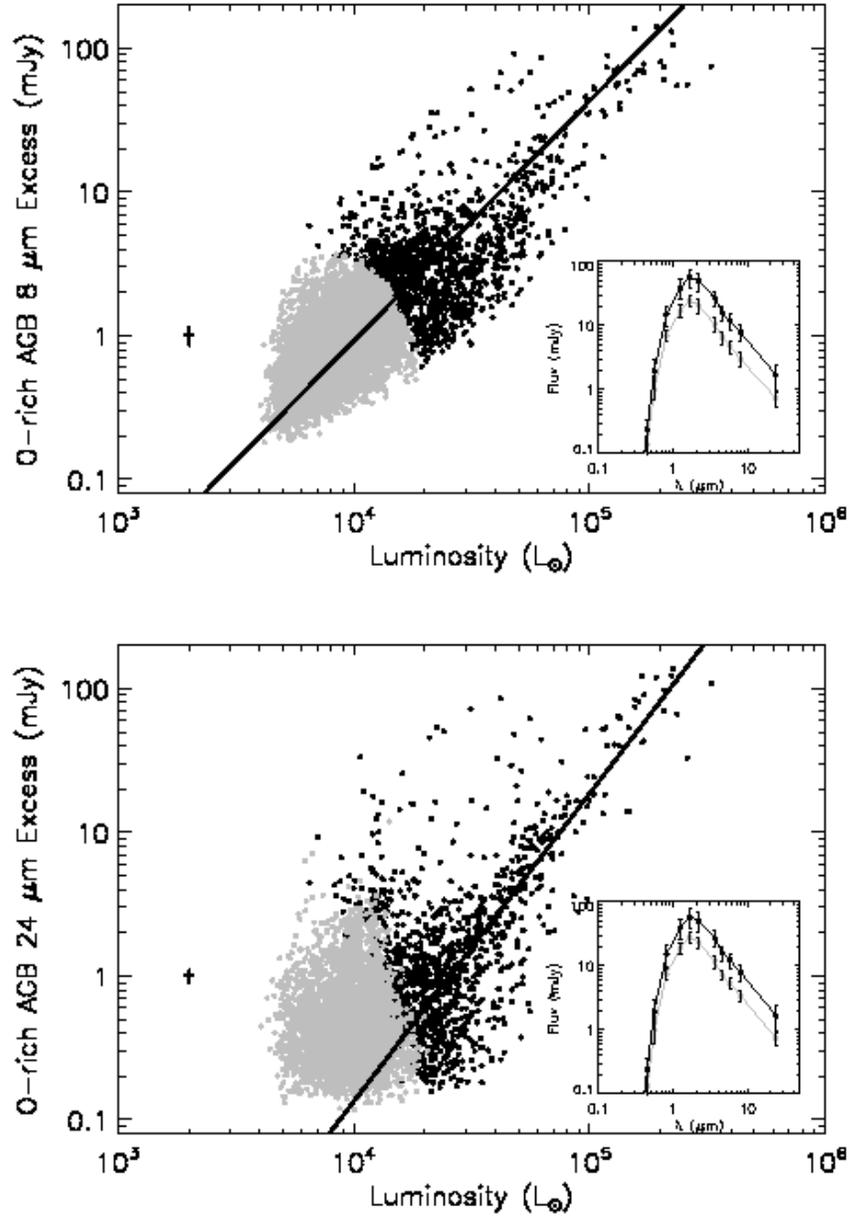}
\ssp{\caption[8 \mic\ excess versus luminosity for O--rich AGB stars]{The 8 \mic\ (top) and 24 \mic\ (bottom) excesses plotted against luminosity for the O--rich sources. Sources from the bright and faint populations (light gray and dark gray circles respectively) are shown.  The solid lines show, respectively, power-law fits to all the sources with reliable 8 \mic\ excesses (top) and sources with reliable 24 \mic\ excesses brighter than $3\times 10^4 $ L$_\odot$ (bottom). Representative error bars are also shown. Inset: the median SEDs of the bright and faint populations.\label{fig:Oxs}}}
\end{figure}

\begin{figure}[!htb]
\epsscale{0.8}\plotone{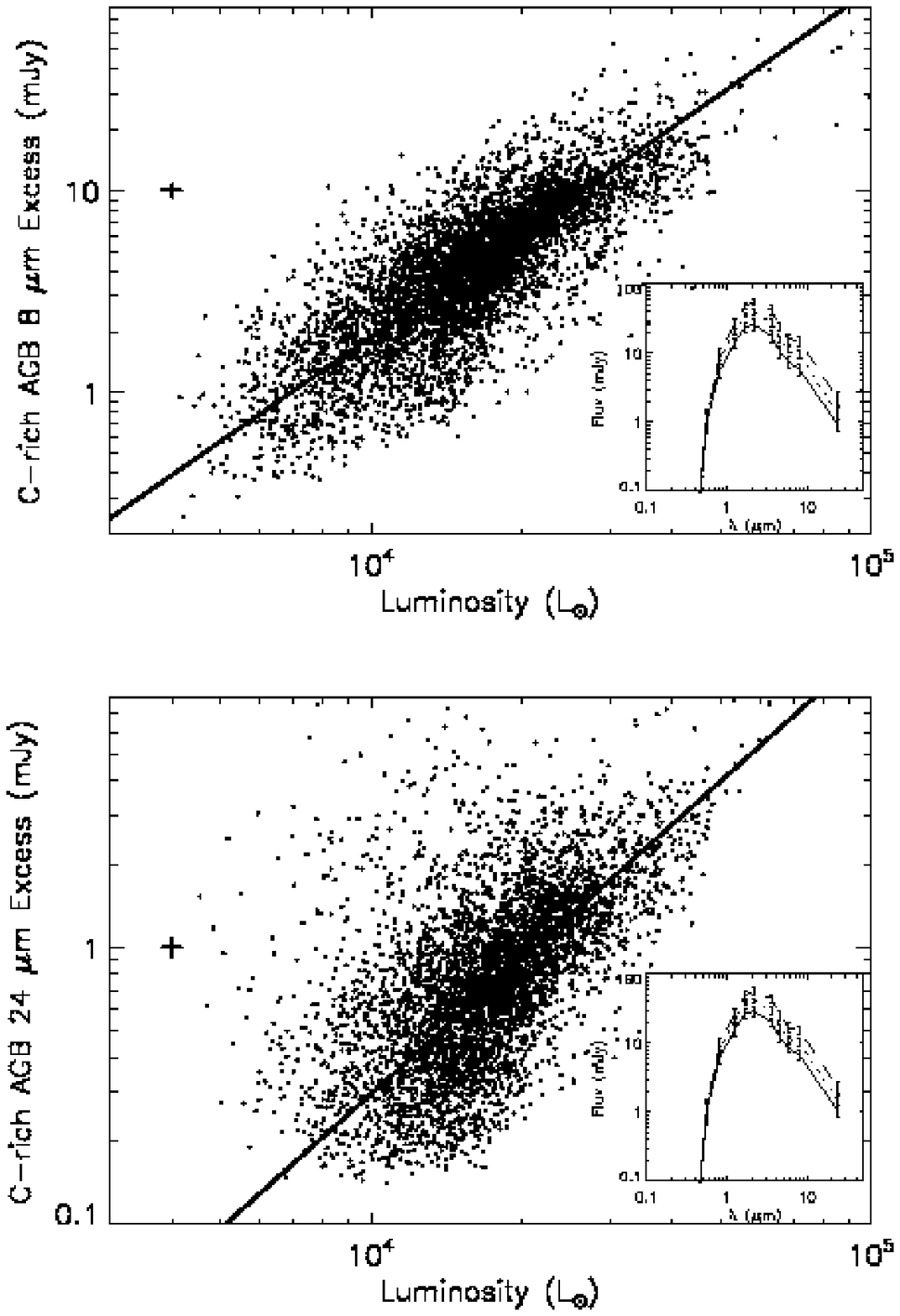}
\ssp{\caption[8 \mic\ excess versus luminosity for C--rich AGB stars]{The 8 \mic\ (top) and 24 \mic\ (bottom) excesses plotted against luminosity for the C-rich sources. The solid lines are power-law fits to the excess--luminosity relations. (The fit is very poor in the 24 \mic\ case) Representative error bars are also shown. Inset: the median SEDs for three equally-populated luminosity bins (solid line: least luminous one-third of the sample, thin dashed line: intermediate luminosity, thick dashed line: most luminous).\label{fig:Cxs}}}
\end{figure}

\begin{figure}[!htb]
\epsscale{0.8}\plotone{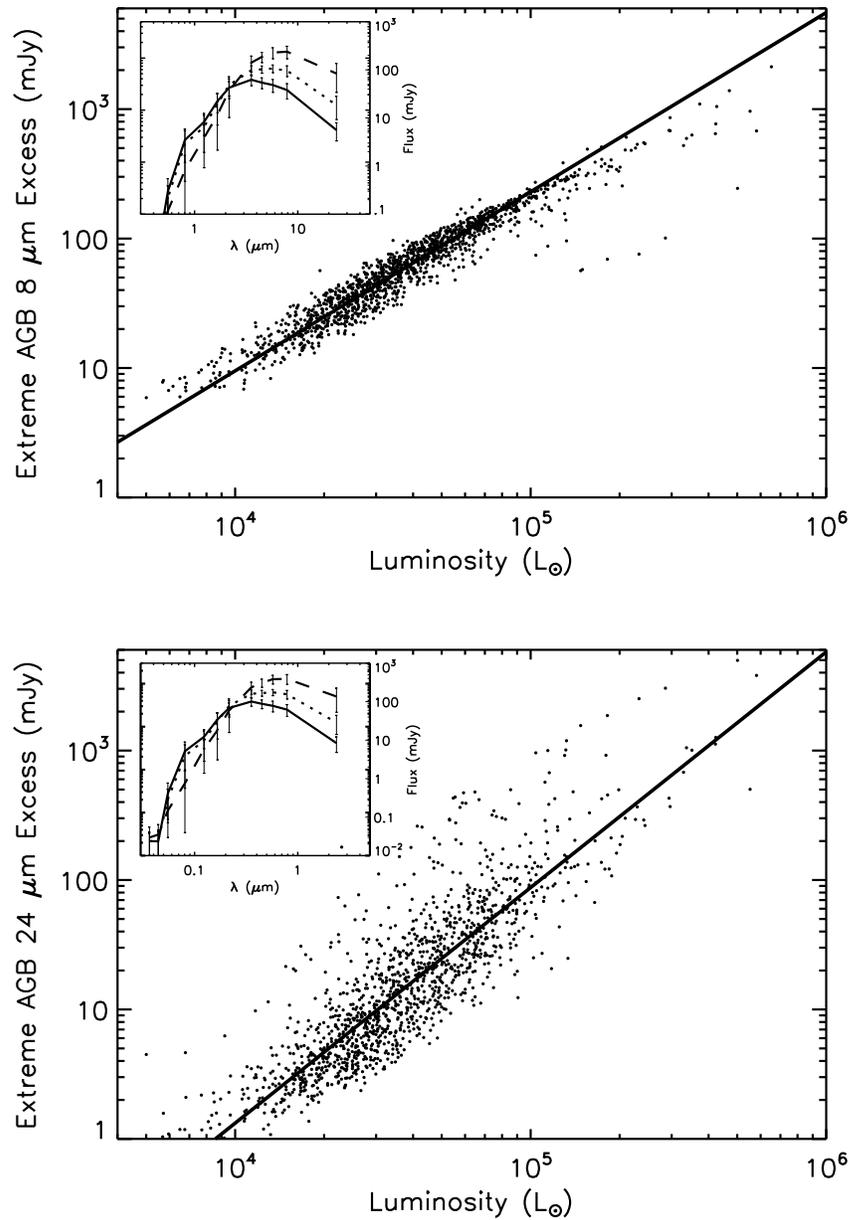}
\ssp{\caption[8 \mic\ excess versus luminosity for extreme AGB stars]{The 8 \mic\ (top) and 24 \mic\ (bottom) excesses plotted against luminosity for the extreme sources. The solid lines are power-law fits to the excess--luminosity relations. Representative error bars are also shown. Inset: the median SEDs for three equally-populated luminosity bins (solid line: least luminous third of sample, thin dashed line: intermediate luminosity, thick dashed line: most luminous).\label{fig:Xxs}}}
\end{figure}

\noindent The calculated IR excesses show a general increasing trend with luminosity at all IRAC wavelengths for O-- and C--rich AGB sources. This trend is most obvious at 8 \mic\ (Figures \ref{fig:Oxs} and \ref{fig:Cxs}). While the correlation is similar at the shorter IRAC wavelengths, the number of sources with significant reliable excesses is considerably smaller. This is expected for the warmer, moderately obscured non-extreme sources. The 8 and 24 \mic\ excesses of the extreme AGB candidates also correlate well with their luminosities, as is evident from Figure \ref{fig:Xxs}. There is a huge spread in 24 \mic\ excesses for the O--rich and C--rich candidates overall, but an increasing trend is apparent for the O--rich AGB candidates at luminosities above $\sim 3\times 10^4$ L$_\odot$. The large spread at low luminosities may be caused by significant variations in the MLRs during the early stages of AGB stars. The plots of O--rich sources show the bright and faint population sources in light and dark grey respectively. We also display the median SED for each population as an inset. For the C--rich and extreme sources, we show the median SED for three equally-populated bins in luminosity. Relations of the form
\ben
\label{powerlaw}
\log{X_\nu({\rm mJy})}=A+B\log{L ({\rm L}_\odot)}
\een

\clearpage
\begin{landscape}
\thispagestyle{empty}
\begin{deluxetable}{rrrrrrrr}
\tablewidth{0pt}
\tablecolumns{8}
\tablecaption{Color temperatures and Optical Depths \label{TempTauList}\tablenotemark{1}}
\tabletypesize{\scriptsize}
\tablehead{
\colhead{Identifier}& \colhead{Type}& \colhead{T\tablenotemark{a}}& \colhead{$\delta$T}& \colhead{$ \tau$(8 \mic)\tablenotemark{b}}& \colhead{$\delta\tau$(8 \mic)}& \colhead{$\tau$(24 \mic)}& \colhead{$\delta\tau$(24 \mic)}\\
&&\colhead{(K)}&\colhead{(K)}&&&&
}
\startdata
SSTISAGE1A J054938.72--683458.2&O--rich&393&28&0.813&0.017&0.938&0.003\\
SSTISAGE1A J055530.35--684647.5&O--rich&352&37&0.806&0.028&0.935&0.005\\
SSTISAGE1A J055420.11--680449.5&O--rich&422&36&0.885&0.012&0.962&0.002\\
SSTISAGE1A J055729.20--684444.2&O--rich&502&48&0.916&0.008&0.972&0.002\\
SSTISAGE1A J055321.17--683114.7&O--rich&401&21&0.743&0.016&0.914&0.003\\
SSTISAGE1A J054522.57--684244.5&C--rich&525&26&0.922&0.004&0.974&0.001\\
SSTISAGE1A J055650.80--675030.5&C--rich&814&84&0.982&0.001&0.994&0.000\\
SSTISAGE1A J055835.31--682009.7&C--rich&705&32&0.953&0.002&0.985&0.000\\
SSTISAGE1A J055311.71--684720.9&C--rich&559&22&0.931&0.002&0.977&0.000\\
SSTISAGE1A J055036.67--682852.3&C--rich&446&53&0.949&0.007&0.983&0.001\\
SSTISAGE1A J052742.48--695251.5&Extreme&521&10&0.830&0.003&0.944&0.001\\
SSTISAGE1A J052714.19--695524.3&Extreme&495&9&0.793&0.004&0.931&0.001\\
SSTISAGE1A J053239.06--700157.5&Extreme&559&11&0.908&0.002&0.970&0.000\\
SSTISAGE1A J052950.52--700000.1&Extreme&456&5&0.601&0.005&0.867&0.001\\
SSTISAGE1A J053441.38--692630.7&Extreme&511&9&0.808&0.003&0.936&0.001
\enddata
\tablenotetext{1}{The electronic version of this table contains all the AGB candidates with valid 8 \mic\ excesses.}
\tablenotetext{a}{Color temperatures and related uncertainties, derived from the 8 \mic\ and 24 \mic\ excesses. The uncertainties are calculated by propagating the photometric errors.}
\tablenotetext{b}{Optical depths and related uncertainties at 24 \mic\ and 8 \mic\ derived from the color temperature. The uncertainties are calculated by propagating the photometric errors.}
\end{deluxetable}
\end{landscape}
\clearpage

are fit to the 8 and 24 \mic\ excesses, resulting in the following best-fit parameters:

\ben
\nonumber
A=-6.7\pm 0.5,\ B=1.7\pm 0.1~\rm{(O-rich, 8\ \mu m)}\\
\nonumber
A=-9.4\pm 3.2,\ B=2.1\pm 0.7~\rm{(O-rich, 24\ \mu m, L> 3\times 10^4~L_\odot)}
\een

\ben
\nonumber
A=-6.6\pm 6.2,\ B=1.7\pm 1.5~\rm{(C-rich, 8\ \mu m)}\\
\nonumber
A=-7.0\pm 15, \ B=1.6\pm 3.5~\rm{(C-rich, 24\ \mu m)}\\
\nonumber
A=-5.0\pm 0.3,\ B=1.5\pm 0.1~\rm{(Extreme, 8\ \mu m)}\\
\nonumber
A=-8.3\pm 0.7,\ B=2.1\pm 0.2~\rm{( Extreme, 24\ \mu m)}
\een

The poor fit to the 24 \mic\ excess for the C--rich candidates reflects the huge spread of 24 \mic\ excesses for these sources.

\subsection{Color Temperature and Opacity}
\begin{figure}
\epsscale{0.8}\plotone{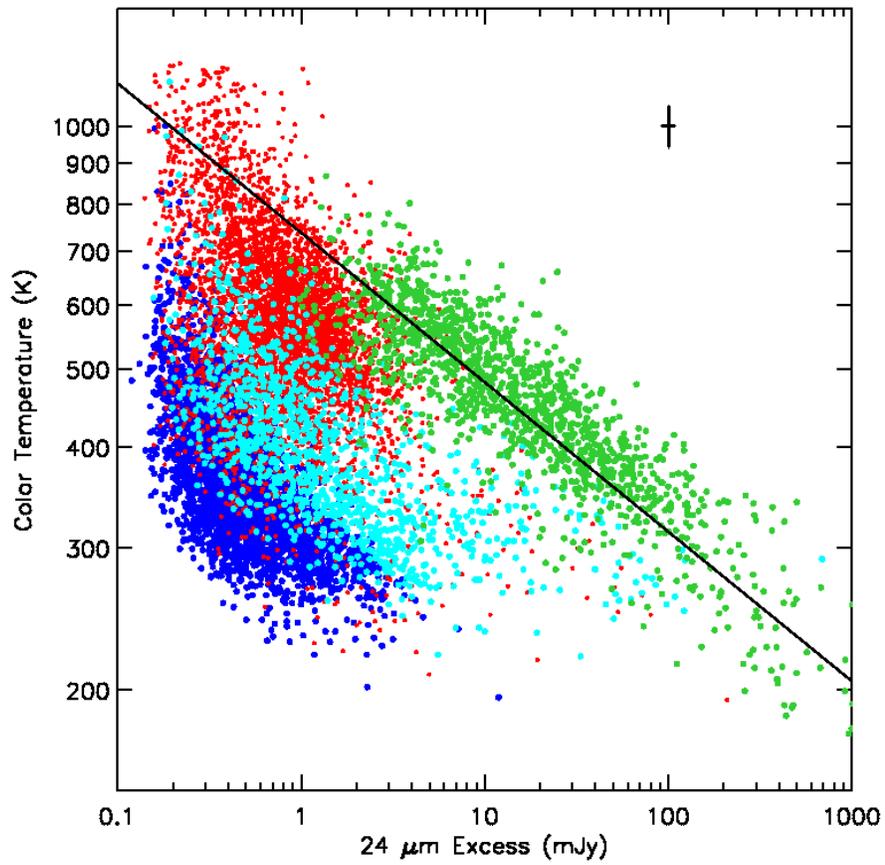}
\ssp{\caption[Color temperature versus 24 \mic\ excess for all three AGB star types]{The color temperature derived from the ratio of the 8 \mic\ and 24 \mic\ excesses plotted against the 24 \mic\ excess for all three types of AGB candidates (blue: faint O--rich, cyan: bright O--rich, red: C--rich, green: Extreme). The cross is a representative error bar.\label{fig:temp}}}
\end{figure}

\begin{figure}
\epsscale{0.8}\plotone{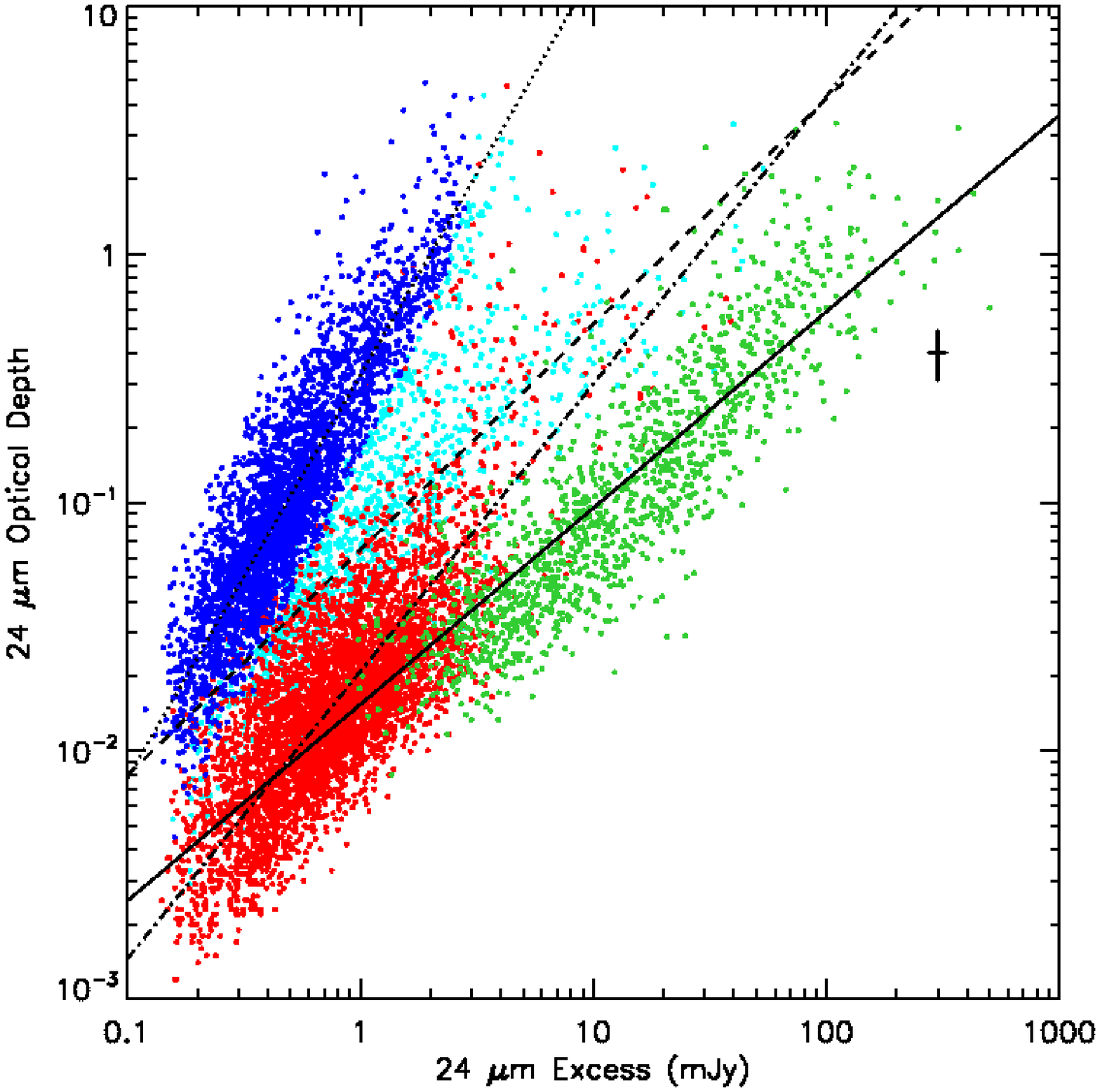}
\ssp{\caption[24 \mic\ optical depth versus 24 \mic\ excess for all three AGB star types]{The 24 \mic\ optical depth derived from the 8 \mic\ and 24 \mic\ excesses for all three types of AGB candidates. The color coding is the same as for Figure \ref{fig:temp}.  Power-law fits to the optical depth--excess relation are shown (dotted: faint O--rich, dashed: bright O--rich, dot-dashed: C--rich, solid: Extreme). The cross is a representative error bar.\label{fig:tau}}}
\end{figure}

\begin{figure}
\epsscale{0.8}\plotone{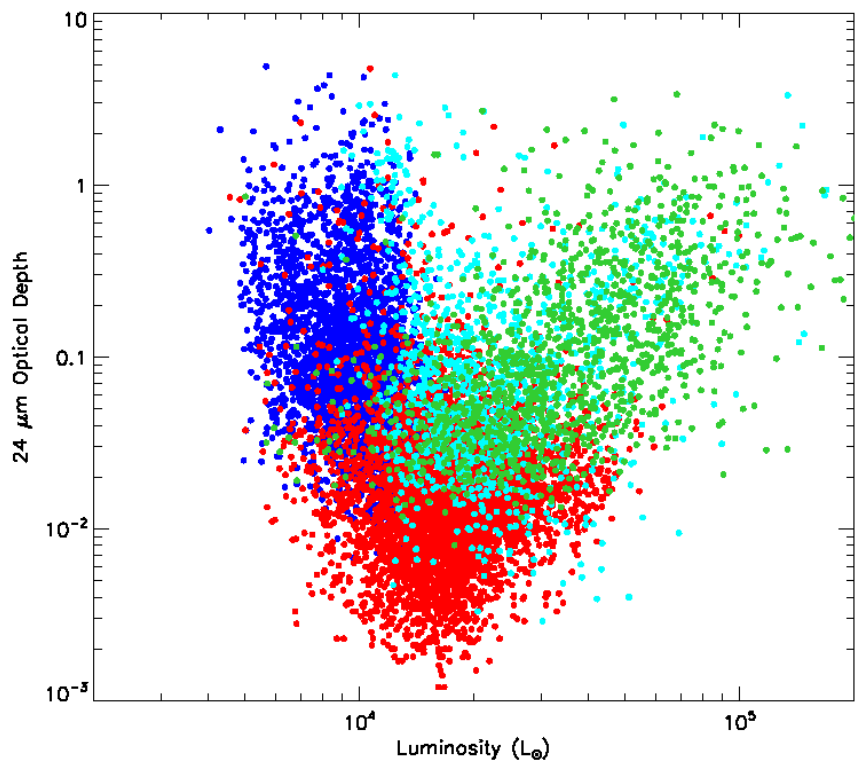}
\ssp{\caption[24 \mic\ optical depth versus luminosity for all three AGB star types]{The 24 \mic\ optical depth as a function of luminosity for all three types of AGB candidates. The color coding is the same as for Figure \ref{fig:temp}.\label{fig:tauvsl}}}
\end{figure}

\begin{figure}
\epsscale{0.8}\plotone{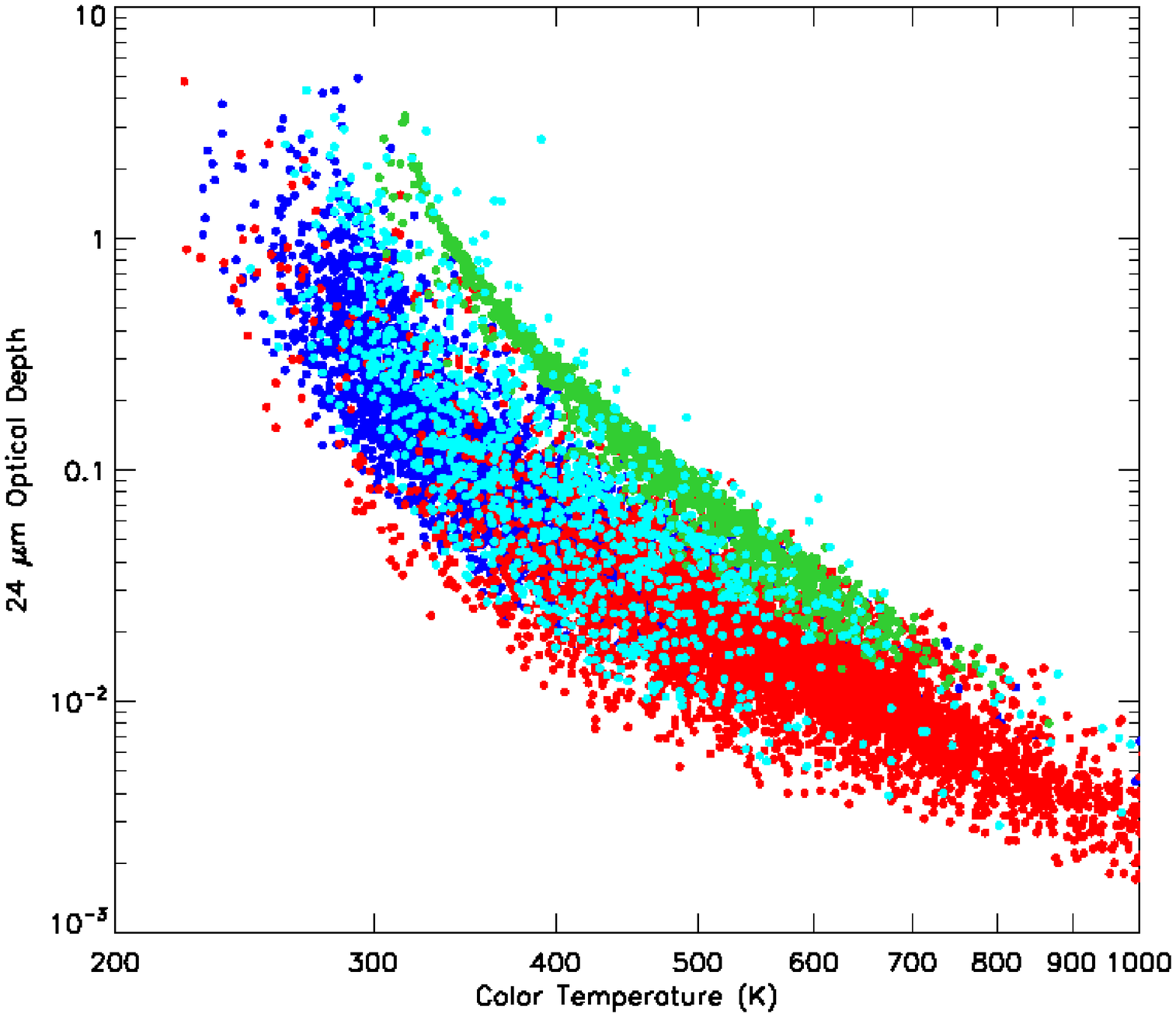}
\ssp{\caption[24 \mic\ optical depth versus color temperature for all three AGB star types]{The 24 \mic\ optical depth plotted against color temperature for all three AGB types. The color coding is the same as for Figure \ref{fig:temp}.\label{fig:tauvstemp}}}
\end{figure}

\noindent Figures \ref{fig:temp} and \ref{fig:tau} show the variation of the color temperature and 24 \mic\ optical depth with 24 \mic\ excess for all three types of AGB candidates. As expected from their redder [8]$-$[24] color, the members of the faint O--rich population tend to have cooler temperatures and higher optical depths than the bright population, and they also show a more pronounced variation with excess. There is a huge spread in the C--rich AGB color temperatures. The maximum color temperature is higher than in the O--rich case, suggesting that the carbonaceous dust grains become hotter than the silicates. AGB winds composed of oxygen-rich compounds are less efficient at absorbing visible photons than those that are carbon-rich \citep{WK1998}. Carbonaceous grains are efficient absorbers of optical photons and are highly emissive at IR wavelengths in comparison to silicates, thus they are more likely to reach higher temperatures. We fit the color temperatures for the extreme AGB candidates with a power-law relation of the form
\ben
\nonumber
\log{T_d({\rm K})}=A+B\log{X_\nu({\rm mJy})} 
\een
and obtain the following best-fit values:
\ben
\nonumber
A=2.9\pm 0.4,\ B=-0.19\pm 0.02~\rm{(Extreme)}
\een
while a similar relation,
\ben
\log{\tau_{24 \mu {\rm m}}}=A+B\log{X_\nu({\rm mJy})} 
\een
when fit to the 24 \mic\ optical depths, gives
\ben
\nonumber
A=-0.49\pm 0.12,\ B=1.6\pm 0.1~\rm{(O-rich, faint)}\\
\nonumber
A=-1.2\pm 0.3,\ B=0.9\pm 0.2~\rm{(O-rich, bright)}\\
\nonumber
A=-1.7\pm 0.3,\ B=1.2\pm 0.2~\rm{(C-rich)}\\
\nonumber
A=-1.8\pm 0.1,\ B=0.79\pm 0.09~\rm{(Extreme)}
\een
Figures \ref{fig:tauvsl} and \ref{fig:tauvstemp} show the variation of the 24 \mic\ optical depth with luminosity and color temperature respectively. There is no trend in the optical depth of the O--rich candidates with luminosity. For the C--rich sources, there is no correlation, but there appears to be an absence of sources with low optical depth at higher luminosities. For the extreme AGB stars, there appears to be a positive correlation of higher optical depth with higher luminosities. The optical depth decreases monotonically with the color temperature for all three types of AGB candidates. CS dust shells have temperature gradients from $\sim$ 1000 K in the interior to $\ltapp$ 100 K in the outer regions. With increasing optical depth, therefore, we are probing the progressively cooler, outer regions of the AGB star candidates.

\section{Discussion}
\label{sec:excesses:discuss}
\noindent The 8 and 24 $\mu$m band excess fluxes show an increase with increasing luminosity of the central star. For each type of AGB candidate, the slope of the excess-luminosity relation is similar across the four IRAC and MIPS24 bands, though in the IRAC 8 \mic\ and MIPS24 bands, excesses are ``reliable" only beyond $\sim$0.1 mJy (see Figure \ref{fig:xshisto}). The increase of excess with luminosity is consistent with the MLR--luminosity relation found by \citet{vL1999}. The mass loss-rate is roughly proportionate to $\tau$$L$ \citep[see, {\it e.g.},][]{IE1995}, so we expect that the MLR increases with increasing luminosity as long as the optical depth does not decrease faster than $\sim$L$^{-1}$. This suggests that the excess is a good reflection of the MLR. More quantitative comparisons of MLR for this whole sample with luminosity is beyond the scope of this study, but this will be addressed in a future paper.

The uncertainties in photometry alone cannot account for the considerable spread in the calculated excesses. Stars on the AGB suffer from variability and episodic mass loss, and our observations of these stars may be capturing their fluxes at different epochs in their variability cycle, or during episodes of increased or diminished mass loss. At a given luminosity, these effects introduce a variation in the MLR and hence also affect the excess. \citet{Vijhetal} show a median variability index of about 3 for the O--rich and C--rich AGB sources and about 5 for the extreme sources. This would correspond to a change in the excess fluxes of about the same factor. The color temperature depends on the ratio of the excesses, and the effect of variability is probably weaker. The effects of grain chemistry, metallicity of the environment, and the C/O ratio of these sources have not been accounted for in this study, as this requires spectroscopic determination. At a given luminosity, there is also a dependence on progenitor mass -- a degeneracy between a less evolved, more massive star and a more evolved, less massive star arises. Such degeneracy effects increase the spread at the lower luminosity end (only the most massive stars evolve to high luminosities).

Observations for Galactic AGB stars \citep{Guandalinietal06,Bussoetal07} show that C--rich AGB stars are in general more obscured than their O--rich counterparts, but this is partly due to the higher opacity of carbonaceous dust. Note that extreme O--rich AGB stars ({\it i.e.}, OH/IR stars) are even more obscured than these C--rich sources. In the case of the extreme sources, there is a high optical depth regardless of chemistry of the CSE. As an AGB star evolves, more mass in the form of dust is deposited into the surrounding shell, increasing its opacity and causing further reddening of stellar light and stronger emission at longer wavelengths, corresponding to cooler color temperatures. The color temperature and optical depth variations for extreme AGB stars in Figures \ref{fig:temp} and \ref{fig:tau} support these claims. The existence of a redder, fainter population of O--rich sources first seen in the IRAC CMD is also apparent in plots of the excess, color temperature and optical depth. The faint population is in general cooler and more obscured than the bright population. The SAGE survey is able to differentiate between these two types of sources for the first time. The bright population corresponds to the young, most massive AGB stars that prevent the dredge-up of carbon by undergoing hot-bottom burning (part of sequence G in \citet{2MASSLMC}). This population also emerges in the model isochrones of \citet{Marigoetal} -- they find that the thermally-pulsing (TP) phase in their isochrone of age $\log({t/{\rm yr}})=8.2$ is well-developed and populated by O--rich stars undergoing HBB. The faint population consists of AGB stars in which HBB does not occur, resulting in a smooth transition from O--rich to C--rich chemistry. The $\log({t/{\rm yr}})=9.1$ isochrone of \citet{Marigoetal} (sequence J in \citet{2MASSLMC}) represents the typical evolutionary phase corresponding to this population. 

\begin{figure}
\epsscale{0.5}\plotone{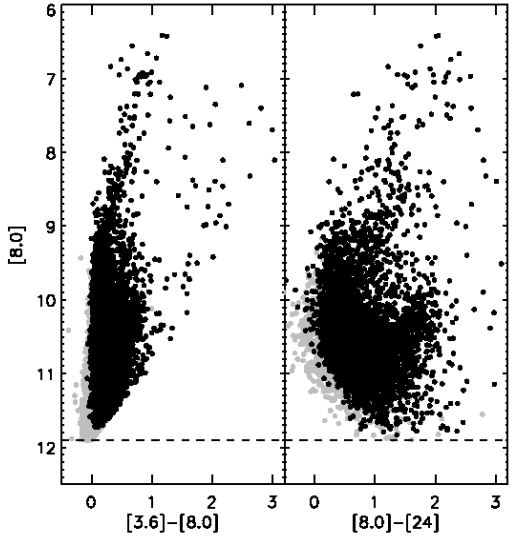}
\ssp{\caption[Discovery space for O--rich AGB candidates]{[8.0] vs [3.6]--[8.0] and [8.0] vs [8.0]--[24] CMDs for the O--rich AGB stars in our list with 8 and 24 $\mu$m detections (gray circles). The sources with ``reliable'' excesses are superimposed (black circles). The tip of the RGB is at [8.0]=11.9 (dashed line)\label{fig:discspace_Orich}}}
\end{figure}

\begin{figure}
\epsscale{0.5}\plotone{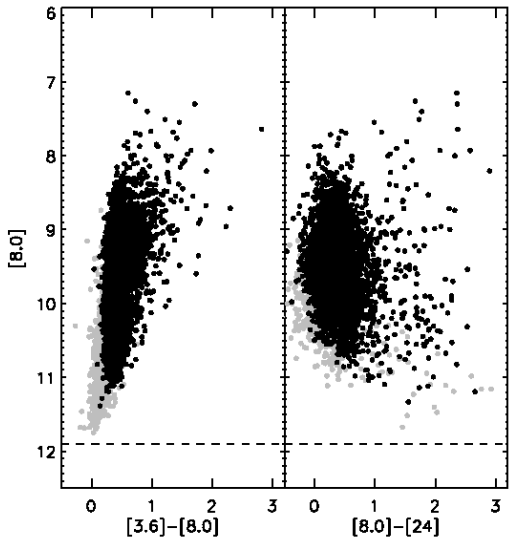}
\ssp{\caption[Discovery space for O--rich AGB candidates]{Same as Figure \ref{fig:discspace_Orich} for our C--rich AGB candidates. \label{fig:discspace_Crich}}}
\end{figure}

\begin{figure}
\epsscale{0.5}\plotone{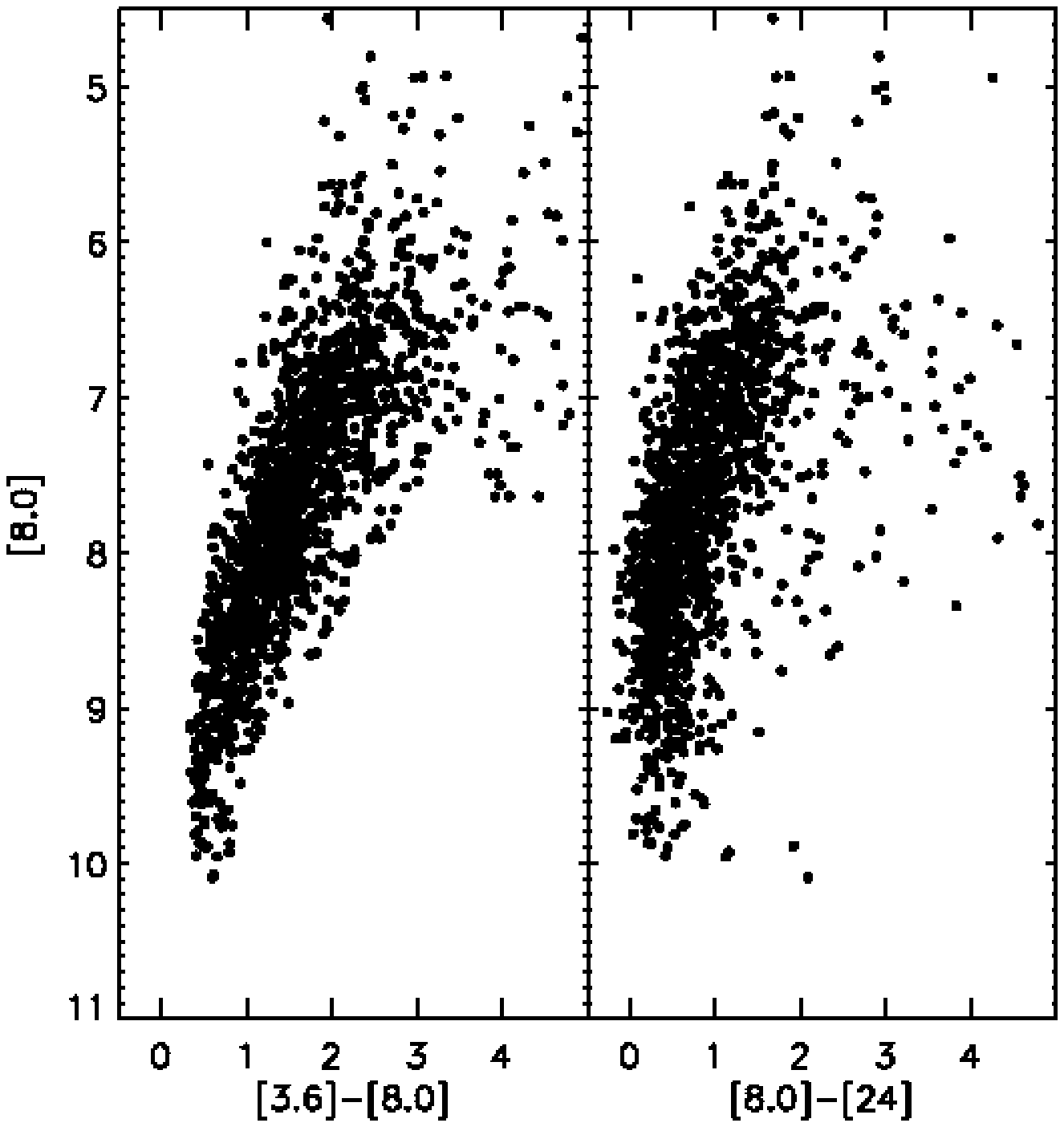}
\ssp{\caption[Discovery space for O--rich AGB candidates]{Same as Figure \ref{fig:discspace_Orich} for our extreme AGB candidates.\label{fig:discspace_xAGB}}}
\end{figure}

Our $\sigma$-clipping method of recovering sources with reliable excesses can be used to compare our results with the expected lower-limit to measurement of MLRs by the SAGE survey \citep{sage1}. The mid-infrared CMDs in Figures \ref{fig:discspace_Orich}, \ref{fig:discspace_Crich}, and \ref{fig:discspace_xAGB} show the sources with reliable 8 \mic\ excesses as black circles. We detect O--rich and C--rich sources with reliable excesses up to within $\sim$ 0.1 mag of the tip of the RGB ([8.0]$\approx$11.9). which is below the detection limit mentioned in \citet{sage1} ([8.0]=11.0) for measuring significant mass loss. The extreme AGB stars in our sample are all well above this detection limit, showing that the SAGE survey detects all the extreme mass-losing AGB sources in the LMC. 

\begin{figure}
\epsscale{0.5}\plotone{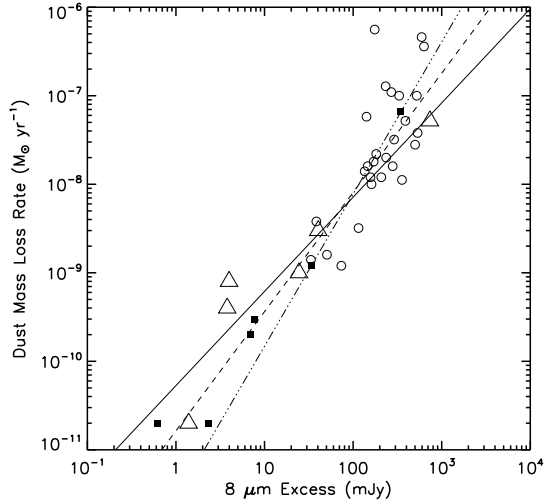}
\ssp{\caption[\citet{vL1999} mass-loss rates versus 8 \mic\ excess]{The mass-loss rates from \citet{vL1999} for O--rich (triangles), C--rich (filled squares) and extreme (circles) AGB sources plotted against their 8 \mic\ excesses. The lines show power-law fits for the dust mass-loss rate--excess relation for all three types of AGB candidates (solid line: O--rich, dashed line: C--rich, dot-dashed line: Extreme).\label{fig:vLMdotvsX8}}
}
\end{figure}

\begin{figure}
\epsscale{0.8}\plotone{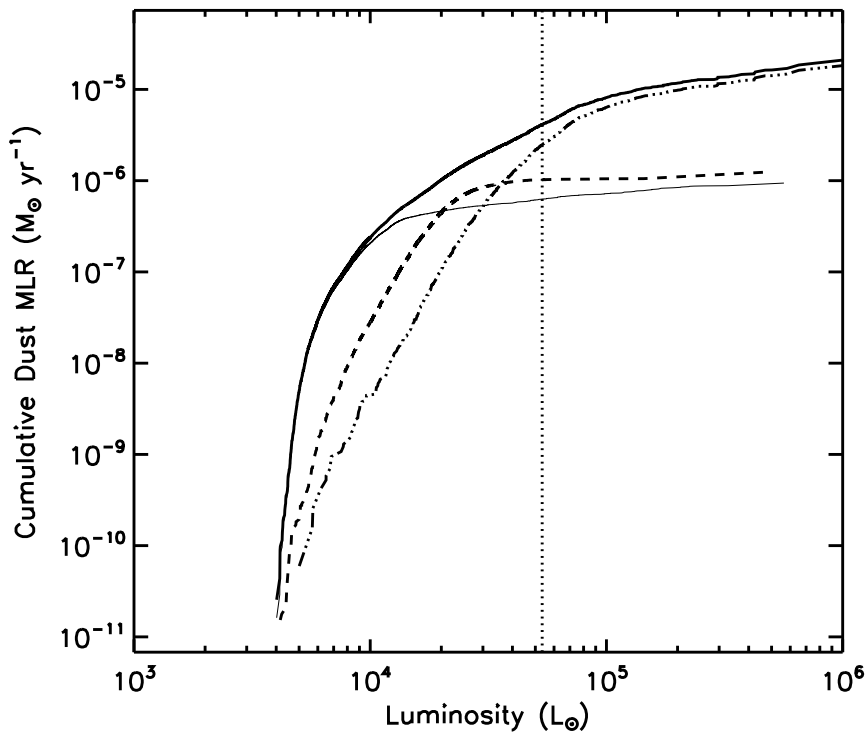}
\ssp{\caption[Cumulative dust injection rate versus luminosity for all three AGB star types]{The cumulative dust mass-loss rate as a function of luminosity for the O--rich (thin solid line), C--rich (dashed line) and Extreme (dot-dashed line) AGB stars in our lists. The thick solid line is the total AGB dust mass-loss rate as a function of luminosity. The extreme AGB stars are the major contributors to the mass-loss rate, at 2.36$\times 10^{-5}$ \msunperyr. The total dust mass-loss rate is 2.74$\times 10^{-5}$ \msunperyr. The vertical dotted line is the classical AGB luminosity limit. As is evident from the figure, there are very few, if any, carbon--rich AGB candidates above this limit. However, deeply embedded extreme AGB stars and O--rich stars undergoing hot-bottom burning can exceed this luminosity limit.\label{fig:mdotcontrib}}}
\end{figure}

We estimate the current LMC AGB star mass-loss budget using a method analogous to \citet{Blumetal07}. The MLRs obtained by \citet{vL1999} for spectroscopically identified AGB stars are plotted against their SAGE 8 \mic\ excess fluxes in Figure \ref{fig:vLMdotvsX8}. The figure also shows power-law fits of the form
\ben
\nonumber
\log{\dot{M} (10^{-6} ~{\rm M}_\odot {\rm yr}^{-1})}=A+B\log{X_\nu({\rm mJy})} 
\een
to all three types of sources with the following best-fit parameters:
\ben
\nonumber
A=-1.5,\ B=1.0~\rm{(O-rich)}\\
\nonumber
A=-1.7,\ B=1.1~\rm{(C-rich)}\\
\nonumber
A=-2.8,\ B=1.7~\rm{(Extreme)}
\een
These relations are then used to derive MLRs for all the sources in our lists with reliable excesses. Figure \ref{fig:mdotcontrib} shows the total dust MLR as a function of LMC AGB luminosity. The extreme AGB stars, despite their comparatively low numbers, are the most significant contributors to the AGB mass-loss budget in the LMC. We find the dust injection rates from O--rich, C--rich and extreme AGB stars to be 0.14, 0.24, and 2.36 $\times 10^{-5}$ \msunperyr\ respectively, which puts the estimate for the total dust injection rate from AGB stars at $2.74\times 10^{-5}$ \msunperyr. While stars with luminosities between $M_{bol}=-7.1$ and $M_{bol}\approx -7.8$ may be highly embedded AGB stars, sources brighter than $M_{bol}> -7.8$ are definitely supergiants (Sloan et al., in preparation)
The luminosity distribution (Figure \ref{fig:AGBLF}) shows that there are very few sources in this range, but it is clear from Figure \ref{fig:mdotcontrib} that the contribution of the red supergiants to the dust injection rate is comparable to that of the AGB stars. This introduces a significant uncertainty in the value for the AGB dust injection rate derived in this section. 

Calculation of the total (gas+dust) injection rate requires knowledge of the gas:dust ratio, $\psi$, in the circumstellar shells of AGB stars. The lower metallicity of the LMC will result in a higher silicate gas:dust ratio but the carbon gas:dust ratio may be similar to Galactic values \citep{Habing1996}. Keeping this in mind, we choose gas:dust ratios of 200 and 500 for C--rich and O--rich stars respectively\footnote{The work of \citet{vL2008} suggests that $\psi$ scales in proportion to the metallicity for both O--rich and C--rich AGB stars, but this effect can not be accounted for in our simple estimate.}. The C--rich value is similar to that obtained by \citet{Skinneretal1999} for the Galactic carbon star, IRC+10 216. The gas MLR from the O--rich and C--rich AGB candidates is then 7$\times 10^{-4}$ \msunperyr\ and 4.8$\times 10^{-4}$ \msunperyr\ respectively. The chemical identification of the extreme AGB sources is not possible from our data, but assuming $\psi$=200 and $\psi$=500 for these stars will provide lower and upper limits. The gas MLR for the extreme AGB population is in the range 4.7--11.8$\times 10^{-3}$ \msunperyr. Thus, the total AGB gas MLR is (5.9--13)$\times 10^{-3}$ \msunperyr. We will improve on our estimate with our follow-up radiate transfer modeling of the dust shells around these stars using the 2DUST code \citep{UetaMeixner03}. The results of the present study will help constrain the range of optical depths and grain temperatures. The optical depth is an important input parameter for the code.

Taking into account the effects of variability on the excesses as discussed previously, we estimate that the MLR of an individual source can vary on average by a factor of $\sim$3 for the O--rich and  C--rich sources and $\sim$15 for the extreme AGB candidates. Due to the large number of O--rich and C--rich sources, the errors of the flux variations on the cumulative MLR estimates would probably cancel out. For the extreme AGB candidates, the larger effects of the variability and the smaller numbers may result in a significant change in the cumulative MLR. The magnitude of such an effect could be calibrated with an IR variability study of the LMC AGB stars to determine their periods and hence phase-correct the fluxes and excesses, but no such study exists for our complete sample.

Our C--rich AGB star gas MLR of 4.8$\times 10^{-4}$ \msunperyr\ is comparable to the value, 6$\times 10^{-4}$ \msunperyr, found by \citet{Matsuuraetal2009}. However, we note that \citet{Matsuuraetal2009} base their measurement on MLR versus color relations and include the extreme C--rich AGB stars, identified by IRS spectroscopy, as well as the color classified C--rich AGB stars. Our excess versus MLR relation method is two orders of magnitude more sensitive to lower MLR sources (dust MLRs of $\sim 10^{-11}$ \msunperyr) compared to the \citet{Matsuuraetal2009} method (dust MLRs of $\sim 10^{-9}$ \msunperyr). It is therefore fortuitous that our estimates are so similar and our exclusion of the highest MLR extreme C--rich AGB stars is balanced by the inclusion of more low MLR C--rich AGB stars. Furthermore, the total mass injection rate of C--rich mass loss is probably higher than both these estimates, but definitive numbers will require a clear identification of what fraction of the extreme AGB stars are C--rich.

Our estimate for the total LMC AGB gas MLR is higher than those obtained for two local group dwarf irregulars, WLM \citep[][(0.7--2.4)$\times 10^{-3}$ \msunperyr]{Jacksonetal200701} and IC 1613 \citep[][(0.2--1.0)$\times 10^{-3}$ \msunperyr]{Jacksonetal200702}. As their values for the MLRs of individual stars are in good agreement with the estimates of individual stars by \citet{vL1999} for the LMC, the disparity between our results and the values for WLM and IC 1613 could arise due to various reasons. The total MLR return due to AGB stars depends on the total number of AGB stars in a galaxy. The total stellar mass of the LMC \citep[$\sim 3 \times 10^9$ M$_\odot$,][]{vdM02} is two orders of magnitude greater than those estimated by \citet{Jacksonetal200701} and \citet{Jacksonetal200702} for WLM and IC 1613 ($1.1\times 10^7$ M$_\odot$ and $1.7\times 10^7$ M$_\odot$ respectively). Moreover, the age of the LMC bar is 4--6 Gyr \citep{Smecker-Haneetal2002}, which is optimal for an enhanced AGB population at present times, compared to the slightly older stellar population of WLM \citep[9 Gyr,][]{Jacksonetal200701}. Based on the differences in stellar masses and ages, we expect that the LMC has considerably more AGB stars than either WLM or IC 1316, leading to a higher total MLR from AGB stars. However, the differences in the MLR determination may also increase our MLR compared to theirs. Our IR excess method is able to detect sources with small MLR, while \citet{vL1999} and \citet{Jacksonetal200701} derive their estimates from the highest mass-losing sources alone. In this sense, ours is a more complete estimate of the total MLR. On the other hand, our interpolation of the \cite{vL1999} MLRs may not be appropriate for sources with very low excess. Thus, while the excess method is able to identify sources with low MLRs, the derived MLRs may be overestimates.

The current star formation rate (SFR) of the LMC is estimated to be about 0.1 \msunperyr\ \citep{sage3}, which is an order of magnitude higher than our calculated AGB MLR. Thus, the current star formation rate is not sustainable, assuming the LMC is a closed box system, and that AGB stars are the only means of replenishing the ISM. However, the most massive stars ({\it e.g.}, red supergiants, supernova explosions) may also contribute to the mass-loss return, and these need to be accounted for to complete the census of the mass budget return to the LMC. In addition, interactions with the Small Magellanic Cloud may cause infall or loss of mass that would need to be considered in the mass budget of the ISM.

\section{Summary and conclusions}
\label{sec:excesses:fin}
\noindent We classify evolved stars in the SAGE Epoch 1 Archive based on their 2MASS and IRAC colors, and use photospheric models to estimate the infrared excess emission due to circumstellar dust. We obtain about 8200 O--rich, 5800 C--rich, and 1400 extreme AGB sources with reliable 8 \mic\ excesses (SNR $\geq$ 3). The corresponding numbers in the 24 \mic\ band are about 4700, 4900, and 1300 respectively. The excesses increase with an increase in luminosity in all four IRAC bands as well as the MIPS24 band. We use the 8 and 24 \mic\ excess fluxes to derive dust color temperatures and optical depths, and we observe that higher excess fluxes correspond to cooler temperatures and optically thicker dust shells.  These quantities for our list of AGB candidates with valid 8 \mic\ excesses are available in the form of electronic tables. We also estimate the present day AGB mass-loss budget in the LMC by comparing modeled mass-loss rates with our excesses estimates to find that the extreme AGB stars are the most significant contributors to mass loss in the LMC.  Our data also suggests that the rate of dust injection from red supergiants to the LMC ISM is comparable to the total dust injection rate from the AGB stars, which we calculate to be about $(5.9-13)\times 10^{-5}$ \msunperyr.


%


\chapter{Modeling The Circumstellar Envelope of \ogles}
\label{ch:cagbmodel}
	        

\section{Introduction}


In order to develop a grid of LMC carbon star models, it is essential to first identify the composition of the dust driving the mass loss in C--rich AGB stars and validate the properties of carbonaceous dust produced. In this chapter, I present a \twodust\ radiative transfer model for the circumstellar dust around the variable carbon star \ogles. This source has been studied as part of many recent LMC surveys \citep[{\it e.g.},][]{Zebrunetal2001, Epchteinetal1999, Cutrietal2003, IRSFLMC, sage1}. This object has been listed as an obscured AGB candidate in \citet{Groenewegen2004}, while \citet{sage2} and \citet{excesses} classified this star as an ``extreme'' AGB candidate (J--[3.6]$>$3.1). The SAGE-Spec spectrum for this source showed signatures of carbonaceous dust, confirming that \ogles\ is an extreme AGB with C-rich chemistry. 

The choice of \ogles\ for this study was motivated by the availability of both SAGE photometry and SAGE-Spec spectroscopic data, and the good agreement ($\ltapp$10\%) between the two. The carbon star candidates chosen for the SAGE-Spec study were required to be infrared-bright. Thus, \ogles\ is redder than most carbon stars in the SAGE sample. Despite this fact, the absence of peculiar features in its spectrum allow us to consider it as a good testbed for the dust properties around LMC C--rich AGB stars.

I begin by describing observational data for \ogles\ from various studies in \S\ref{sec:cagbmodel:obs}. In \S\ref{sec:cagbmodel:analysis} I detail the method employed to create a radiative transfer model for the circumstellar dust as well as a model for the molecular features observed in the mid-infrared (MIR) spectrum. I present the results of these models for \ogles\ in \S\ref{sec:cagbmodel:results} and discuss the implications of these results for the dust properties to be used in the model grid in \S\ref{sec:cagbmodel:discuss}.

\section{Observations}
\label{sec:cagbmodel:obs}

Simultaneous measurements in the optical, NIR and MIR wavelengths are essential for studying the spectral energy distribution (SED) of a variable source. Although SAGE data has been combined with photometry from the optical Magellanic Clouds Photometric Survey \citep[MCPS;][]{MCPS} as well as 2MASS \citep{2MASS}, these observations were not concurrent with SAGE. In order to constrain the optical and near-infrared (NIR) variability, I discuss OGLE and IRSF observations for \ogles\ in \S\ref{subsec:cagbmodel:obs:variability}. In \S\ref{subsec:cagbmodel:obs:sagephot} and \S\ref{subsec:cagbmodel:obs:sagespec}, I describe {\it Spitzer Space Telescope} \citep[{\it Spitzer},][]{Gehrzetal2007} IRAC/MIPS photometry and IRS spectroscopy for \ogles\ (\sages) obtained as part of the SAGE and SAGE-Spec studies, respectively (See \citet{sage1} and Kemper et al. (2009, in preparation) for program overviews).

\subsection{Variability in optical and NIR data}
\label{subsec:cagbmodel:obs:variability}
\ogles\ was observed as part of the Optical Gravitational Lensing Experiment \citep[OGLE-II,][]{Zebrunetal2001} survey of the Magellanic Clouds. The catalog contains BVI light curves for 68\,000 variable objects. \citet{Itaetal2004} and \citet{Groenewegen2004} crossmatched these data with the IRSF LMC Survey \citep{IRSFLMC}, DENIS \citep{Epchteinetal1999} and 2MASS All-Sky Release \citep{Cutrietal2003} NIR catalogs and fit the light curves to obtain variability periods and amplitudes. The period determined by \citet{Itaetal2004} for \ogles\ (361 d) was in good agreement with that of \citet{Groenewegen2004} (358.6$\pm$0.136 d). \citet{Itaetal2004} also reported a mean I magnitude of 17.483 mag and an amplitude of 1.865 mag.


SAGE data is also matched with 2MASS as well as the IRSF survey of the LMC \citep{IRSFLMC}, which contain JH\ks\ photometry for \ogles. While the NIR information from these surveys is useful for qualitative confirmation, the fact that they are not coeval with SAGE observations leads us to expect a slight disagreement between the spectral energy distributions (SEDs) in the NIR and MIR. We model the source using SAGE Epoch 1 photometry due to its agreement with the observed spectrum, and use the optical and NIR data as to constrain the shape of the resulting SED.


\subsection{SAGE photometry}
\label{subsec:cagbmodel:obs:sagephot}
The SAGE \citep[Surveying the Agents of a Galaxy's Evolution,][]{sage1} survey imaged a $\sim 7^\circ \times 7^\circ$ area of the Large Magellanic Cloud (LMC) with the {\it Spitzer} Infrared Array Camera \citep[IRAC;][3.6, 4.5, 5.8 and 8.0 \mic]{IRAC} and Multiband Imaging Photometer \citep[MIPS;][24, 70 and 160 \mic]{MIPS}. Two epochs of observations separated by three months were obtained to constrain source variability. Spectroscopic follow-up observations using the Infrared Spectrograph (IRS) aboard {\it Spitzer} were performed as part of the SAGE-Spectroscopy survey (Kemper et al. 2009, in preparation) on the brightest objects observed in the SAGE survey.

\begin{figure}[t] 
  \epsscale{0.6}
  \plotone{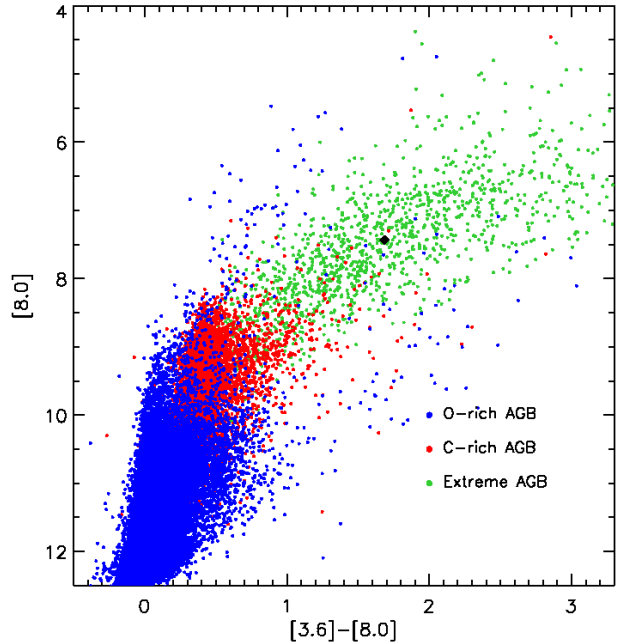}
 \ssp \caption[\ogles\ on the IRAC--IRAC CMD]{The [8.0] vs [3.6]--[8.0] CMD with \ogles\ (filled diamond) superimposed on the O--rich (blue), C--rich (red) and extreme (green) AGB candidates identified by \citet{excesses}. \label{fig:cagbmodel:cmd}}
\end{figure}

IRAC and MIPS epoch 1 archive photometry for \ogles\ is available as part of SAGE data delivered to the Spitzer Science Center\footnote{Data version S13 available on the SSC website, http://ssc.spitzer.caltech.edu/legacy/sagehistory.html} (SSC). \citet{sage2} presented an overview of the evolved star population in the SAGE data of the Large Magellanic Cloud (LMC) and identified asymptotic giant branch (AGB) star candidates based on their locations in NIR and MIR color-magnitude diagrams (CMDs). \citet{excesses} extracted AGB star candidates from the SAGE IRAC Epoch 1 Archive list using similar color cuts (see paper for details). Fig. \ref{fig:cagbmodel:cmd} shows the location of \ogles\ on a [8.0] vs. [3.6]--[8.0] CMD. By comparing data from both epochs, \citet{Vijhetal} identified variable sources
in the LMC and found that AGB candidates make up $\sim81$\% of this population. I use both epochs of SAGE photometry for \ogles\ for the present study, as shown in Fig. \ref{fig:cagbmodel:variability}.


\subsection{SAGE-Spec data}
\label{subsec:cagbmodel:obs:sagespec}
{\it Basic Calibrated Data} products for \ogles\ were obtained from the SSC. The spectra were reduced using techniques described by Kemper et al. (2009, in preparation). \ogles\ was observed in both spectral orders of the Short-Low (SL; 5.2--14.3 \mic, R$\sim$60--120) and Long-Low (LL; 14.3--13.7 \mic, R$\sim$60--120) instruments on board {\it Spitzer} at two nod positions. Data from the S15.3 and S17.2 pipelines for SL and LL, respectively, were obtained from the SSC for \ogles\ (AOR \# 22415360). SL sky subtraction was performed by subtracting the observations in one order from the other while keeping the nod position fixed, while LL sky subtraction involved subtracting one nod position from the other keeping the order fixed. Bad pixels on the detector arrays were replaced with values determined by comparing to the local point-spread function\footnote{For details, please refer to the IRS Data Handbook, available at\\ http://ssc.spitzer.caltech.edu/IRS/dh/dh32.pdf}. 

Point source signal extraction used apertures at each wavelength whose width increased in proportion to the wavelength; this involved using the profile, ridge and extract modules\footnote{These modules are available in SPICE, the {\it Spitzer} IRS Custom Extraction package.}. Flux densities were obtained by calibrating the raw signal extractions using observations of the standard stars HR 6348 (K0 III), HD 166780 (K4 III) and HD 173511 (K5 III). While all three standard stars were used for LL, only the first was used for SL. Where available, the mean flux from multiple data collection events (DCEs) for a single module/order/nod was computed, with the standard deviation from the mean providing a measure of the corresponding uncertainty.  Spikes 
not eliminated by replacement of bad pixel values were effectively removed by replacing any uncertainty larger than the average in the local wavelength range by the uncertainty from the nod position closer to the local average. Bonus-order spectra (corresponding to a small segment of the first-order spectrum obtained by the second-order slit) were averaged with spectra from the second and first orders over wavelengths they shared in common.

\begin{figure}[!htb]
\epsscale{1.0}
\plotone{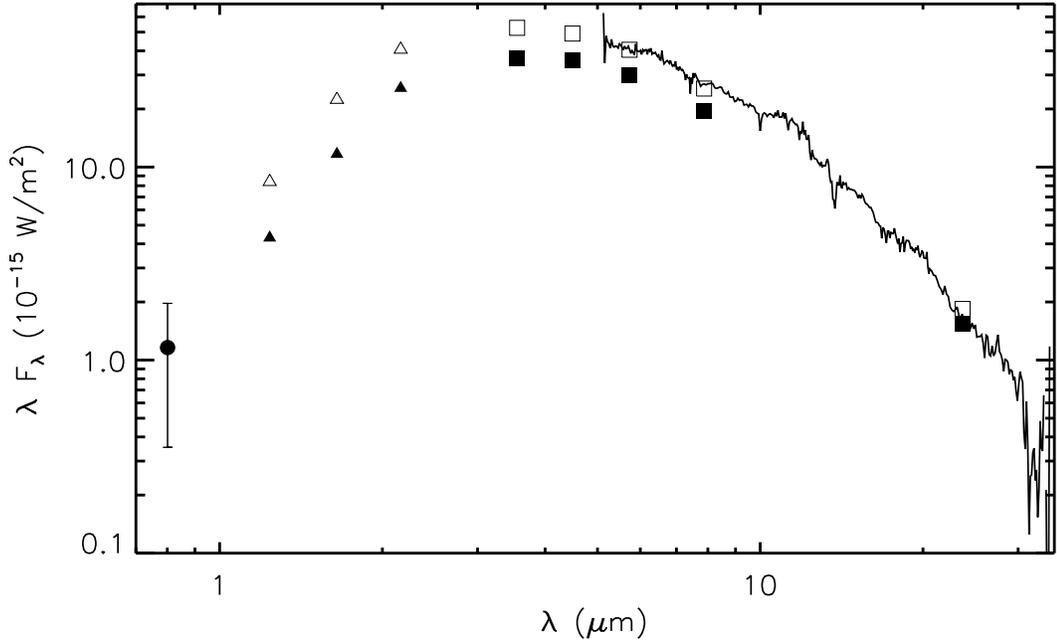}
\ssp{\caption[Multi-band multiphase information for \ogles]{The SED for \ogles\, showing the mean OGLE I band flux (solid circle) and its range of variation (error bar), 2MASS (open triangles), IRSF (filled triangles) and SAGE (Epoch 1: open squares, Epoch 2: filled squares) photometry. The SAGE-Spec IRS spectrum (solid curve) is scaled down to fit the 5.8 \mic\ SAGE Epoch 1 photometry point.\label{fig:cagbmodel:variability}}}
\end{figure}

The IRS spectrum for \ogles\ (Fig. \ref{fig:cagbmodel:variability}) shows typical features of carbonaceous dust -- a featureless continuum at infrared wavelengths combined with a broad 11.3 \mic\ SiC emission feature. There is also a 13.7 \mic\ \acet\ feature. The strength of this feature cannot be reproduced by a model photosphere alone, it is circumstellar in origin \citep[see][]{Matsuuraetal2006} . We model both the circumstellar dust as well as the molecular gas in the extended atmosphere in \S\ref{sec:cagbmodel:analysis}.

The optical, NIR and MIR photometry as well as MIR spectroscopy described above are incorporated into Fig. \ref{fig:cagbmodel:variability}. The two-epoch photometry shows that there is no significant change in the shape of the MIR SED. We have taken advantage of this fact and scaled the SAGE-Spec spectrum down to the SAGE Epoch 1 5.8 \mic\ flux to enable easy comparison to photometry. This scaling changes the flux predicted for the spectrum in the 5.8 \mic\  band by $\sim$5\%. 

\section{Analysis}
\label{sec:cagbmodel:analysis}
\subsection{2Dust Radiative Transfer Models}
\label{sec:cagbmodel:analysis:dust}
The observed spectrum of an AGB star contains contributions from the central star as well as the circumstellar dust shell. We model the AGB star+envelope system using \twodust\ \citep{UetaMeixner03}, a radiative transfer code for axisymmetric systems, with the simplifying assumption of a spherically symmetric dust shell. A comparison of the \twodust\ output SED with the observed spectrum enables us to determine the AGB star mass-loss rate.

As input, \twodust\ needs information about the central star (effective temperature, size, distance and SED), the shell geometry (inner and outer radius, density variation), and the dust grains (species, optical depth at a reference wavelength, grain size distribution). We use the C--rich model photospheres of \citet{Gautschy2004} to represent the central star. To simplify the process, we choose a solar-mass, solar-metallicity model with an effective temperature of 3000 K and surface gravity $\log{g}=-0.4$ and C/O ratio of 1.3 and place it at 50 kpc, the approximate distance to the LMC \citep{Feast99}. 

We assume a constant mass-loss rate and a constant outflow velocity $v_{\rm exp}$, leading to an inverse-square density distribution in the shell. While AGB outflow velocities are well-studied in the Galaxy \citep[see, {\it e.g.},][]{Loupetal1993, Olofssonetal1993}, no measurements of outflow velocities exist for LMC carbon stars. While the lower metallicity of the LMC suggests a lower expansion velocity for O--rich AGB outflows than for their Galactic counterparts, the outflow velocities of carbon stars are less sensitive to metallicity \citep{Habing1996}. For simplicity, we assume a value of $v_{\rm exp}$=10 km s$^{-1}$ and ignore these complications. The ratio of the outer radius of the dust shell to the inner radius is kept fixed at 1000. The outer radius determines the total amount of mass in the shell, which in turn is related to the timescale of the mass loss. While this timescale is an important quantity, we are more interested in obtaining the AGB mass-loss rate, which is only weakly sensitive to changes in the outer radius as long as it is large enough to include contributions to the flux at the longest wavelengths of interest (in our case, the 24 \mic\ band) from the coldest dust in the shell. The mass-loss rate is, however, very sensitive to the value chosen for the inner radius. Observations of Galactic carbon stars \citep[{\it e.g.}, IRC+10216,][]{Danchietal1990,Danchietal1995} and results of radiative transfer modeling \citep[{\it e.g.},][]{vL1999, Groenewegenetal1998} suggest that the inner radius varies in the range $\sim$2--20 stellar radii, while simple estimates based on energy balance suggest that amorphous carbon dust should form within a few stellar radii \citep{Hofner2007}. We vary the inner radius within these bounds until the desired SED shape is obtained.


\twodust\ also requires us to specify the optical depth of the dust at a reference wavelength, which we choose to lie near the center of the SiC emission (11.3 \mic). This optical depth is adjusted until a good fit to the SiC feature is produced. The spectrum of \ogles\ has a strong 11.3 \mic\ SiC feature, and a long-wavelength continuum. We model the dust around this star with amorphous carbon \citep[optical constants from][]{Zubkoetal1996} and silicon carbide \citep[$\alpha$-SiC, optical constants from][]{Pegourie1988} grains. \citet{Pitmanetal2008} discuss the limitations of using the $\alpha$-SiC optical constants and provide newer, more reliable values, but they also require non-spherical grains. This is a complication we would like to avoid for the time being.

The code requires a distribution of grain sizes from which it calculates cross-sections for an ``average" grain. Our grain sizes are distributed according to the \citep{Kimetal1994} (KMH) prescription, with sizes between 0.01 \mic\ and $\sim$1 \mic\ (There is no strict upper bound on the KMH grain sizes; the latter number is an exponential scale height in the distribution.) Given this distribution of sizes as input, \twodust\ then internally calculates absorption and scattering cross-sections and asymmetry factors for an ``average" spherical dust grain from Mie theory \citep{Mie1908} assuming isotropic scattering. We run the code in the Harrington averaging \citep{Harringtonetal1988} mode, which gives an average grain size of $\sim$0.1 \mic, which is typically the single size chosen for dust grains in radiative transfer models \citep[{\it e.g.},][]{vL1999, Groenewegen2006}.

\twodust\ is now run for different values of the inner radius and 11 \mic\ optical depth until the observed shape  is reproduced. The flux expected in the 5.8 \mic\ band from this model SED is then scaled to the 5.8 \mic\ flux of the SAGE Epoch 1 data for \ogles. The scaling changes the luminosity of the model photosphere as well. At constant effective temperature, this implies a change $\log{g}$. Such a change would in reality also change the shape of the model photosphere slightly, but this scaling is unavoidable given the small range of available models. The \twodust\ input and output parameters are listed in Table \ref{tab:cagbmodel:twodustparam}. We will summarize the results of the RT modeling in \S\ref{sec:cagbmodel:results}.

\subsection{Modeling The Circumstellar Gas}
\label{sec:cagbmodel:analysis:gas}
\begin{figure}[!htb]
  \epsscale{0.7}
  \plotone{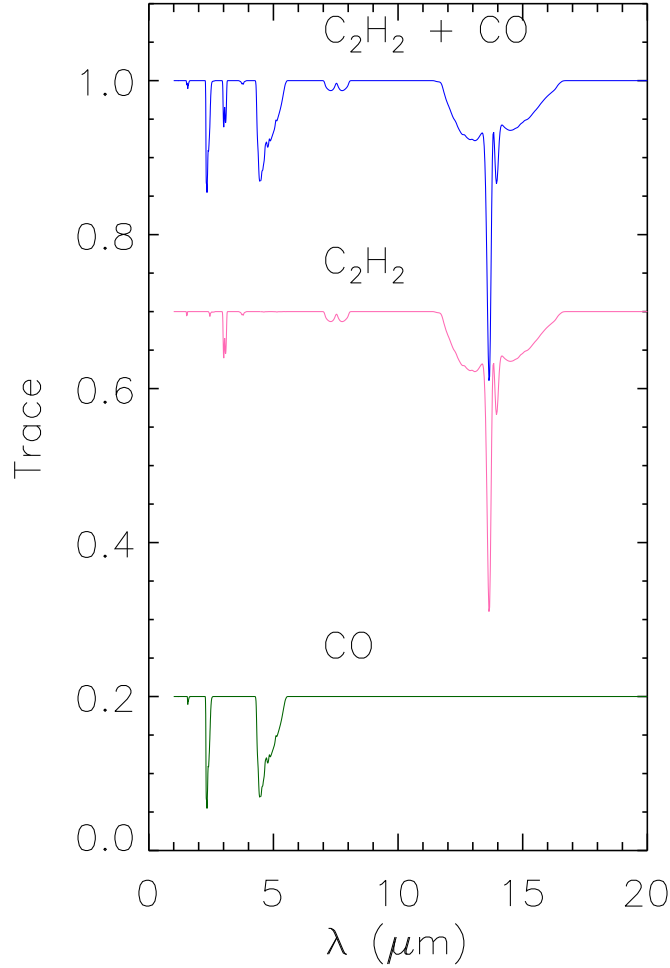}
  \ssp\caption[Molecular traces used to model the circumstellar gas]{The molecular features used to model the gas in \ogles. The \acet\ and CO traces are shifted down for clarity.\label{fig:cagbmodel:gastrace}}
\end{figure}

\begin{figure}[!htb] 
  \epsscale{1.0}
  \plotone{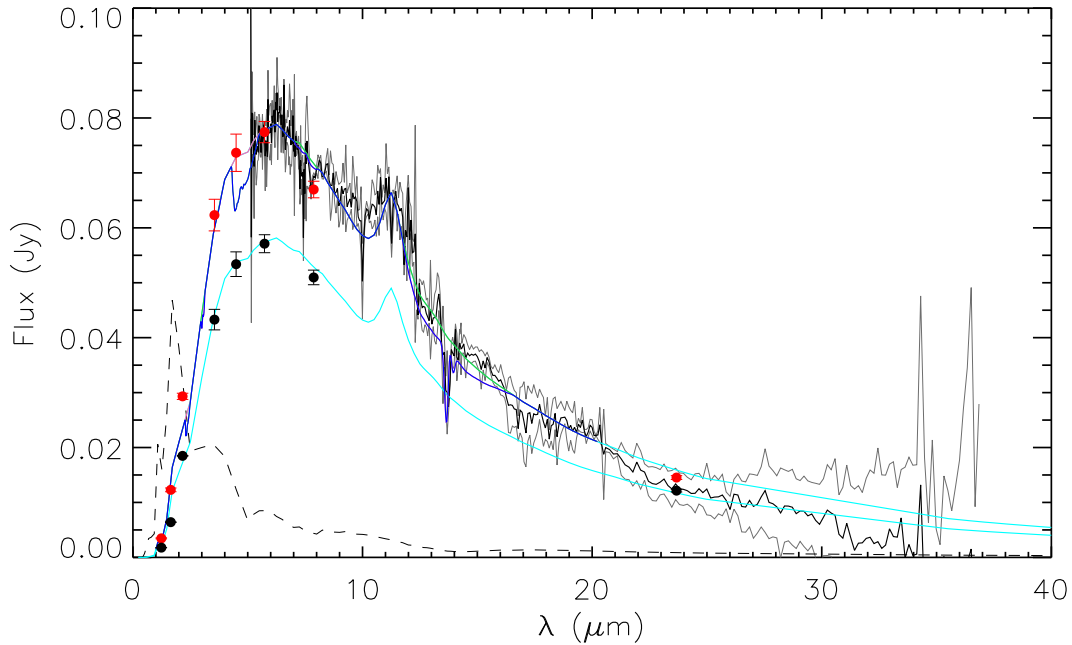}
  \ssp\caption[The dust+gas model fit to the \ogles\ data]{The results of modeling the gas and dust around \ogles. The SAGE Epoch 1 (red dots) and 2 (black dots) photometry as well as SAGE-Spec spectrum (black curve, scaled to Epoch 1 photometry) shown with the best-fit dust RT model, scaled to fit Epoch 1 and Epoch 2 photometry (cyan curves). The model SED is convolved with the molecular features shown in Fig. \ref{fig:cagbmodel:gastrace} to show the effect of adding \acet\ only (pink), CO only (green) and the combined effect (blue).
  \label{fig:cagbmodel:totalfit}}
\end{figure}

Our model is also unable to reproduce some absorption features in the spectrum, despite the fact that our photospheric model takes into account the absorption due to molecules forming near the photosphere, leading us to believe that the strong absorption features seen in the data are a result of gas in the extended atmosphere and the circumstellar shell. We present a treatment of the circumstellar gas in this section. \ogles\ has a few molecular signatures in its spectrum, such as \acet\ absorption at $\sim$7 and 13.7 \mic. The most prominent among these is the 13.7 \mic\ absorption. We simulate the molecular bands of CO and \acet\ in \ogles\ using the models of \citet{Matsuuraetal2002} with recent line lists published by \citet{Rothmanetal2009}. Assuming that the majority of the molecular absorption comes from the circumstellar shell, the excitation temperatures are chosen to be cooler than the temperature at the inner shell radius, {\it i.e.}, $T_{\rm ex}<T_{\rm in}$. Based on these considerations, we derive excitation temperatures of 1000 K for both species, with column densities of $10^{19}$ (\acet) and $3\times 10^{21}$ (CO) cm$^{-2}$ respectively. Fig. \ref{fig:cagbmodel:gastrace} shows the locations of the spectral features for the parameters above. The resulting (gas+dust) fit to the spectrum is shown in Fig. \ref{fig:cagbmodel:totalfit}.

\clearpage
\thispagestyle{empty}
\begin{landscape}
\begin{deluxetable}{lllllllllllllllll}
\setlength{\tabcolsep}{0.05in}
\tablewidth{0pt}
\tablecolumns{17}
\tablecaption{\twodust\ parameters for \ogles}
\tabletypesize{\footnotesize}
\tablehead{\multicolumn{2}{l}{\bf Central star\tablenotemark{a}} &&&&
\multicolumn{2}{l}{\bf Dust shell} &&&& \multicolumn{2}{l}{\bf Dust grains} &&&& \multicolumn{2}{l}{\bf Fit results}
}
\startdata
T$_{\rm eff}$ & 3000 K &&&& $R_{\rm in}$ & 6 $R_*$ &&&&Species & AmC\tablenotemark{b}\ + SiC\tablenotemark{c}&&&& $L$ & 5670 \lsun\\
$\log{g}$ & -0.4 &&&& $R_{\rm{out}}$ & 1000 $R_{\rm in}$ &&&&Abundance & 15\% SiC&&&&$\dot{M}_d$ & 2.6$\times$10$^{-9}$ \msunperyr\\
C/O & 1.3 &&&& density profile & $\rho(r)\sim r^{-2}$ &&&&$\tau$(11.3 \mic) & 0.25&&&&$T_{\rm in}$ & 1150 K\\
&&&&&  $v_{\rm{exp}}$ & 10 km s$^{-1}$&&&&Size distribution & KMH\tablenotemark{d}&&& &\\
&&&&&&&&&&&($a_{\rm min}$=0.01 \mic,&&&&\\
&&&&&&&&&&& $a_{0}$=1 \mic, $\gamma$=3.5)&&&&\\[0.25em]


\enddata
\sspf{\tablenotetext{a}{Photosphere model from \citet{Gautschy2004}.}
\tablenotetext{b}{Amorphous carbon grains, $\rho=1.8$ g cm$^{-3}$, optical constants from \citet{Zubkoetal1996}.}
\tablenotetext{c}{$\alpha$-SiC grains, $\rho=3.22$ g cm$^{-3}$, optical constants from \citet{Pegourie1988}.}
\tablenotetext{d}{Size distribution from \citet{Kimetal1994}, $n(a)\sim a^{-\gamma}\exp{\left(-a/a_0\right)}$.}}
\label{tab:cagbmodel:twodustparam}
\end{deluxetable}
\end{landscape}
\clearpage

\section{Results}
\label{sec:cagbmodel:results}
We find that the best-fit model SED is produced for an inner radius six times the stellar radius and an optical depth of 0.25 at 11.3 \mic. The luminosity of the model photosphere is scaled up by a factor of 1.29 to 5670 \lsun\ to fit the observations. As mentioned in \S\ref{sec:cagbmodel:analysis:dust}, a higher luminosity at constant effective temperature can only be obtained by changing $\log{g}$, which will affect the photospheric spectrum.  We disregard this complication for this study. Assuming that the mass-loss rate scales as $\sqrt{L}$ \citep[see, \examp,][]{Groenewegen2006}, we find the dust mass-loss rate to be $2.6\times 10^{-9}$ \msunperyr. Since the shell geometry is defined in terms of the stellar radius, we do not expect the dust temperature to change as a result of scaling. The temperature $T_{\rm in}$ of the dust at the distance corresponding to the inner radius $R_{\rm in}$ of the dust shell is 1160 K, which is close to the condensation temperature of $\alpha$-SiC. We also fit the \twodust\ output to the second epoch of SAGE photometry with good agreement. In this case, we obtain a luminosity of 5330 \lsun and a corresponding mass-loss rate of $2.2\times 10^{-9}$ \msunperyr.

Fig. \ref{fig:cagbmodel:totalfit} shows the \twodust\ output SED (cyan curve) superimposed on the SAGE-Spec spectrum and Epoch 1 photometry (red dots). While the fit is acceptable overall, we need to incorporate the effects of circumstellar gas to fit the 13.7 \mic\ feature. The figure also shows a convolution of the \twodust\ output SED with the molecular features (\acet\ only: pink; CO only: green; \acet\ + CO: blue), which is in excellent agreement with the SAGE-Spec spectrum.

We obtain an estimate for the gas mass-loss rate from our results for \acet\ (the CO band is at the edge of our spectrum). Depending on the location of the \acet\ line forming region, the mass-loss rate ranges between $1\times 10^{-7} - 2\times 10^{-6}$ \msunperyr. Here, we have adopted a \acet\ abundance of $10^{-5}$ compared to H$_2$ \citep[see][]{Matsuuraetal2006}, but some of the \acet\ will have condensed out as dust.

\section{Discussion}
\label{sec:cagbmodel:discuss}
\ogles\ is a long-period variable (LPV) with a period of about 1 yr \citep{Itaetal2004,Groenewegen2004}. The two epochs of SAGE observations have thus captured the star at different stages in its pulsation. While 2MASS and IRSF observations in the NIR also offer multiple-epoch data, these were not taken in conjunction with SAGE data. The amplitude of variation is expected to be higher in the NIR magnitudes than in the optical \citep[][estimated the I-band amplitude to be about 0.8 mag]{Groenewegen2004}.
The dust mass-loss rate predicted by our \twodust\ model, $2.6\times 10^{-9}$ \msunperyr, is on the low end of the rates derived by \citet{vL1999} for the brightest, most obscured carbon stars in the LMC. The derived dust MLR can be converted to gas MLR of $5.2\times 10^{-7}$ \msunperyr, assuming a gas:dust ratio for C--rich AGB stars of 200. This is comparable to the gas MLR estimated from the period vs MLR relations of \citet{VW93} and \citet{Groenewegenetal1998} ( $10^{-7}$ and $7.6 \times 10^{-7}$ \msunperyr respectively, assuming a period of 358 d) and the MLR predicted by the color vs MLR relation of \citet{Matsuuraetal2009} ($4-6\times 10^{-7}$ \msunperyr\ using \ks--[8.0]=--0.395 mag for \ogles). The 8 \mic\ excess -- MLR relation from \citet{excesses} gives $2\times 10^{-6}$ \msunperyr\ and $1.26\times 10^{-6}$ \msunperyr\ for the Epoch 1 and 2 observations respectively. Despite the fact that the \ks\ and 8 \mic\ data were obtained for different phases, the MLR predicted by the \ks--[8.0] color agrees well with the MLR determined from other methods.

The long-term goal of this study is to use \twodust\ modeling of carbon star circumstellar shells to fit the photometry of the $\sim$7000 C--rich AGB candidates in the SAGE survey. In this section, we weight the simplifying assumptions used in our modeling of \ogles\ against this future goal. Our combined gas+dust model improves on the spectrum predicted by \twodust\ modeling of the dust shell alone. There are, however, many issues associated with our simple assumptions in this work. 
For instance, our model requires a significant amount of SiC (15\%). Furthermore, the value of $T_{\rm in}$ is very close to the condensation temperature of $\alpha$-SiC. \citep{Specketal2009} have shown that the use of graphite over amorphous carbon can result in lower amounts of SiC needed to fit the observed 11 \mic\ feature, as well as lower temperatures for the dust. \\

From Fig. \ref{fig:cagbmodel:totalfit}, it is clear that our AmC+SiC dust composition reproduces the 11.3 \mic\ feature and the continuum in overall observed spectrum well. However, the SiC feature is observed to be broader than the model prediction; this is in part a problem with the \citet{Pegourie1988} optical constants for $\alpha$-SiC, as discussed by \citet{Pitmanetal2008}, who suggest the usage of $\beta$-SiC optical constants for non-spherical grains. 
The observed spectrum also looks broader than those of known Galactic and LMC carbon-rich spectra. A spike noise may be increasing the shape artificially.
For the purpose of modeling the SAGE photometry, however, the details of the SiC feature are irrelevant as the SiC feature does not fall in any of the {\it Spitzer} bands.

The model also systematically overestimates the flux beyond $\sim$ 20\mic. Although the quality of data in the long-wavelength range is poor, this disagreement may also be due to the simplifying assumptions in our model. 
Our assumption of spherical grains is also an idealization, and the \citet{Kimetal1994} prescription may not represent the true distribution of dust grain sizes in the circumstellar shells of AGB stars, but these details do not affect the overall properties \citep{Specketal2009}. Our main interest is to derive dust composition and mass-loss properties for the entire sample of SAGE C--rich AGB star candidates using a baseline dust description, and for these purposes the assumption of spherical grains is sufficient.\\
\section{Summary and conclusions}
\label{summary}
We have modelled the circumstellar shell a carbon-rich asymptotic giant branch star using the radiative transfer code \twodust\ to fit SAGE photometry and SAGE-Spec spectroscopy for the source \ogles. We find a dust mass-loss rate of 2.6$\times 10^{-9}$ \msunperyr\ for the source, corresponding to an inner radius for the dust shell of six times the stellar radius. The optical depth at 11.3 \mic\ is found to be 0.25 and the dust temperature at the inner radius of the shell is 1160 K. We model the molecular features in the circumstellar shell with slab models of \acet\ and CO. The gas mass-loss rate estimated from the column density of \acet\ is in the range $1\times 10^{-7} - 2\times 10^{-6}$ \msunperyr and it depends heavily on the location of the line-forming region. The model SED when combined with \acet\ and CO models shows excellent agreement with the observed spectrum and photometry. We are successful in our aim of defining a baseline set of dust properties for use in our model grids, which will be the focus of the following chapter.
\chapter{A Grid of Carbon Star Models}
\label{ch:modelgrid}

\section{Introduction}
Asymptotic giant branch (AGB) stars are among the brightest IR objects in star-forming galaxies. The $\sim 45\,000$ AGB candidates \citep[see, \examp ,][]{sage2,excesses} inject a significant fraction of their mass (30--80\%) in the form of gas and dust into the LMC ISM at a total rate of $\sim 10^{-2}$ \msunperyr\ \citep{excesses}, thereby increasing its metallicity and driving the chemical evolution of the galaxy. AGB stars are thus essential ingredients in modeling galaxy evolution. 

Despite being outnumbered by the O--rich AGB stars, carbon stars are probably responsible for the majority of the AGB mass ejection in the LMC because most extreme AGB stars are C--rich. \citet{excesses} estimated the total mass loss return from the near-IR identified C--rich AGB stars to be $4.8\times 10^{-4}$ \msunperyr\ and the extreme AGB stars to be $(4.7-11.8)\times 10^{-3}$ \msunperyr. However, their estimate was an empirical one, based on excess mid-IR flux over the photosphere and using an excess versus mass-loss rate relation derived from data for a handful of C--rich and extreme AGB stars. Fits to spectral energy distributions for the AGB stars using radiative transfer model grids provides a more direct and quantitative measure of the mass-loss rate return by these stars.  

Recent work has resulted in many model grids for carbon stars: \citet{MarigoGirardi2007} presented models of synthetic thermally pulsing (TP) AGB evolution and derived theoretical isochrones from these models in \citet{Marigoetal}. \citet{Mattssonetal2009} improved upon the dynamical wind models of \citet{Mattssonetal2007} and \citet{Hofneretal2003} and computed 900 carbon star models to study their dependence on stellar parameters. \citet{Gautschy2004} and, more recently, \citet{Aringeretal2009} constructed a grid of hydrostatic photosphere models for C--rich AGB stars. 

In this chapter, I introduce a grid of carbon-star models that calculates the effect of circumstellar dust on the  \citet{Aringeretal2009} model photospheres using the dust radiative transfer code \twodust\ \citep{UetaMeixner03}. The models span a wide range of stellar and dust-related parameters. I also present synthetic photometry for these models in the 2MASS, {\it Spitzer}, AKARI and WISE bandpasses. In \S\ref{sec:modelgrid:input} I discuss the various input parameters that went into the construction of the grid. I briefly describe the details of the method for computing the models in \S\ref{sec:modelgrid:computation}, and discuss the resulting model grid and synthetic photometry in \S\ref{sec:modelgrid:results}. I present my conclusions in \S\ref{sec:modelgrid:conclusions}.

\section{Input to the radiative transfer code}
\label{sec:modelgrid:input}



\subsection{Model photospheres}
\label{subsec:modelgrid:photospheres}
As input to \twodust, I utilize the \citet{Aringeretal2009} photosphere models, which contain 132 models with metallicity of $Z=0.33$, close to that of the LMC \citep[$Z\sim 0.3-0.5$ \zsun,][p. 234]{Westerlund1997}. Their LMC grid was calculated completely for 2 \msun\ models, with a subset of 1 \msun\ models for \teff=2600 K. The choice of C/O ratios (1.4, 2.0 and 5.0) reflected the fact that this ratio can increase above unity in low-metallicity environments such as the LMC after a single third dredge-up event. C/O ratio for more evolved stars can rise significantly above unity \citep{MarigoGirardi2007}. While \citet{Aringeretal2009} considered a temperature grid of 2600 K to 4000 K in increments of 100 K, we ignored the models with \teff\ values higher than 3500 K as these temperatures correspond to the post-AGB phase. This restriction on \teff\ reduces the number of available photosphere models to 120. Carbon stars may have effective temperatures as low as 2400 K \citep{Mattssonetal2009}. However, the lower temperature limit of 2600 K for the \citet{Aringeretal2009} models provides adequate coverage for the LMC C--rich sample \citep{Groenewegenetal2009}.

The surface gravity ranged from $\log{g}$=0 down to $\log{g}$=--1 in steps of 0.1, but not all values were available for each \teff\ value because of convergence issues. For \teff=2600 K, the only $\log{g}$=0 was considered. The corresponding luminosity is about 1000 \lsun, which is significantly less than that for typical carbon stars \citep[see, \examp , Fig. 4 in][]{excesses}. The lower limit of $\log{g}$ depends on \teff. It was chosen to cover the region of AGB carbon stars predicted by \citet{MarigoGirardi2007} and \citet{Marigoetal}. For the range of model parameters corresponding to each temperature range, see Table 1 in \citet{Aringeretal2009}. The luminosities of the selected models range from $10^{3}$ \lsun to $3\times 10^{4}$ \lsun. The most luminous optically visible carbon stars are at luminosities slightly higher than this upper limit, and the extreme carbon stars are perhaps as luminous as $\sim 10^5$ \lsun. The lack of more luminous model photospheres means that in order to model the most evolved, highly enshrouded stars, we will have to scale the luminosities of the available models. For the optically thick dust shells, the scaling of luminosities should provide adequate matches to the available data which measures the total luminosity of the star+envelope system.

\subsection{Circumstellar envelope}
In chapters \ref{ch:excesses} and \ref{ch:cagbmodel}, I probed the parameter space for modeling by first estimating color temperatures and optical depths of the dust, followed by the radiative transfer modeling of a star for which photometry and spectroscopic data were available. In this section, I will use these results to constrain the parameter space for the circumstellar envelope.
\subsubsection{Shell geometry}
With the simplification of spherical symmetry, we are left to specify the inner and outer radii of the shell, as well as the density structure in the shell and the expansion velocity of the dust. The outer radius is kept fixed at a thousand times the inner radius. As mentioned in chapter \ref{ch:cagbmodel}, this is acceptable as long as we do not use it to estimate the timescale of mass loss. Four values are chosen for the inner radius, at 1.5, 4.5, 7 and 12 times the radius of the star. This range of radii samples the theoretically expected range of dust production in carbon star winds \citep{Hofner2007}.

The expansion velocity is fixed at 10 \kms. The velocity of a radiation-driven wind should in general be luminosity-dependent \citep{GailSedlmayr1986} and this dependence can be accounted for by an appropriate scaling relation with luminosity \citep[see, \examp ,][]{vanLoon2007}. However, in this first attempt at generating a model grid, we will ignore this complication.

We also choose an inverse-square density drop-off for the shell. This implies a constant mass-loss rate, which may be unrealistic over the course of AGB evolution or during rapid thermal pulsing. Some authors \citep{Bedijn1986, Zijlstraetal1992, GroenewegenThesis} have considered time-varying mass-loss rates to account for such situations. Since our main interest is fitting a large set of photometric data in order to derive global mass injection rates, this is beyond the scope of our present study. \twodust\ is capable of accepting user-defined density functions, however, and an investigation of density profiles different from $r^{-2}$ can be undertaken if desired.

\subsubsection{Dust properties}
As discussed in chapter \ref{ch:cagbmodel}, we choose a mixture of amorphous carbon \citep[optical constants from][]{Zubkoetal1996} and silicon carbide \citep[$\alpha$-SiC, optical constants from][]{Pegourie1988}, with 15\% SiC. The grain sizes are distributed according to \citet[KMH]{Kimetal1994} with a minimum grain size of 0.01 \mic\ and a ``maximum" size of 1 \mic. We vary the optical depth at the center of the SiC feature at 11.3 \mic\ from $\tau_{\rm 11.3~\mu m} = 10^{-3}$ to 1 in equally spaced logarithmic steps with five values per decade. Furthermore, we also consider four $\tau_{\rm 11.3~\mu m}$=2, \teff=2600 K models.
 
This range covers most of the optically visible carbon-rich AGB stars, as well as the lower luminosity extreme AGB stars. We will probe higher optical depths in future studies in order to model the most extreme carbon stars. The dust temperature at the inner radius is another parameter that we constrain in our modeling, we only accepted models that resulted in a value of $T(R_{\rm in})$ lesser than or equal to the condensation temperature of $\alpha$-SiC is $\sim$1800--2000 K.

With the above constraints on the \citet{Aringeretal2009} photosphere parameters and the various dust parameters, about 7750 models are available to run through \twodust. The parameter ranges for these models are summarized in Table \ref{tab:modelgrid:params}. Placing an added constraint on the dust temperature brings the number of models down to about 7100, as discussed in the following section. 

\begin{deluxetable}{ll}
\setlength{\tabcolsep}{0.05in}
\tablewidth{0pt}
\tablecolumns{2}
\tablecaption{Input parameters for model grid}
\tabletypesize{\footnotesize}
\tablehead{\colhead{Parameter} & \colhead{Value or Range of values}
}
\startdata
\hline
\hline
Luminosity & $\sim 10^3$ to $3 \times 10^5$ \lsun\\
Inner shell radius & 1.5, 4.5, 7, 12\\
Outer shell radius & 1000\\
Density profile & $r^{-2}$\\
Expansion velocity & 10 \kms \\
Dust species & 85\% AmC\tablenotemark{1} + 15\% SiC\tablenotemark{2}\\
Optical depth at 11.3 \mic & $10^{-3}$ to $1$ (five values per decade)\tablenotemark{3}\\
\enddata
\sspf{
\tablenotetext{1}{Optical constants from \citet{Zubkoetal1996}}
\tablenotetext{2}{Optical constants from \citet{Pegourie1988}}
\tablenotetext{3}{Additionally, four \teff=2600 K, $\tau$=2 models.}
}
\label{tab:modelgrid:params}
\end{deluxetable}

\section{Computational details}
\label{sec:modelgrid:computation}
The models were run on the Royal Linux computing cluster at the Space Telescope Science Institute. The cluster consists of 23 computing nodes, each with 2 x 2 x 2.4 GHz processors, 8 GB RAM, 160 GB local disk space and  
17 TB of total disk space. Royal enables parallel processing at high computing speeds, and also automates the job submission and scheduling. \twodust\ was therefore run in the non-interactive mode, with an initial run to ensure that the number of radial grid points was sufficient. The 120 photospheres were split into at most 19 photospheres per job submission, totaling up to $\sim$ 1200 individual models submitted simultaneously. The time taken for successful convergence ranged from $\sim$ 10 min for the optically thin models to a few hours for the models with moderate optical depth. The Royal cluster had an 8 hour time limit on individual submissions, which was insufficient for optical depths higher than unity. A smaller subset of $\tau=2$, \teff=2600 K models with $\log{g}$=--0.9, --0.8, --0.6 and --0.5 were executed successfully for two values of inner radius ($R_{\rm in}/R_*$ = 7 and 12) on a Mac Pro Desktop with a 2 x 2.8 GHz Quad-Core Intel Xeon CPU and 10 GB RAM. In general, the code required an increase in the number of radial grid points to deal with higher optical depths, which eventually resulted in memory allocation errors. At present, therefore, our models are reliable up to $\tau=2$. Once the models converged, only those runs that produced dust temperatures lesser than 2000 K were considered as successful. This resulted in the rejection of some $R_{\rm in}=1.5 R_*$ models, which for hotter stars naturally produced warmer dust temperatures.

I then produced synthetic photometry for the remaining models in the optical UBVI\footnote{The MCPS magnitudes were placed on the Johnson-Kron-Cousins UBVI system. The detector quantum efficiency curve was obtained from the Las Campanas Observatory website, http://www.lco.cl/lco/index.html. Filter profiles for the Johnson U, Harris B, V, and Cousins I filters were obtained from the references in Table 9 of \citet{Fukugita}.}, 2MASS\footnote{The 2MASS filter relative spectral responses (RSRs) derived by \citet{2MASSRSRs} were obtained from the {\it 2MASS All-Sky Data Release Explanatory Supplement}, available at http://www.ipac.caltech.edu/2mass/releases/allsky/doc/sec6$_{}$4a.html.} and {\it Spitzer}\footnote{The IRAC RSRs are plotted in \citet{IRAC}, and were obtained from the {\it Spitzer Science Center} IRAC pages at http://ssc.spitzer.caltech.edu/irac/spectral$_{}$response.html; The MIPS \citep{MIPS} RSRs were obtained from the {\it Spitzer Science Center} MIPS pages at http://ssc.spitzer.caltech.edu/mips/spectral$_{}$response.html} bands in order to compare the results with SAGE data. Synthetic photometry was also derived for the AKARI 
\citep{Murakamietal2007} and WISE \citep{Wrightetal2004} passbands\footnote{Filter response curves available at\\ http://www.ir.isas.jaxa.jp/ASTRO-F/Observation/RSRF/IRC$_{}$FAD/index.html\\ and http://www.astro.ucla.edu/$\sim$wright/WISE/passbands.html respectively}.
	
\section{Results and discussion}
\label{sec:modelgrid:results}
In this section, I demonstrate the success of our model grid in covering the range of luminosities and mass-loss rates expected for carbon stars. I use synthetic photometry to demonstrate the coverage of our model grid in color-color and color-magnitude diagrams (CMDs) and compare it to SAGE data for AGB star candidates.  I also provide color-color diagrams for the AKARI and WISE bands, and I revisit the model for \ogles\ and compare my results in Chapter \ref{ch:cagbmodel} to the best-fit model produced by using the model grid generated in this chapter. Finally, I show preliminary results of fitting the model grid to SAGE C--rich AGB candidates.

\subsection{Dust mass-loss rates}
\begin{figure}[!htb]
  \epsscale{1}
  \plotone{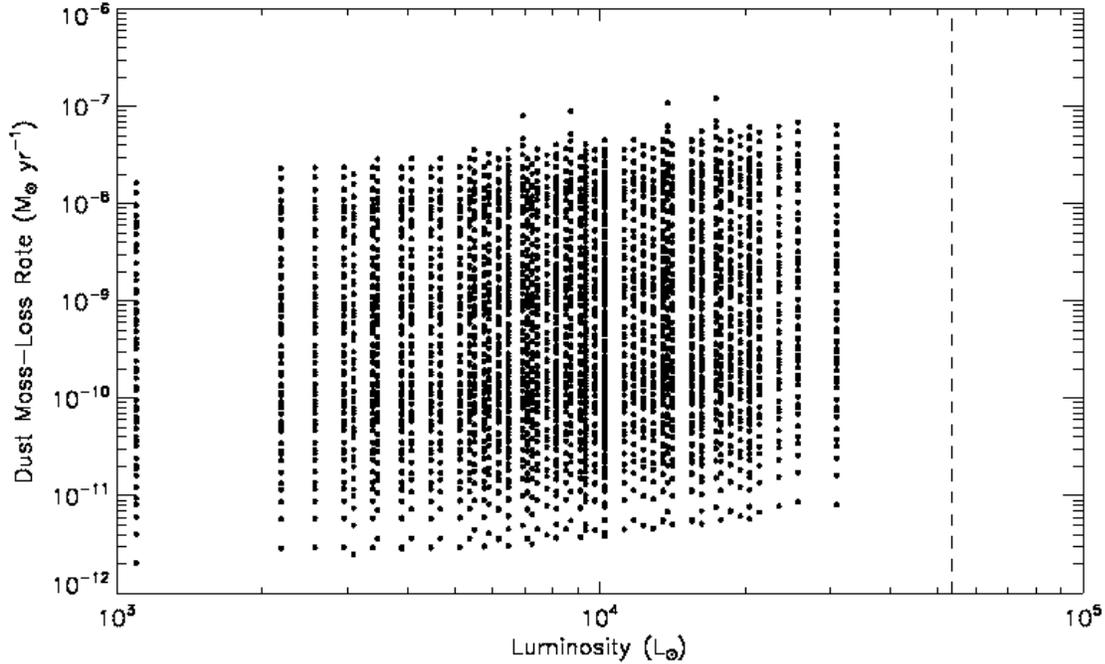}
  \ssp\caption[Dust mass-loss rate]{The range of values covered in luminosity-mass-loss rate space by the model grid. The four $\tau$=2 models and the $\tau$=6, 8 and 10 models can be clearly seen.The dashed line represents the classical AGB luminosity limit.\label{fig:modelgrid:mlrvsl}}
\end{figure}

Fig. \ref{fig:modelgrid:mlrvsl} shows the range of luminosities and mass-loss rates explored by the current model grid.  The available luminosities span the range of C--rich AGB star luminosities observed in the SAGE survey (see Fig. \ref{fig:AGBLF} in Chapter \ref{ch:excesses}). The highest mass-loss rates are $\sim 10^{-7}$ \msunperyr, corresponding to a total rate of $2\times 10^{-4}$ \msunperyr\ (assuming a gas:dust ratio for carbon dust of 200). This covers the range of mass-loss rates expected for the low- and moderately-obscured carbon stars. More models for higher optical depths will be required to reproduce the mass-loss rates of the extreme AGB stars.

\subsection{SAGE synthetic photometry}
\begin{figure}[!htb]
  \epsscale{1}
  \plotone{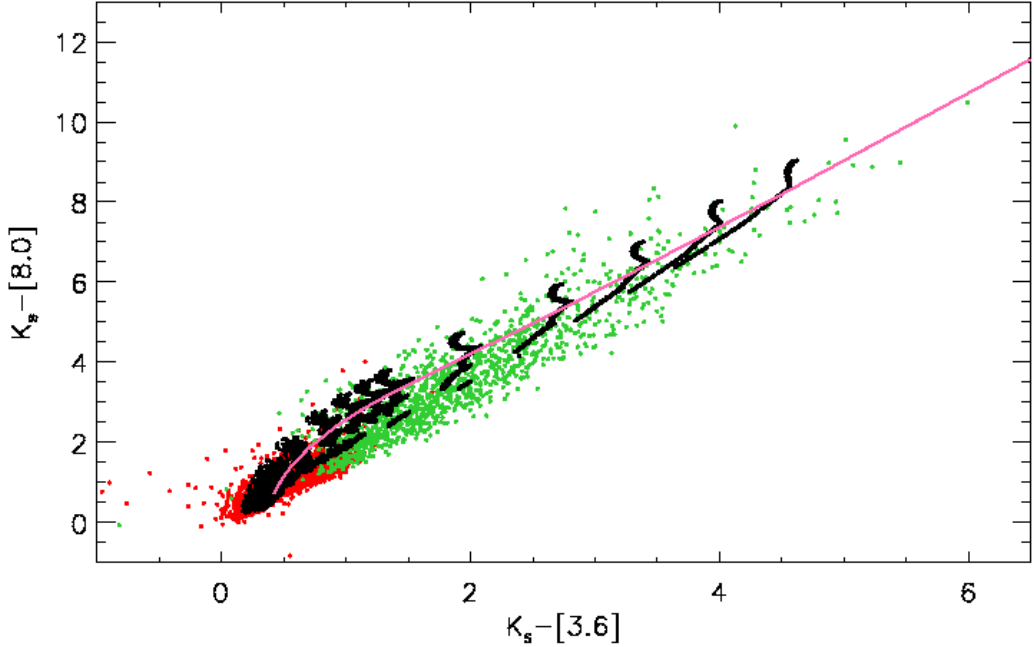}
  \ssp\caption[2MASS--IRAC color-color diagram]{The \ks--[8.0] vs \ks--[3.6] color-color diagram showing the model grid (black dots) superimposed on the C-rich (red dots) and extreme (green dots) AGB candidates. The pink line shows the track for a single model (\teff=2600 K, $\log{g}=-0.9$, $R_{\rm in}=7R_{\rm star}$) with increasing optical depth.\label{fig:KOvsKL}}
\end{figure}

\begin{figure}[!htb]
  \epsscale{1}
  \plotone{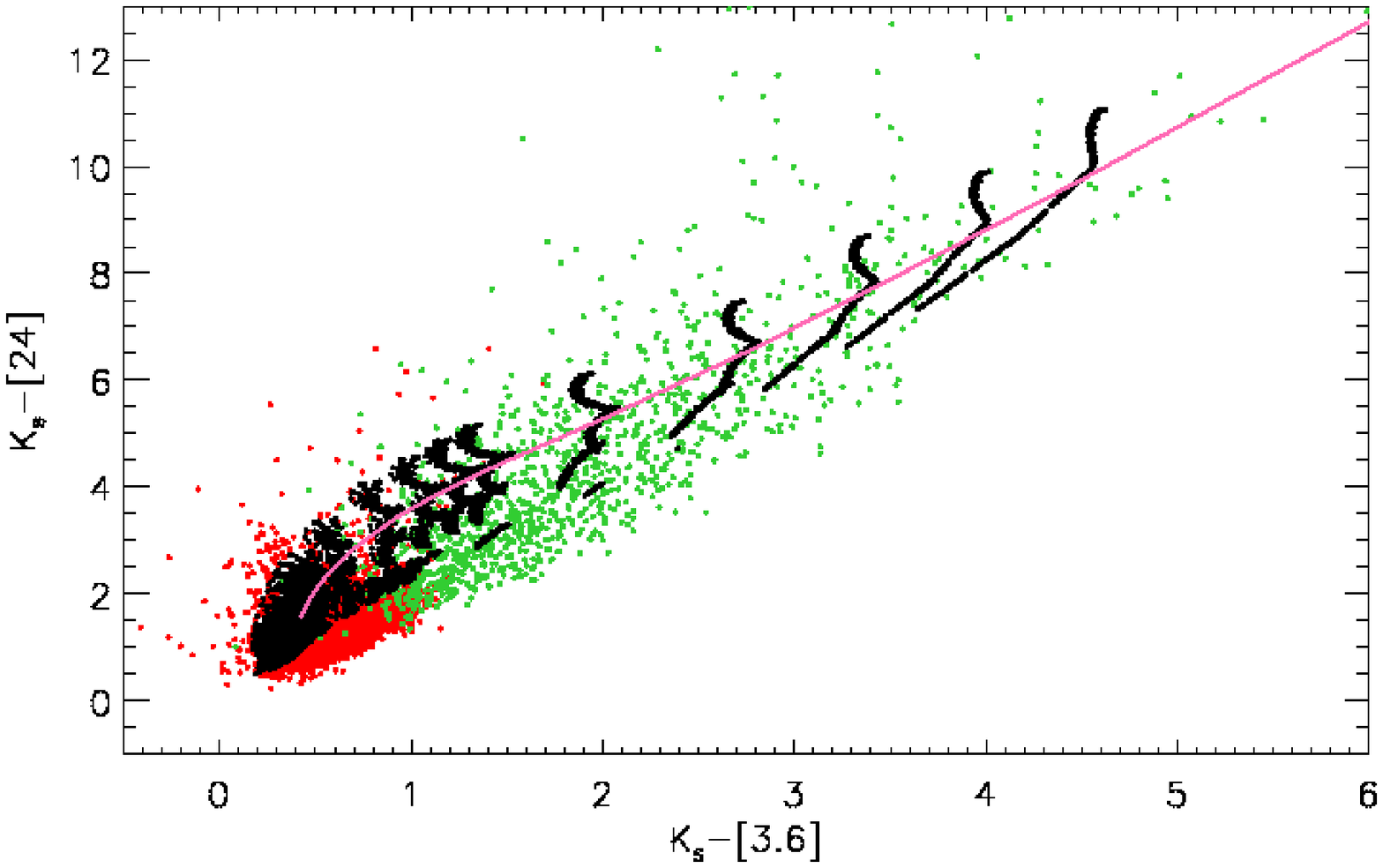}
  \ssp\caption[2MASS--IRAC--MIPS color-color diagram]{The \ks--[24] vs \ks--[3.6] color-color diagram showing the model grid (black dots) superimposed on the C-rich (red dots) and extreme (green dots) AGB candidates. The pink line shows the track for a single model (\teff=2600 K, $\log{g}=-0.9$, $R_{\rm in}=7R_{\rm star}$) with increasing optical depth.\label{fig:KPvsKL}}
\end{figure}

\begin{figure}[!htb]
  \epsscale{1}
  \plotone{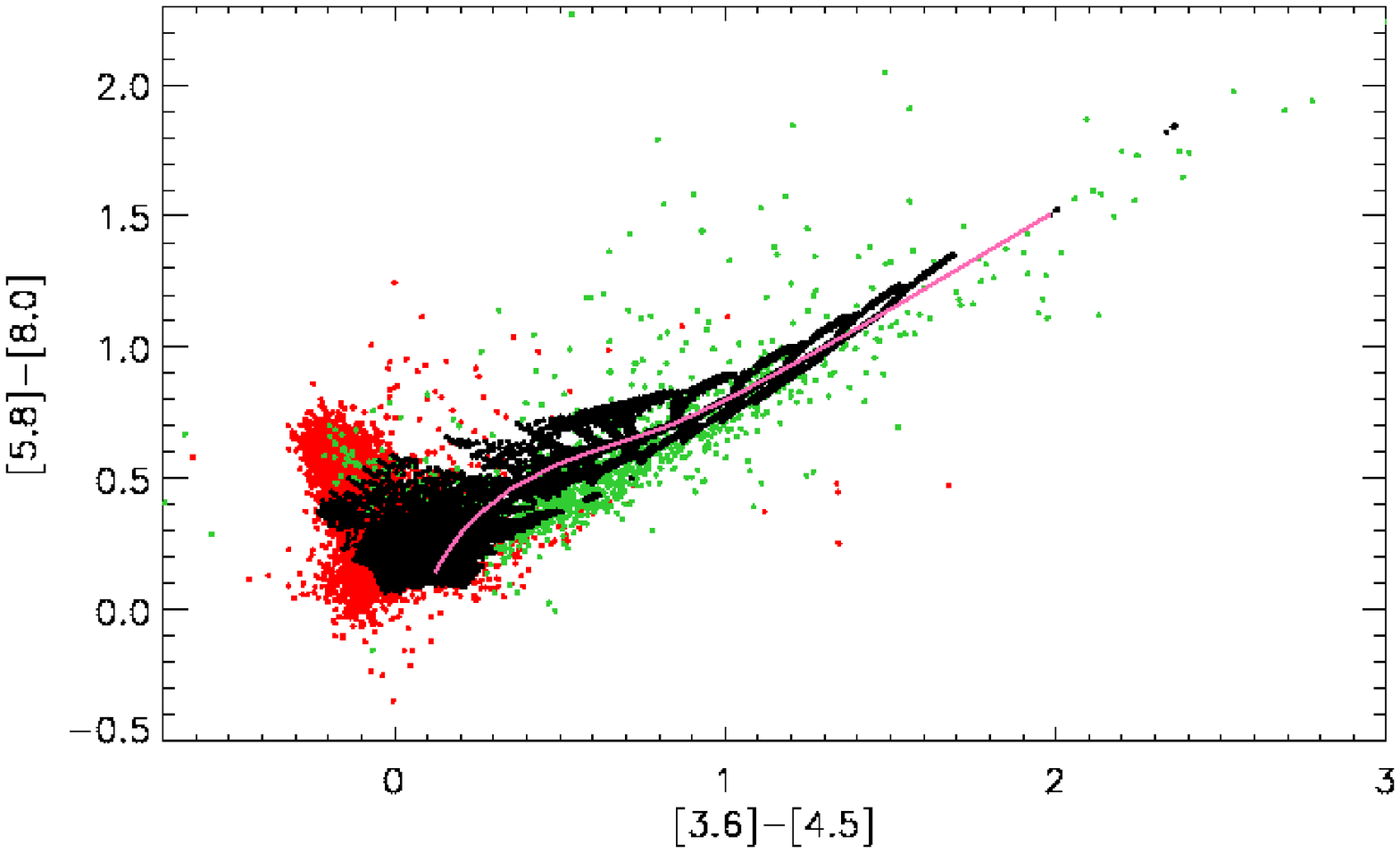}
  \ssp\caption[IRAC--IRAC color-color diagram]{The [5.8]--[8.0] vs [3.6]--[4.5] color-color diagram showing the model grid (black dots) superimposed on the C-rich (red dots) and extreme (green dots) AGB candidates. The pink line shows the track for a single model (\teff=2600 K, $\log{g}=-0.9$, $R_{\rm in}=7R_{\rm star}$) with increasing optical depth.\label{fig:NOvsLM}}
\end{figure}

The models are superimposed on SAGE C--rich and extreme AGB candidates in Fig. \ref{fig:KOvsKL}. The pink line shows how the \teff=2600 K, $\log{g}=-0.9$, $R_{\rm in}=7R_{\rm star}$ model would evolve with increasing optical depth, assuming that the rest of the parameters stay fixed. Fig. \ref{fig:KPvsKL} shows a similar variation. The IRAC--IRAC color-color diagram (Fig. \ref{fig:NOvsLM}) shows that we underestimate the [5.8]--[8.0] colors of the optically thin C--rich AGB stars or lack coverage of bluer C--rich AGB stars. This could be due to the presence of small grains in our models -- the minimum grain size was set to 0.01 \mic, whereas meteoritic evidence points to the fact that $\sim$70\% of SiC grains have sizes in the range 0.3--0.7 \mic\ \citep[see][and references therein]{Specketal2009}. We will investigate the effect of grain sizes on such deviations in detail in follow-up studies. The figures show that the 
models reproduce the colors of the low- and moderately-obscured carbon stars to a great extent, despite the simplifying assumptions that went into creating the grid. The grid allows us to readily model the mass-loss rates of carbon stars of low and moderate optical depths.

\begin{figure}[!htb]
  \epsscale{1}
  \plotone{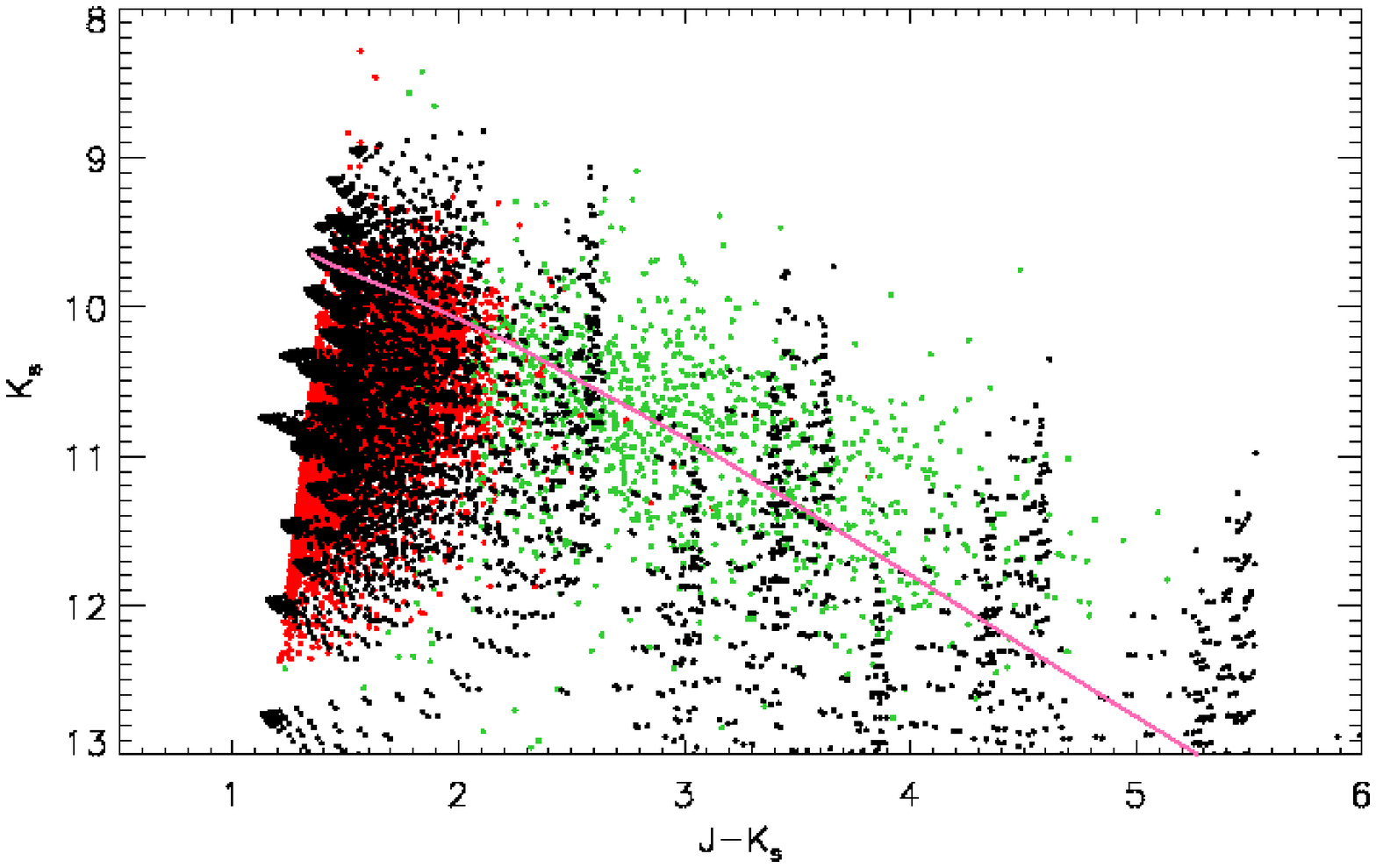}
  \ssp\caption[2MASS color-magnitude diagram]{The \ks\ vs J--\ks\ CMD showing the model grid (black dots) superimposed on the C-rich (red dots) and extreme (green dots) AGB candidates. The pink line shows the track for a single model (\teff=2600 K, $\log{g}=-0.9$, $R_{\rm in}=7R_{\rm star}$) with increasing optical depth.\label{fig:KvsJK}}

\end{figure}

\begin{figure}[!htb]
  \epsscale{1}
  \plotone{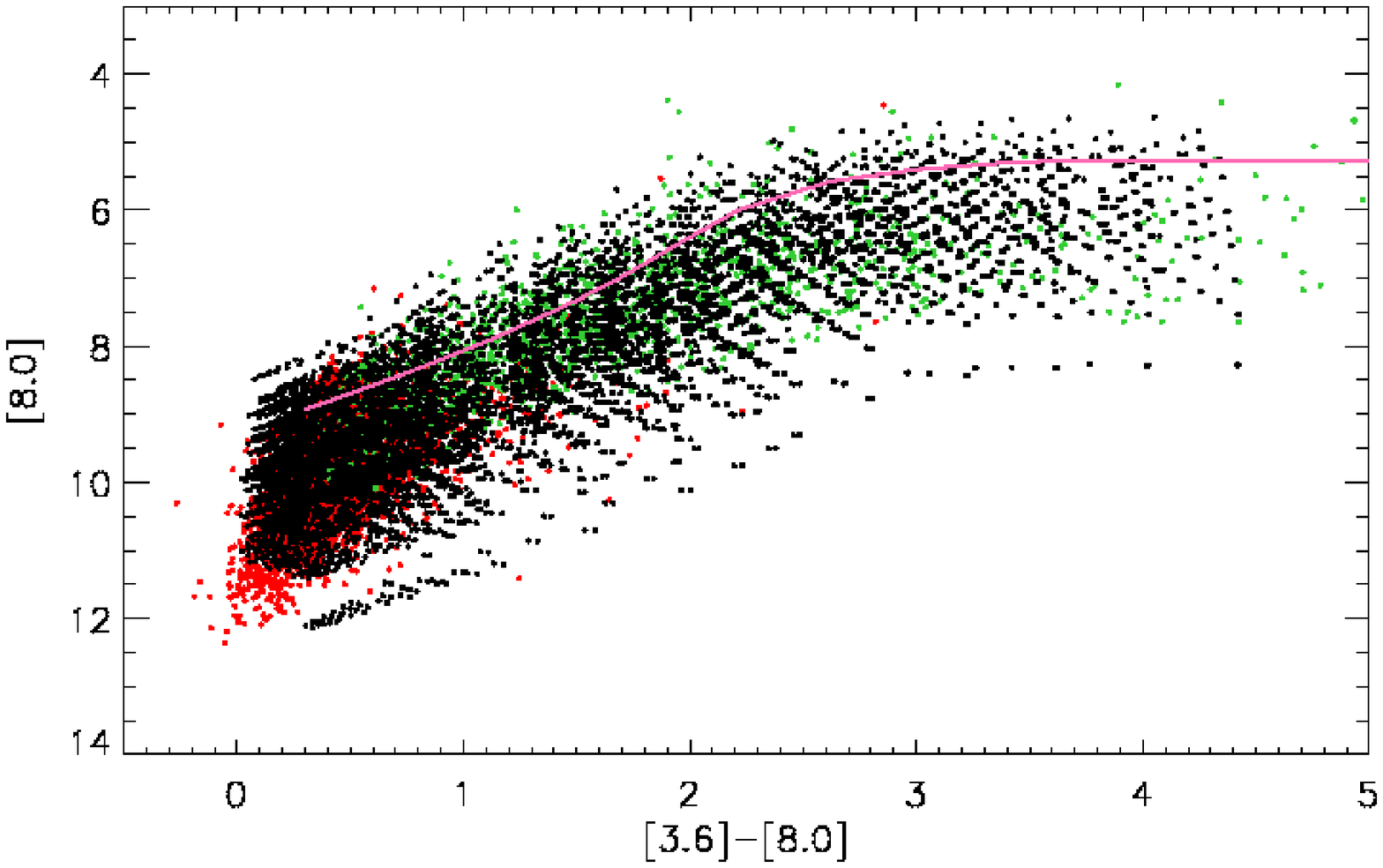}
  \ssp\caption[IRAC color-magnitude diagram]{The [8.0] vs [3.6]--[8.0] CMD showing the model grid (black dots) superimposed on the C-rich (red dots) and extreme (green dots) AGB candidates. The pink line shows the track for a single model (\teff=2600 K, $\log{g}=-0.9$, $R_{\rm in}=7R_{\rm star}$) with increasing optical depth.\label{fig:OvsLO}}
\end{figure}

\begin{figure}[!htb]
  \epsscale{1}
  \plotone{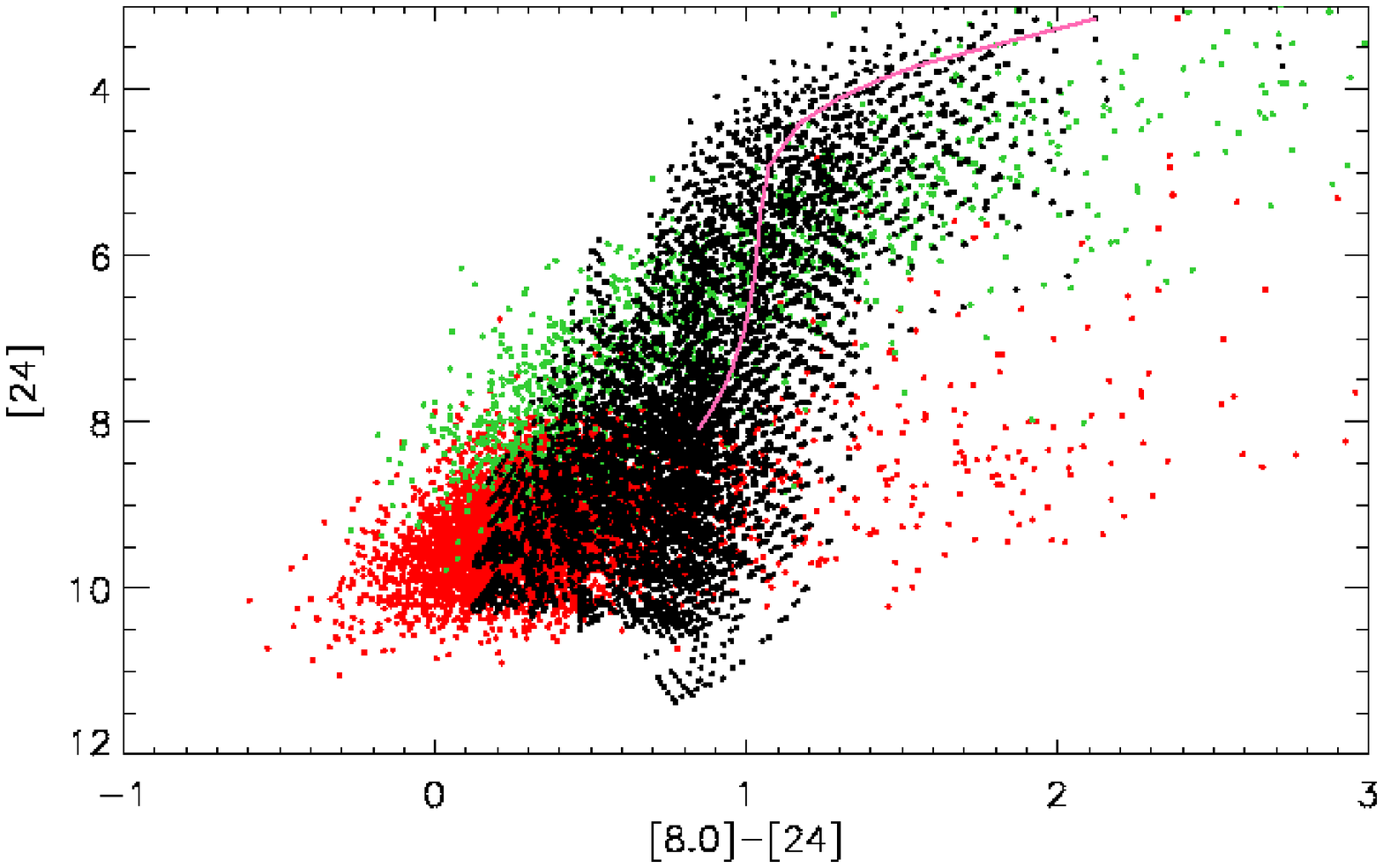}
  \ssp\caption[IRAC--MIPS color-magnitude diagram]{The [24] vs [8.0]--[24] CMD showing the model grid (black dots) superimposed on the C-rich (red dots) and extreme (green dots) AGB candidates. The pink line shows the track for a single model (\teff=2600 K, $\log{g}=-0.9$, $R_{\rm in}=7R_{\rm star}$) with increasing optical depth.\label{fig:PvsOP}}
\end{figure}

Knowledge of the distance to the LMC allows us to also construct color-magnitude diagrams (CMDs) for our models. Fig. \ref{fig:KvsJK}, \ref{fig:OvsLO} and \ref{fig:PvsOP} show the 2MASS, IRAC and IRAC--MIPS CMDs comparing the models to SAGE C--rich and extreme AGB candidates. Once again, we see that the models reproduce the data well. The [8.0] vs [3.6]--[8.0] CMD shows that beyond a certain optical depth, the 8 \mic\ magnitude is constant for a large range of colors. This corresponds to the spectral energy distribution (SED) peaking in the 8 \mic\ band. The 24 \mic\ flux for the single track considered in Fig. \ref{fig:PvsOP} undergoes a sudden drastic increase at almost constant [8.0]--[24] color. This probably corresponds to the onset of appreciable mass-loss.


\subsection{Synthetic photometry for other surveys}
\begin{figure}[!htb]
  \epsscale{1}
  \plotone{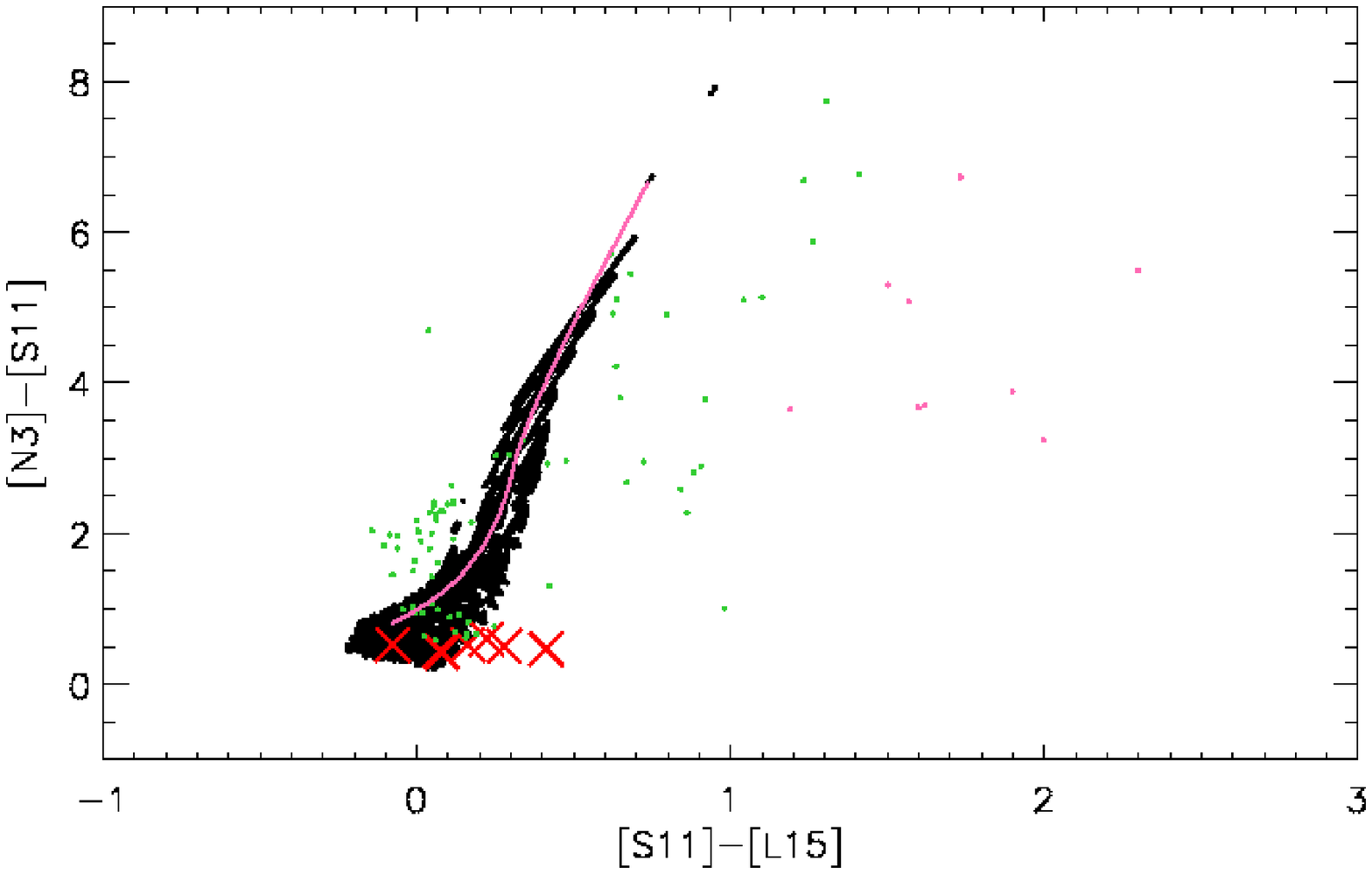}
  \ssp\caption[AKARI color-color diagram]{The AKARI [N3]--[S11] color vs [S11]--[L15] color, comparing the coverage of the model grid to the locations of spectroscopically classified stars from \citet{Kraemeretal2002} showing carbon--rich features (red crosses: NC, green dots: CE/CR, pink dots: CN/CT). The pink line shows the track for a single model (\teff=2600 K, $\log{g}=-0.9$, $R_{\rm in}=7R_{\rm star}$) with increasing optical depth.\label{fig:AKARICCD}}
\end{figure}

In this section, I provide color-color diagrams for the AKARI and WISE passbands generated from synthetic photometry of our model grid. \citet{Itaetal2008} presented the AKARI survey of the LMC and discussed color-color and color-magnitude diagrams. The AKARI survey is important for AGB stars because the S11 band directly samples the 11.3 \mic\ SiC feature in C--rich AGB stars as well as the $\sim$ 10 \mic\ silicate emission in O--rich AGB stars. The increase in feature strength, therefore, is reflected in the [S11]-[L15] color. \citet{Itaetal2008} found by generating synthetic photometry in the AKARI bands for the ISO-SWS spectra of \citet{Kraemeretal2002} that the change in the SiC feature strength manifests itself as first a blueward shift of the [S11]--[L15] color followed by a reddening. Fig. \ref{fig:AKARICCD} shows the \citet{Kraemeretal2002} sources classified as stars with carbonaceous molecular features (NC), stars showing carbon dust (CE/CR), and carbon--rich post-AGB objects (CN/CT) plotted on the grid of models. The model grid does not show the same blueward-redward shift as seen in the colors derived from the \citet{Kraemeretal2002} spectra. We note that in future studies, LMC data from AKARI will perhaps help constrain the role of SiC in our models better than SAGE data because of the direct detection of the 11.3 \mic\ feature. The [S11]--[L15] color is also a tracer of mass-loss.

\begin{figure}[!htb]
  \epsscale{1}
  \plotone{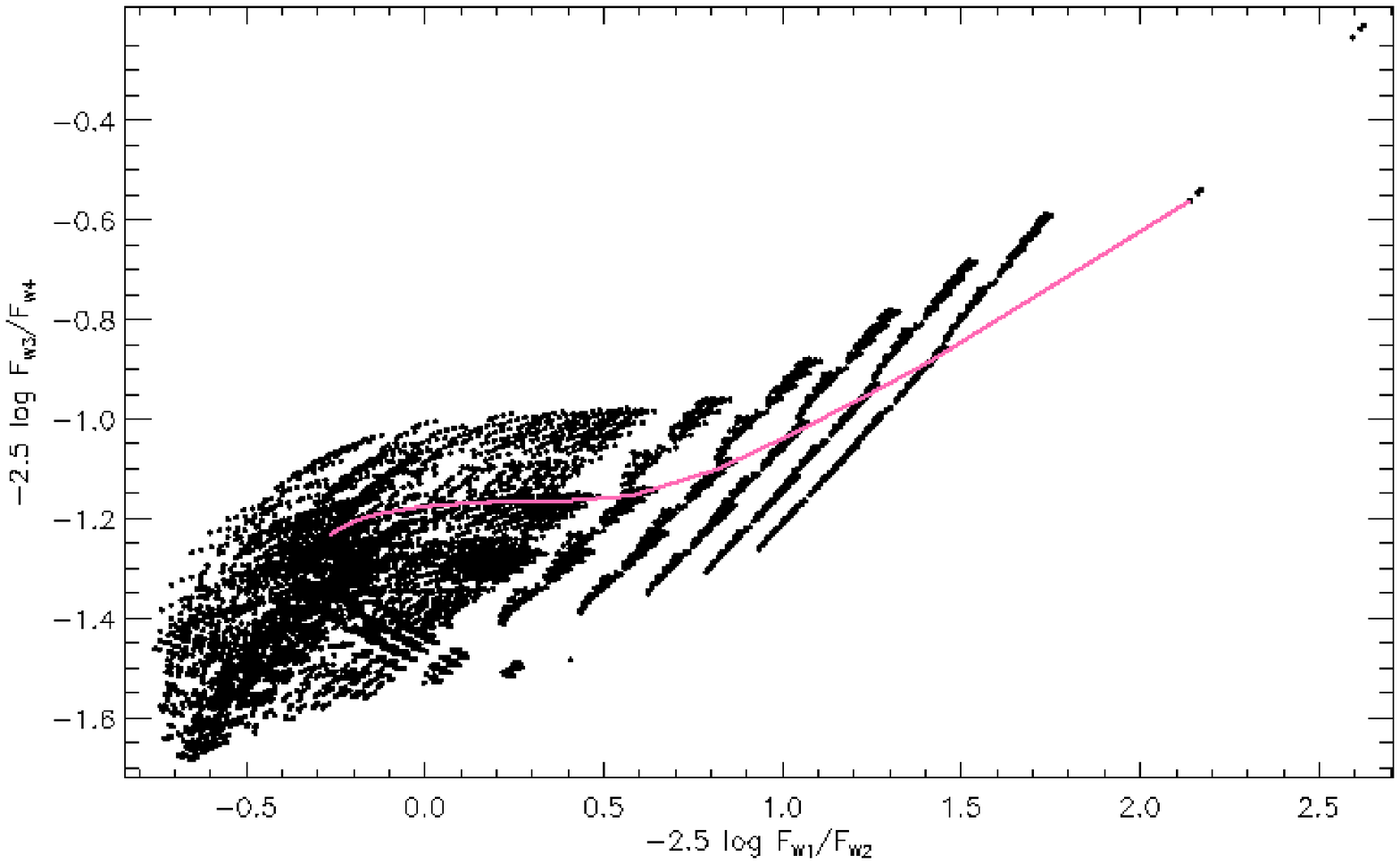}
  \ssp\caption[WISE color-color diagram]{The ratio of the fluxes in the W3 and W4 bands as a function of the flux ratio in the W1 and W2 bands. The pink line shows the track for a single model (\teff=2600 K, $\log{g}=-0.9$, $R_{\rm in}=7R_{\rm star}$) with increasing optical depth.\label{fig:WISECCD}}
\end{figure}

Fig. \ref{fig:WISECCD} shows the ratio of the model fluxes in the WISE W3 (12 \mic) and W4 (22 \mic) bands as a function of the corresponding ratios in the W1 (3.4 \mic) and W2 (4.6 \mic) bands. The WISE W3 band can directly detect the 11.3 \mic\ SiC feature and is therefore an important band for AGB studies. Similar to the AKARI S11 band, when WISE data is available, information from the W3 filter will help improve the treatment of SiC dust in our model grid.

From these diagrams, it is clear that overall, our models provide good coverage over a wide range of parameters. The phenomenon of variability can also be studied using our model grids. The availability of information about the SiC feature in the AKARI and WISE surveys will help constrain the behavior of the dust shell over the course of the pulsation, which will assist us improving our model prescriptions.

\subsection{SED fits}
In order to calculate mass-loss rates for all the SAGE AGB star candidates, we need to fit their photometry to that synthesized from the model grid. In order to verify this, I fit all four IRAC magnitudes and the MIPS 24 \mic\ magnitude of the SAGE data to the models as a first attempt at obtaining best-fits. I minimized $\chi^2$ for each source where
\ben
\chi^2 \equiv \displaystyle\sum_i{\left(m_i-m_{{\rm mod},i}\right)^2 \over \delta m_i^2}
\een
here, $m_i$ and $\delta m_i$ are the magnitude and error of the AGB candidate in the ith wavelength band, $m_{{\rm mod},i}$ are the corresponding magnitudes for the models, and the sum is over all four IRAC bands and the MIPS 24 band. In this section, I show some results from this simple fitting technique in order to demonstrate the utility of the model grid.

\subsubsection{Revisiting the model for \ogles}
In Chapter \ref{ch:cagbmodel}, I used the \citet{Gautschy2004} model photospheres to calculate radiative transfer models for the variable carbon star \ogles. In this chapter, I compare the predictions of this model with those obtained from our model grid using the \citet{Aringeretal2009} photospheres. The results are summarized in Table \ref{tab:modelgrid:ogle}. The model grid is able to predict a mass-loss rate very similar to that predicted by the \citet{Gautschy2004} models, and it also predicts a similar temperature for the dust at the inner radius. Both best-fit models correspond to the same \teff but different surface gravities. This latter effect is perhaps due to the fact that there are a limited number of $\log{g}$ values available for the \citet{Gautschy2004} models. The model in Chapter \ref{ch:cagbmodel} was constructed with due attention paid to the spectrum of \ogles. Despite simply fitting to the photometry of the source in this chapter, we have obtained a similar result for the mass-loss rate of the source. This demonstrates the effectiveness of the model grid.

\begin{deluxetable}{lll}
\setlength{\tabcolsep}{0.05in}
\tablewidth{0pt}
\tablecolumns{2}
\tablecaption{Comparison of \twodust\ fits to \ogles}
\tabletypesize{\footnotesize}
\tablehead{\colhead{Quantity\tablenotemark{1}} & \colhead{Chapter \ref{ch:cagbmodel}} & \colhead{This chapter}
}
\startdata
Photosphere model & \citet{Gautschy2004} & \citet{Aringeretal2009}\\
\teff	&	3000		& 3000\\
$L$ (\lsun) &	5670 & 6166 {\bf(C)}\\
$\log{g}$ & -0.51 {\bf(Ltd.)} & -0.2\\
$M$ (\msun) & 1.0 {\bf(Ltd.)} & 2.0 {\bf(Ltd.)}\\
$R_{\rm in}$ ($R_*$) &		6	&	4.5 {\bf(C)}\\
$\tau_{\rm 11.3~\mu m}$ &		0.25 &		0.2 {\bf(C)}\\
$\dot{M}$ (\msunperyr) & 2.57$\times 10^{-9}$	&	2.16$\times 10^{-9}$\\
$T(R_{\rm in})$ (K) &	1160	&	1140
\enddata
\sspf{\tablenotetext{1}{We denote by {\bf(Ltd.)} or {\bf(C)} the model parameter values that were not constrained well owing to limited availability of models or coarseness of grid, respectively.}}
\label{tab:modelgrid:ogle}
\end{deluxetable}


\section{Conclusions}
\label{sec:modelgrid:conclusions}
I constructed a grid of carbon star models using the radiative transfer code \twodust\ and for a range of various stellar and dust shell parameters. At present, the model grid only covers optical depths at 11.3 \mic\ up to 2, so it can not yet be used to model the most obscured stars. In the future the coverage will be extended to higher optical depths. We will also address the effects of grain chemistry, sizes and shapes and investigate the dependence of the models on the metallicity of the environment.

I provide synthetic photometry in the optical, 2MASS and {\it Spitzer} bands as well as in the AKARI and WISE bandpasses. The latter are chosen for the purpose of directly studying the variation of model parameters with the observed strength 11.3 \mic\ SiC feature. When AKARI and WISE data becomes available, this model grid will prove to be very helpful in determining the dust properties of and the mass-loss return from the evolved stars in these data sets. One of the aims of this thesis is to devise a general-use fitter that can be used for large sets of photometric data.

A concurrent effort to model O--rich AGB and red supergiant stars is being undertaken by Sargent et al. (in preparation). When combined with the C--rich models presented in this thesis, these models will enable the assessment of mass loss return from galaxy-wide point source catalogs from projects such as SAGE. While most extreme AGB stars are probably carbon stars, some are very bright OH/IR stars. As of yet, it is not possible with photometry alone to distinguish between these two chemical types. The two grids will be used in conjunction to define the separation of C--rich and O--rich sources in this extreme AGB star regime.

\chapter{Concluding Remarks} 
\label{ch:conclusions}
The main aim of this thesis was to be able to quantify the dust mass-loss return from asymptotic giant branch (AGB) stars to the Large Magellanic Cloud (LMC). This determination requires that we model the circumstellar dust shells around these stars individually, which is not easy considering the sheer number of AGB stars in the LMC. One way to achieve this is to generate a grid of models spanning a range of stellar and dust shell parameters typical of stars on the AGB which can then be fit to observed photometry in order to derive mass-loss rates. I have presented such a model grid in this thesis. These models are in progress of being extended as well as improved based on comparisons to data.

Using data from the SAGE survey \citep{sage1}, I first identified O--rich, C--rich and extreme AGB star candidates based on their locations in various near- and mid-infrared (IR) color-magnitude diagrams (CMDs). Knowledge of the distance to the LMC \citep[$\sim$50 kpc,][]{Feast99} enables us to also estimate the luminosities of these AGB candidates, allowing a study of the luminosity function of the sample. A future paper will address this topic in detail. SAGE data was also able, for the first time, to identify a faint, red population of O--rich AGB stars. 
The theoretical isochrones of \citet{Marigoetal} explain that the bright population corresponds to massive sources that undergo hot-bottom burning and stay O--rich, while stars in the faint population (which consists of about 80\% of the O--rich stars) undergo third dredge-up and eventually turn C--rich. I compare the spectral energy distributions (SEDs) of the SAGE AGB candidates with model photospheres in order to calculate their mid-IR excesses, and use these excesses in turn to estimate the color temperature and optical depth of their dust shells. I find that the IR excess increases with luminosity for all three types of sources in all four IRAC bands as well as the MIPS24 band. The color temperature shows a decrease with increasing excess, and the optical depth shows an increasing trend. I conclude that the IR excess is a good proxy for the mass-loss rates of AGB stars. I quantify this relation by deriving a fit to the excess--mass-loss rate relation for a handful of well-studied AGB stars and using this relation to calculate mass-loss rates for all the sources in order to estimate the total dust mass loss return. I find that the dust injection rate from AGB stars is about (5.9--13)$\times 10^{-5}$ \msunperyr.

In order to constrain the dust properties that will be incorporated into the carbon star model grid, I perform \twodust\ radiative transfer modeling of the dust around \ogles, a variable carbon star for which both photometric and spectroscopic data is available through the SAGE and SAGE-Spec (Kemper et al. 2009, in preparation) studies, respectively. The source has a prominent 11.3 \mic\ SiC emission feature, which I use to constrain the ratio of SiC to amorphous carbon used in the modeling. I find that with 15\% of SiC, I find a good fit to the spectrum of \ogles, resulting in a dust mass-loss rate of $2.6\times 10^{-9}$ \msunperyr. We also model the CO and the 13.7 \mic\ \acet\ feature in the circumstellar shell with slab models \citep{Matsuuraetal2002}. From the strength of the \acet\ feature, we estimate a gas mass-loss rate in the range $1\times 10^{-7}$--$2\times 10^{-6}$ \msunperyr, which depends heavily on the location of the line-forming region.

Finally, I present a grid of carbon star models constructed with the radiative transfer code \twodust\ for a range of various stellar and dust shell parameters. I provide synthetic photometry in the 2MASS and {\it Spitzer} bands as well as the filters used in the AKARI and WISE surveys. The color-color and color-magnitude diagrams show that despite our simplifying assumptions we are able to reproduce the general features of the SAGE photometry well. A preliminary fitting scheme is then used to find the best-fit model to the photometry of \ogles, and I find that this best-fit model agrees well with that predicted in the Chapter \ref{ch:cagbmodel}. At present, the model grid reaches to an optical depth of 2 at 11.3 \mic, so it can not be used to fit the most obscured extreme AGB stars. In the future this coverage will be extended to higher optical depths. Along with a companion grid of O--rich AGB models (Sargent et al., in preparation), these models should help distinguish O--rich and C--rich features in extreme AGB stars. We will also consider the effect of changing the grain chemistry, grain size and non-spherical shape. The AKARI and WISE bandpasses contain filters that can directly detect the 11.3 \mic\ SiC emission, making them important for AGB star studies. When these data become available, we will be able to better constrain the dust properties used in the model grid. The model grid will also incorporate a general-use fitter that can be used for large sets of photometric data, enabling accurate determinations of mass loss return from photometric data of entire AGB populations in galaxies.

\addcontentsline{toc}{chapter}{Bibliography}
\bibliographystyle{apj}
\ssp{\bibliography{thesis}}

\noindent{\bf\huge Acknowledgements}\\

\vspace{0.4in}
\ssp{
I express my sincere gratitude to my advisor, Dr. Margaret Meixner, without whom this thesis would not have come into being. Margaret has offered me immense support and guidance throughout the duration of this project. She has always provided words of encouragement when I hit a roadblock, making sure I don't get bogged down by the details or become distracted from the big picture. Her perseverance and zeal will continue to inspire me in the years to come. I would also like to thank my co-advisor, Prof. Holland Ford, for keeping track of my academic progress and wading through the mountains of paperwork I had him sign every semester.

Thanks to the members of the Mega-SAGE evolved star group for helpful advice and discussions along the way -- 
Uma Vijh for getting me started with the SAGE data, Kevin Volk for his insight on the model grid, Bob Blum for guiding me through my first observing run, Ciska Kemper, Angela Speck and Xander Tielens for valuable comments on dust properties, Mikako Matsuura for her help with modeling the circumstellar gas and Maria-Rosa Cioni for discussions on the AGB luminosity function. I could not have developed some of the techniques in this thesis without advice and comments from Ben Sargent, whose experience as a recent Ph.D. has also aided my thesis preparation. I have enjoyed interacting with other members of the SAGE team from around the world, and I hope to continue working with them in the future. Thanks to Bernie Shiao, the ``lifeguard" on the shores of the ocean of SAGE data, for always bring on duty and helping me with all of my SQL queries with a speed and clarity that I haven't encountered elsewhere. Thanks to Toshi Ueta for answering emails about the \twodust\ code, and to Megan Sosey for always being an email away if I needed urgent \twodust\ advice, to Justin Rogers and Seamus Riley for gracious use of their high-RAM computers and to Lynn Barlson and Dave Riebel, my SAGE grad student buddies, for sharing complaints about the latest crazy STScI rules and/or regulations we encountered.

I cannot begin to thank my parents for all that they have done for me, not least of which was to give me the freedom to choose my career. I am also indebted to my science teacher in middle school for instilling the basics of the scientific method, and to my chemistry teacher in high school for keeping me interested in science. My brothers have helped me become the person I am today, and I thank them for their continued support.

The transition into a foreign country is usually difficult, but it wasn't for me, thanks to Janet, Pam, Carm, Norma, Connie, Brian and everyone else in the department office, who have always treated me like family. My classmates also played a huge part, so I would like to thank them all, especially my housemates Wass and Soo.~~~~~~~~~~~This sentence is intentionally ``Boyered" for Mike. Thanks to Xuemei, Binquan, Kerry and Yawei for putting up with our shenanigans in the office, to Urmila for helping me through the first two weeks of JHU paperwork and to Marci for helping us Indians with our first snowball fight. Special shout-outs to some fellow 2714 survivors: DJ Rowdy Reid for passing the baton on to me, Dan for being himself and for the drums,  
Marin for tolerating my cooking the first year, Sean for GTP and Annie for being our indispensable fifth roommate. Thanks to Miriam for sharing a family song, and to Michele for the Italian lessons.

No IG acknowledgement section would be complete without a huge thank-you to all the members of IGwAD for making grad school fun. Thanks especially to Erik (who, besides creating IGwAD, has also shared more houses, microphones and stories with me than he'd have liked to); Kevin for his Monday morning theme emails and enthusiastic return trips to Baltimore and to the ``newer" IGwAD members -- ``Derek" for Cappallo's Theorem, Slahmahs for being a Leroy-magnet, Greg for not making it onto the A-list, and to our aggressively loyal manager, Jenny, a single stare from whom reduces a man bad-mouthing IGwAD to ashes (Or forces him to pay the cover twice, I don't remember). Thanks to Gabe for being an Awful Little Boy and to Jane for her perfect piano recitals... during practice. Big props to former Aussieboy ({\em name redacted}) for playing fiddle, for Egypt \& Iceland, and for being the 24-hour human Google conduit. Thanks to everyone at Fraziers for their support and for being interesting -- Marty, Joel The Godfather, Josh The Bartender, PCC, The OMK, Morgan and Andrew. Thanks to our fans for the t-shirts and support!

Finally, thanks to the extended P\&A family for being the best collection of grads \& friends of grads. Thanks especially to Hilbert for many things, including the printing of a huge poster in 24 Letter-sized pieces; Jim ``Meal Ticket" Felton for innumerable rides to dinner and for all the \LaTeX\ help; Ricardo for the non-stop apologies; Jess for ``Silly Boy", Stuart for the Family Guy quotes and to Brigid for the temporary housing. I have made many good friends over the years, and I've probably left a few out, but I'm going to miss all of you. 

\vspace{.2in}
\begin{quotation}
\ssp{
{\it "Well, I've really enjoyed forgetting. When I first come to a place, I notice all the little details. I notice the way the sky looks. The color of white paper. The way people walk. Doorknobs. Everything. Then I get used to the place and I don't notice those things anymore. So only by forgetting can I see the place again as it really is."}\\

\centerline{ -- David Byrne, {\bf{\em True Stories}}}
}
\end{quotation}
}
\begin{vita}
\ssp 
Sundar Srinivasan was borned on July 23, 1976 in Bombay, India, where he attended Veekays English School in Vikhroli and St. Joseph's School in Malad. He attended high school in Madras, graduating in 1994 from The Hindu Senior Secondary School, Indra Nagar, Adyar. Returning to Bombay for college, he joined The Guru Nanak Khalsa College of Arts, Science \& Commerce, Matunga, choosing in his second year to major in physics much to the surprise of his high school chemistry teacher. Graduating with a Bachelor of Science (Honors, First Class) in Physics in 1997, he was accepted into the Integrated Ph.D. program in Physics at the Indian Institute of Science, Bangalore, where he received his Master's degree in Physics in 2000.

Sundar then came to the United States for graduate studies, joining The Henry A. Rowland Department of Physics and Astronomy at The Johns Hopkins University, Baltimore, MD, as a graduate student in astronomy and astrophysics. While at JHU, his work as a teaching assistant was rewarded with the Rowland Prize for Innovation and Excellence in Teaching in 2001. He was awarded the Masters of Arts degree in 2005, and he received his Doctor of Philosophy degree in Physics in 2009 for his thesis work on estimating the mass-loss return from asymptotic giant branch stars to the Large Magallanic Cloud, supervised by Dr. Margaret Meixner (Space Telescope Science Institute).

Starting November 2009, Sundar will be working as a postdoctoral research fellow with the Herschel Telescope KINGFISH Survey Team at the Institut d'Astrophysique de Paris.
\end{vita}
\end{document}